\documentclass[10pt,journal,compsoc]{IEEEtran}
\ifCLASSOPTIONcompsoc
  \usepackage[nocompress]{cite}
\else
  \usepackage{cite}
\fi

\usepackage{graphicx}
\usepackage{balance}  % for  \balance command ON LAST PAGE  (only there!)
\usepackage{defpaper2e}
\usepackage{multirow}
\usepackage{amssymb}
\usepackage{xcolor}
\usepackage[outdir=./]{epstopdf}
\usepackage{amsmath}

\newcommand{\Wendy}[1]{{{\textcolor{black}{\textbf{Wendy:}}}{\textcolor{red}{\textbf{#1}}}}}
\newcommand{\Bo}[1]{{{\textcolor{black}{\textbf{Bo:}}}{\textcolor{green}{\textbf{#1}}}}}

\begin{document}

% ****************** TITLE ****************************************

\title{Integrity Authentication for SQL Query Evaluation on Outsourced Databases: A Survey}

\author{Bo~Zhang,~Boxiang~Dong,~Wendy~Hui~Wang
\IEEEcompsocitemizethanks{\IEEEcompsocthanksitem B. Zhang and H.W. Wang are with the Department
of Computer Science, Stevens Institute of Technology, Hoboken, NJ, USA.\protect\\
Email: bzhang41@stevens.edu, hwang4@stevens.edu
\IEEEcompsocthanksitem B. Dong is with the Department of Computer Science, Montclair State University, Montclair, NJ, USA. \protect\\
Email: dongb@montclair.edu}% <-this % stops an unwanted space
}

\nop{
\author{\IEEEauthorblockN{Bo Zhang}
\IEEEauthorblockA{Department of Computer Science \\
Stevens Institute of Technology\\
New Jersey, Hoboken 07030\\
Email: bzhang41@stevens.edu}
\and
\IEEEauthorblockN{Boxiang Dong}
\IEEEauthorblockA{Department of Computer Science\\
Montclair State University\\
New Jersey, Montclair, USA\\
Email: dongb@montclair.edu}
\and
\IEEEauthorblockN{Wendy Hui Wang}
\IEEEauthorblockA{Department of Computer Science\\
Stevens Institute of Technology\\
New Jersey, Hoboken 07030\\
Email: hwang4@stevens.edu}}
}

\IEEEtitleabstractindextext{
\begin{abstract}
Spurred by the development of cloud computing, there has
been considerable recent interest in the {\em Database-as-a-Service} (DaaS) paradigm. Users lacking in expertise or computational resources can outsource their data and database management needs to a third-party service provider. Outsourcing, however, raises an important issue of {\em result integrity}: how can the client verify  with lightweight overhead that the query results returned by the service provider are correct (i.e., the same as the results of query execution locally)?
This survey focuses on categorizing and reviewing the progress on the current approaches for result integrity of SQL query evaluation in the DaaS model. 
The survey also includes some potential future research directions for result integrity verification of the outsourced computations.
\end{abstract}

}
\maketitle
\IEEEdisplaynontitleabstractindextext
\IEEEpeerreviewmaketitle

\section{Introduction}

The amount spent by corporations, non-profit organizations, and government agencies in implementing and supporting data management and analytics is considerable. Globally 3500 enterprises spend, on average, \$664,000 annually on data management  \cite{weisman-2001}. 
Due to the fast growth of data volumes, the scale of data management systems 
 is increasingly crossing the petabyte barrier \cite{Abadi-IEEEDEbulltin09,yahoo,Monash-barrier}. 
Unfortunately, the ability
of organizations to support effective and efficient  data management typically lags behind their ability to collect and store the data. 
The solution of in-house data analytics software is not satisfactory, as it either may not be able to address users' specific data analysis needs or fail to deliver data management services that are easily deployable. On the other hand, hiring in-house data management professionals is not affordable by small and medium-sized organizations that have limited financial budget. 

Explosive development of the Internet and advances in networking technology have fueled a new computing paradigm called Database-as-a-Service (DaaS) \cite{hacigumus2002providing}. In the DaaS paradigm, the clients  who own large volumes of data but lack resources to do data management themselves outsource their data as well as the data  analysis to a third-party database service provider. The service provider offers adequate hardware, software, and network resources to host the clients' databases, and provides technical supports of data management services, including data access, query processing, and dealing with updates. 
%Outsourcing data mining computations to a computationally powerful service provider 
Several industrial organizations such as Amazon, Google
and Microsoft are providing cloud-based database services in various forms. For example, Amazon Web Services (AWS) provides computation capacity and data storages via Amazon Elastic Compute Cloud (EC2) \cite{amazon} and Simple Storage Service (S3) \cite{amazons3}. Google 
provides Cloud SQL \cite{googlecloudsql}, a fully-managed database service for relational PostgreSQL and MySQL databases.  Microsoft provides cloud database services \cite{mscloud} on Azure cloud \cite{windowsazure}. 
By using
these services, the clients can exploit the benefit of mass storage, accelerated processing capacity, and sophisticated data management and analytics at a low cost. 

Outsourcing database management to a computationally powerful service provider enables to achieve sophisticated analysis on large volumes of data in a cost effective way. With this architecture, however, the client no longer has direct control over the outsourced data and computations. 
%Clients are willing to give up their control based on the implicit assumption that the services provide professional levels of security and integrity. However, a careful examination of the customer agreement of those services reveals that in fact clients themselves are responsible for adequate security protection\footnote{The customer agreement for Amazon Web Services \cite{amazon}, a leader of Cloud computing, reveals that Amazon does not provide even the basic guarantee of security protection, let alone the guarantee of correct computations and result integrity: ``We...make no representations or warranties of any kind...that the service or third party content will be uninterrupted, error free or free of harmful components, or that any content...will be secure or not otherwise lost or damaged.''}. 
Without any security guarantee from the service provider, clients have little faith in the security of the received outsourced services. 
One of the important security issues of DaaS paradigm is the {\em result integrity} of the outsourced computations. 
In many
DaaS paradigms, the client typically has a pay-per-use arrangement with the service provider, where the service 
provider charges the client proportional to the effort involved in the computations. However, the service
providers in practice rarely provide sufficient details of the computation efforts that are charged for service
fees\footnote{Amazon Elastic Compute Cloud (EC2) \cite{amazon} charges the users by the number of computing hours and the size of transferred data. It
does not provide any detail of the involved computations (e.g., bandwidth, computing cycles, etc.) in the bills. Many online discussions
\cite{amazon-hiddenfee,amazon-cloudprice} have revealed that there are hidden service fees related to the computation details, e.g., how data is located on the servers.}. Due to the lack of transparency in the current outsourcing services, the service provider is incentivized to improve its revenue by computing with less resource while charging for more. As an example, project managers
from SETI@home \cite{seti}, a well-known volunteer computing grid, have reportedly \cite{setihome} uncovered attempts by
some users “to forge the amount of time they have donated in order to move up on the Web listings of top contributors”.
Indeed, the director for SETI@home is quoted \cite{setihome} saying that such cheaters comprise roughly 1\% of
their users. Besides cheating on computations, there are many other reasons that a service provider may return
incorrect query results. For instance, the data or the computations may be corrupted due to malware or security
break-ins. Or a malicious insider (e.g., a disgruntled employee) could modify the program and/or the query 
results. Given the fact that many data analytics applications (e.g., fraud detection and business intelligence) are 
mission critical, it is important to provide efficient and practical methods to enable the client to verify whether
the service provider returned correct results of the outsourced database services.

%The client  may need to ensure that the results returned by the server are correct. Providing such a guarantee is important if the client does not fully trust the service provider, or even if the client is concerned about the possibility of server errors or external compromise \cite{vSQL}.  

Intuitively, the general-purpose protocols for verifiable computations \cite{gennaro2010non,parno2013pinocchio} can support the verification of any arbitrary query. However, due to the generality, these approaches can incur  excessive proof construction overhead at the 
server side \cite{papadopoulos2014taking}. In the last two decades, a large variety of efficient authentication methods have been proposed for specific types of SQL queries. 
A brief survey of a subset of these methods was presented in \cite{bajaj2013correctdb}. 
In this survey article, we will give a comprehensive overview of
the existing approaches for result integrity authentication of the DaaS paradigm. 
By doing the survey, we hope to provide a useful resource for both database  and security communities. 

The rest of the survey is organized as follows. Section \ref{sc:overview} overviews the DaaS framework, the integrity verification goals, and a categorization of the existing authentication methods for SQL query evaluation. Section \ref{sc:ads} presents two authenticated data structures that are popularly used in authentication. Section \ref{sc:crytobag} introduces the cryptographic background of a set of authentication approaches. Section \ref{sc:vo}  discusses the existing authentication techniques that return {\em deterministic} integrity guarantees of the result correctness. Section \ref{sc:prob} presents the methods that can return {\em probabilistic} integrity guarantee. Section \ref{sc:freshness} discusses the existing authentication methods that verify result freshness. Section \ref{sc:additionalsecurity} presents the works that consider result integrity with additional security features (e.g., access control and privacy). Finally, Section \ref{sc:conclusion} concludes the survey and discusses the possible research directions. 

\nop{
\begin{table*}[!htb]
\begin{center}
\begin{tabular}{ |c|c|c|c|c|c| } 
 \hline
 Paper & Range queries & Join & Aggregation & MapReduce & Authentication method\\ \hline
 \cite{devanbu2003authentic,li2006dynamic}  & X &  &  & & VO construction based on tree-type ADS\\ \hline
 \cite{zhang2015integridb}  & X & X & X & & VO construction based on tree-type ADS\\ \hline
 \cite{pang2004authenticating,narasimha2006authentication,cheng2006authenticating}  &  & X & & & VO construction based on signature aggregation\\ \hline
 \cite{yang2009authenticated}  &  & X &  & & VO construction based on tree-type ADS\\ \hline
 \cite{di2014optimizing}  &  & X &  & & Fake tuples\\ \hline
 \cite{li2010authenticated} &  &  & X & & VO construction based on tree-type ADS\\ \hline
 \cite{ulusoy2015trustmr,wei2009securemr,wang2013result} &  &  &  & X & Replication\\ \hline
\end{tabular}
\end{center}
\caption{\label{table:relatedwork} Related Work}
\end{table*}
}

\vspace{-0.05in}
\section{Overview}
\label{sc:overview}

In this section, we overview the preliminaries of the survey. We also provide a categorization of the existing authentication methods for outsourced SQL query evaluation. 

\vspace{-0.05in}
\subsection{Database-as-a-Service (DaaS) Paradigm}
The concept of the {\em database-as-a-service} (DaaS) paradigm is first defined  in \cite{hacigumus2002providing}. A typical database-as-a-service (DaaS) paradigm consists of three entities: 
(1) the {\em data owner} who possesses a large collection of records $D$. Due to the lack of resources, the data owner outsources $D$ to a service provider; 
(2) the {\em service provider} (SP) who provides storage and data management services for the clients; and 
(3) the {\em client} who sends queries to SP. SP executes the queries on the outsourced data, and returns the query results to  the client. The client may have limited computational power (e.g., she may launch the verification on  mobile devices). 
The data owner and the client may be the same entity. In this survey, we mainly focus on relational databases and SQL queries, including range queries, joins, and aggregate queries.

\vspace{-0.05in}
\subsection{Verification Goal}

One of the serious concerns of the DaaS paradigm is that the SP may return incorrect query evaluation results, due to various reasons such as server errors or external compromise \cite{vSQL}. 
The goal of result integrity authentication is to enable the client to verify that for any query $Q$, if the results $R$ returned by SP is the same as by executing $Q$ on the dataset $D$ locally, i.e., if $R = Q(D)$. There are four different authentication goals: {\em authenticity}, {\em soundness}, {\em completeness}, and {\em freshness}. 
\begin{itemize}
\item {\bf Authenticity.} The results $R$ must be generated from the outsourced dataset $D$ and not being tampered with. 
 
\item{\bf Soundness.} All the records in $R$ must satisfy $Q$. Intuitively, the soundness of $R$ can be measured by the {\em precision} of $R$:
\[Precision(R) = \frac{|R\cap Q(D)|}{|R|}.\] 
 The query is sound if precision is 1. 

\item{\bf Completeness.} $R$ must include all records that satisfy $Q$. The completeness of $R$ 
can be measured by  {\em recall} of $R$:
\[ Recall(R) = \frac{|R\cap Q(D)|}{|Q(D)|}.\] 
The query is complete if recall is 1.
 
\item{\bf Freshness.} When there are updates on the outsourced dataset $D$, the query $Q$ must be
executed against the latest data records, rather than out-of-date versions. The returned results $R$ only includes the records from the latest dataset. 
\end{itemize}

\vspace{-0.05in}
\subsection{A Categorization of Integrity Authentication Techniques}

\begin{table*}[!htb]
 \centering
\begin{tabular}{|*{15}{c|}} 
\hline
\multirow{3}{*}{Paper} & \multicolumn{3}{|c}{Query Type}  
& \multicolumn{4}{|c|}{Verification Goal}
& \multicolumn{6}{|c|}{Authentication Method}
%& \multirow{3}{*}{}
& \multirow{3}{*}{Update} \\ \cline{2-14}

%second row
\multicolumn{1}{|c|}{} & 
\multirow{2}{*}{Range} & \multirow{2}{*}{Join} &
\multirow{2}{*}{Agg.} &
%\multicolumn{1}{|c}{Range} & \multicolumn{1}{|c}{Join} & \multicolumn{1}{|c}{Agg.} & 
\multirow{2}{*}{$\mathcal{A}$} & \multirow{2}{*}{$\mathcal{S}$} &
\multirow{2}{*}{$\mathcal{C}$} & \multirow{2}{*}{$\mathcal{F}$} &
\multicolumn{4}{|c|}{Deterministic Approach} &
\multicolumn{2}{|c|}{Probabilistic Approach} &
%\multicolumn{1}{|c|}{Integrity Guarantee} &
\multicolumn{1}{|c|}{}
\\ \cline{9-14}

%third row
\multicolumn{1}{|c|}{} & 
\multicolumn{1}{|c}{} & \multicolumn{1}{|c}{} & \multicolumn{1}{|c}{} & 
\multicolumn{1}{|c}{} & \multicolumn{1}{|c}{} &
\multicolumn{1}{|c}{} & \multicolumn{1}{|c}{} &
\multicolumn{1}{|c}{Tree} &
\multicolumn{1}{|c}{Sign.} & \multicolumn{1}{|c}{Acc.} & \multicolumn{1}{|c}{Hardware} &
\multicolumn{1}{|c}{Checkpoint} & \multicolumn{1}{|c}{Interactive-proof} &
%\multicolumn{1}{|c}{Tree.} & \multicolumn{1}{|c}{Sign.} & %\multicolumn{1}{|c}{Acc.} & \multicolumn{1}{|c}{FT.} & 
\multicolumn{1}{|c|}{} \\ \hline 

%1
\multicolumn{1}{|c}{\cite{devanbu2003authentic}} 
& \multicolumn{1}{|c}{X} & \multicolumn{1}{|c}{X} & \multicolumn{1}{|c}{} & \multicolumn{1}{|c}{X} & \multicolumn{1}{|c}{X} & \multicolumn{1}{|c}{X} & \multicolumn{1}{|c}{} &
\multicolumn{1}{|c}{X} & \multicolumn{1}{|c}{} & \multicolumn{1}{|c}{} & \multicolumn{1}{|c}{} &
\multicolumn{1}{|c}{} & \multicolumn{1}{|c|}{} & \multicolumn{1}{|c|}{} \\ \hline

\multicolumn{1}{|c}{\cite{yang2009authenticated}} 
& \multicolumn{1}{|c}{} & \multicolumn{1}{|c}{X} & \multicolumn{1}{|c}{} & \multicolumn{1}{|c}{X} & \multicolumn{1}{|c}{X} & \multicolumn{1}{|c}{X} & \multicolumn{1}{|c}{} &
\multicolumn{1}{|c}{X} & \multicolumn{1}{|c}{} & \multicolumn{1}{|c}{} & \multicolumn{1}{|c}{} &
\multicolumn{1}{|c}{} & \multicolumn{1}{|c|}{} & \multicolumn{1}{|c|}{} \\ \hline

%2
\multicolumn{1}{|c}{\cite{pang2004authenticating,4812487,Mouratidis:2009:PMD:1541533.1541543}} 
& \multicolumn{1}{|c}{X} & \multicolumn{1}{|c}{} & \multicolumn{1}{|c}{} & 
\multicolumn{1}{|c}{X} & \multicolumn{1}{|c}{X} &
\multicolumn{1}{|c}{X} & \multicolumn{1}{|c}{} &
\multicolumn{1}{|c}{X} & \multicolumn{1}{|c}{} & \multicolumn{1}{|c}{} & \multicolumn{1}{|c}{} & \multicolumn{1}{|c}{} & \multicolumn{1}{|c|}{} & \multicolumn{1}{|c|}{X} \\ \hline

\multicolumn{1}{|c}{\cite{li2006dynamic}} 
& \multicolumn{1}{|c}{X} & \multicolumn{1}{|c}{} & \multicolumn{1}{|c}{} & 
\multicolumn{1}{|c}{X} & \multicolumn{1}{|c}{X} &
\multicolumn{1}{|c}{X} & \multicolumn{1}{|c}{X} &
\multicolumn{1}{|c}{X} & \multicolumn{1}{|c}{} & \multicolumn{1}{|c}{} & \multicolumn{1}{|c}{} & \multicolumn{1}{|c}{} & \multicolumn{1}{|c|}{} & \multicolumn{1}{|c|}{X} \\ \hline

%3
\multicolumn{1}{|c}{\cite{li2010authenticated}} 
& \multicolumn{1}{|c}{X} & \multicolumn{1}{|c}{} & \multicolumn{1}{|c}{X} & 
\multicolumn{1}{|c}{X} & \multicolumn{1}{|c}{X} &
\multicolumn{1}{|c}{X} & \multicolumn{1}{|c}{} &
\multicolumn{1}{|c}{X} & \multicolumn{1}{|c}{} & \multicolumn{1}{|c}{} & \multicolumn{1}{|c}{} &
\multicolumn{1}{|c}{}  & \multicolumn{1}{|c|}{} & \multicolumn{1}{|c|}{X} \\ \hline

\nop{
%4
\multicolumn{1}{|c}{\cite{yang2009authenticated}} 
& \multicolumn{1}{|c}{X} & \multicolumn{1}{|c}{X} & \multicolumn{1}{|c}{} & 
\multicolumn{1}{|c}{X} & \multicolumn{1}{|c}{X} &
\multicolumn{1}{|c}{X} & \multicolumn{1}{|c}{} &
\multicolumn{1}{|c}{X} & \multicolumn{1}{|c}{} & \multicolumn{1}{|c}{} & \multicolumn{1}{|c}{} &
\multicolumn{1}{|c}{}  & \multicolumn{1}{|c|}{} & \multicolumn{1}{|c|}{} \\ \hline

%5
\multicolumn{1}{|c}{\cite{Wei2014}} 
& \multicolumn{1}{|c}{X} & \multicolumn{1}{|c}{X} & \multicolumn{1}{|c}{} & 
\multicolumn{1}{|c}{X} & \multicolumn{1}{|c}{X} & 
\multicolumn{1}{|c}{X} & \multicolumn{1}{|c}{} &
\multicolumn{1}{|c}{X} & \multicolumn{1}{|c}{} & \multicolumn{1}{|c}{} & \multicolumn{1}{|c}{} & \multicolumn{1}{|c}{}  & \multicolumn{1}{|c|}{} & \multicolumn{1}{|c|}{X} \\ \hline

%7
\multicolumn{1}{|c}{\cite{vSQL}} 
& \multicolumn{1}{|c}{X} & \multicolumn{1}{|c}{X} & \multicolumn{1}{|c}{X} & 
\multicolumn{1}{|c}{X} & \multicolumn{1}{|c}{X} &
\multicolumn{1}{|c}{X} & \multicolumn{1}{|c}{} &
\multicolumn{1}{|c}{} & \multicolumn{1}{|c}{X} & \multicolumn{1}{|c}{X} &
\multicolumn{1}{|c}{} & \multicolumn{1}{|c|}{} & \multicolumn{1}{|c|}{X} \\ \hline
}

%6
\multicolumn{1}{|c}{\cite{Nuckolls2005}} 
& \multicolumn{1}{|c}{X} & \multicolumn{1}{|c}{} & \multicolumn{1}{|c}{} & 
\multicolumn{1}{|c}{X} & \multicolumn{1}{|c}{X} & 
\multicolumn{1}{|c}{X} & \multicolumn{1}{|c}{} &
\multicolumn{1}{|c}{X} & \multicolumn{1}{|c}{} & \multicolumn{1}{|c}{} & \multicolumn{1}{|c}{} & \multicolumn{1}{|c}{} & \multicolumn{1}{|c|}{} & \multicolumn{1}{|c|}{} \\ \hline

\nop{
\multicolumn{1}{|c}{\cite{pang2004authenticating,narasimha2006authentication}} 
& \multicolumn{1}{|c}{} & \multicolumn{1}{|c}{} & \multicolumn{1}{|c}{} & \multicolumn{1}{|c}{} & \multicolumn{1}{|c}{} &
\multicolumn{1}{|c}{X} & \multicolumn{1}{|c}{} &
\multicolumn{1}{|c}{} & \multicolumn{1}{|c}{} & \multicolumn{1}{|c}{} &
\multicolumn{1}{|c}{} & \multicolumn{1}{|c}{} & \multicolumn{1}{|c|}{} & \multicolumn{1}{|c|}{} \\ \hline
}

%9
\nop{
\multicolumn{1}{|c}{\cite{nguyenverification, goodrich2012efficient}} 
& \multicolumn{1}{|c}{X} & \multicolumn{1}{|c}{} & \multicolumn{1}{|c}{} & \multicolumn{1}{|c}{X} & \multicolumn{1}{|c}{X} &
\multicolumn{1}{|c}{X} & \multicolumn{1}{|c}{} &
\multicolumn{1}{|c}{X} & \multicolumn{1}{|c}{} & \multicolumn{1}{|c}{X} &
\multicolumn{1}{|c}{} & \multicolumn{1}{|c}{Deterministic} & \multicolumn{1}{|c|}{} & \multicolumn{1}{|c|}{X} \\ \hline
}

%10
\multicolumn{1}{|c}{\cite{mykletun2006authentication}} 
& \multicolumn{1}{|c}{X} & \multicolumn{1}{|c}{} & \multicolumn{1}{|c}{} & \multicolumn{1}{|c}{X} & \multicolumn{1}{|c}{X} &
\multicolumn{1}{|c}{} & \multicolumn{1}{|c}{} &
\multicolumn{1}{|c}{} & \multicolumn{1}{|c}{X} & \multicolumn{1}{|c}{} & \multicolumn{1}{|c}{} & 
\multicolumn{1}{|c}{} & \multicolumn{1}{|c|}{} & \multicolumn{1}{|c|}{X} \\ \hline

%13
\multicolumn{1}{|c}{\cite{pang2005verifying,Narasimha2006}} 
& \multicolumn{1}{|c}{X} & \multicolumn{1}{|c}{X} & \multicolumn{1}{|c}{} & \multicolumn{1}{|c}{X} & \multicolumn{1}{|c}{X} &
\multicolumn{1}{|c}{X} & \multicolumn{1}{|c}{} &
\multicolumn{1}{|c}{} & \multicolumn{1}{|c}{X} & \multicolumn{1}{|c}{} & \multicolumn{1}{|c}{} &
\multicolumn{1}{|c}{} & \multicolumn{1}{|c|}{} & \multicolumn{1}{|c|}{X} \\ \hline

\multicolumn{1}{|c}{\cite{Pang:2009:SVO:1687627.1687718}} 
& \multicolumn{1}{|c}{X} & \multicolumn{1}{|c}{X} & \multicolumn{1}{|c}{} & \multicolumn{1}{|c}{X} & \multicolumn{1}{|c}{X} &
\multicolumn{1}{|c}{X} & \multicolumn{1}{|c}{X} &
\multicolumn{1}{|c}{} & \multicolumn{1}{|c}{X} & \multicolumn{1}{|c}{} & \multicolumn{1}{|c}{} &
\multicolumn{1}{|c}{} & \multicolumn{1}{|c|}{} & \multicolumn{1}{|c|}{X} \\ \hline

%14
\multicolumn{1}{|c}{\cite{cheng2006authenticating}} 
& \multicolumn{1}{|c}{X} & \multicolumn{1}{|c}{} & \multicolumn{1}{|c}{} & \multicolumn{1}{|c}{X} & \multicolumn{1}{|c}{X} &
\multicolumn{1}{|c}{X} & \multicolumn{1}{|c}{} &
\multicolumn{1}{|c}{} & \multicolumn{1}{|c}{X} & \multicolumn{1}{|c}{} & \multicolumn{1}{|c}{} &
\multicolumn{1}{|c}{}  & \multicolumn{1}{|c|}{} & \multicolumn{1}{|c|}{} \\ \hline

%23
\multicolumn{1}{|c}{\cite{Xie:2008:PFG:1353343.1353384}} 
& \multicolumn{1}{|c}{} & \multicolumn{1}{|c}{} & \multicolumn{1}{|c}{} & \multicolumn{1}{|c}{} & \multicolumn{1}{|c}{} &
\multicolumn{1}{|c}{} & \multicolumn{1}{|c}{X} &
\multicolumn{1}{|c}{} & \multicolumn{1}{|c}{X} & \multicolumn{1}{|c}{} & \multicolumn{1}{|c}{} &
\multicolumn{1}{|c}{X} & \multicolumn{1}{|c|}{} & \multicolumn{1}{|c|}{} \\ \hline

%11
\multicolumn{1}{|c}{\cite{zheng2012efficient}} 
& \multicolumn{1}{|c}{X} & \multicolumn{1}{|c}{X} & \multicolumn{1}{|c}{X} & \multicolumn{1}{|c}{X} & \multicolumn{1}{|c}{X} &
\multicolumn{1}{|c}{X} & \multicolumn{1}{|c}{} &
\multicolumn{1}{|c}{} & \multicolumn{1}{|c}{X} & \multicolumn{1}{|c}{} & \multicolumn{1}{|c}{} &
\multicolumn{1}{|c}{}  & \multicolumn{1}{|c|}{} & \multicolumn{1}{|c|}{X} \\ \hline

%12
\multicolumn{1}{|c}{\cite{papadopoulos2014taking}} 
& \multicolumn{1}{|c}{X} & \multicolumn{1}{|c}{} & \multicolumn{1}{|c}{} & \multicolumn{1}{|c}{X} & \multicolumn{1}{|c}{X} &
\multicolumn{1}{|c}{X} & \multicolumn{1}{|c}{} &
\multicolumn{1}{|c}{} & \multicolumn{1}{|c}{} & \multicolumn{1}{|c}{X} & \multicolumn{1}{|c}{} &
\multicolumn{1}{|c}{}  & \multicolumn{1}{|c|}{} & \multicolumn{1}{|c|}{X} \\ \hline

%8
\multicolumn{1}{|c}{\cite{zhang2015integridb} } 
& \multicolumn{1}{|c}{X} & \multicolumn{1}{|c}{X} & \multicolumn{1}{|c}{X} & \multicolumn{1}{|c}{X} & \multicolumn{1}{|c}{X} &
\multicolumn{1}{|c}{X} & \multicolumn{1}{|c}{} &
\multicolumn{1}{|c}{} & \multicolumn{1}{|c}{} & \multicolumn{1}{|c}{X} & \multicolumn{1}{|c}{} &
\multicolumn{1}{|c}{}  & \multicolumn{1}{|c|}{} & \multicolumn{1}{|c|}{X} \\ \hline

%14
\multicolumn{1}{|c}{\cite{bajaj2013correctdb}} 
& \multicolumn{1}{|c}{X} & \multicolumn{1}{|c}{X} & \multicolumn{1}{|c}{X} & \multicolumn{1}{|c}{X} & \multicolumn{1}{|c}{X} &
\multicolumn{1}{|c}{X} & \multicolumn{1}{|c}{} &
\multicolumn{1}{|c}{} & \multicolumn{1}{|c}{} & \multicolumn{1}{|c}{} & \multicolumn{1}{|c}{X} &
\multicolumn{1}{|c}{}  & \multicolumn{1}{|c|}{} & \multicolumn{1}{|c|}{X} \\ \hline

%15
\nop{
\multicolumn{1}{|c}{\cite{ghosh2016efficient}} 
& \multicolumn{1}{|c}{X} & \multicolumn{1}{|c}{} & \multicolumn{1}{|c}{} & \multicolumn{1}{|c}{X} & \multicolumn{1}{|c}{X} &
\multicolumn{1}{|c}{X} & \multicolumn{1}{|c}{} &
\multicolumn{1}{|c}{X} & \multicolumn{1}{|c}{X} & \multicolumn{1}{|c}{X} &
\multicolumn{1}{|c}{}  & \multicolumn{1}{|c|}{} & \multicolumn{1}{|c|}{} \\ \hline

%16
\multicolumn{1}{|c}{\cite{Narasimha2006}} 
& \multicolumn{1}{|c}{X} & \multicolumn{1}{|c}{X} & \multicolumn{1}{|c}{} & \multicolumn{1}{|c}{X} & \multicolumn{1}{|c}{X} &
\multicolumn{1}{|c}{X} & \multicolumn{1}{|c}{} &
\multicolumn{1}{|c}{} & \multicolumn{1}{|c}{X} & \multicolumn{1}{|c}{} &
\multicolumn{1}{|c}{} & \multicolumn{1}{|c|}{} & \multicolumn{1}{|c|}{X} \\ \hline

%17
\multicolumn{1}{|c}{\cite{Pang:2009:SVO:1687627.1687718}} 
& \multicolumn{1}{|c}{X} & \multicolumn{1}{|c}{X} & \multicolumn{1}{|c}{} & \multicolumn{1}{|c}{X} & \multicolumn{1}{|c}{X} &
\multicolumn{1}{|c}{X} & \multicolumn{1}{|c}{} &
\multicolumn{1}{|c}{} & \multicolumn{1}{|c}{X} & \multicolumn{1}{|c}{} &
\multicolumn{1}{|c}{} & \multicolumn{1}{|c|}{} & \multicolumn{1}{|c|}{X} \\ \hline

%18\cite{4812487,Mouratidis:2009:PMD:1541533.1541543}
\multicolumn{1}{|c}{\cite{4812487}} 
& \multicolumn{1}{|c}{X} & \multicolumn{1}{|c}{} & \multicolumn{1}{|c}{} & \multicolumn{1}{|c}{X} & \multicolumn{1}{|c}{X} &
\multicolumn{1}{|c}{X} & \multicolumn{1}{|c}{} &
\multicolumn{1}{|c}{X} & \multicolumn{1}{|c}{} & \multicolumn{1}{|c}{} &
\multicolumn{1}{|c}{}  & \multicolumn{1}{|c|}{} & \multicolumn{1}{|c|}{X} \\ \hline

%19
\multicolumn{1}{|c}{\cite{Mouratidis:2009:PMD:1541533.1541543}} 
& \multicolumn{1}{|c}{X} & \multicolumn{1}{|c}{} & \multicolumn{1}{|c}{} & \multicolumn{1}{|c}{X} & \multicolumn{1}{|c}{X} &
\multicolumn{1}{|c}{X} & \multicolumn{1}{|c}{} &
\multicolumn{1}{|c}{X} & \multicolumn{1}{|c}{} & \multicolumn{1}{|c}{} &
\multicolumn{1}{|c}{}  & \multicolumn{1}{|c|}{} & \multicolumn{1}{|c|}{X} \\ \hline

%20
\multicolumn{1}{|c}{\cite{Singh:2008:ECO:1353343.1353402}} 
& \multicolumn{1}{|c}{X} & \multicolumn{1}{|c}{X} & \multicolumn{1}{|c}{} & \multicolumn{1}{|c}{X} & \multicolumn{1}{|c}{X} &
\multicolumn{1}{|c}{X} & \multicolumn{1}{|c}{} &
\multicolumn{1}{|c}{X} & \multicolumn{1}{|c}{} & \multicolumn{1}{|c}{} &
\multicolumn{1}{|c}{}  & \multicolumn{1}{|c|}{} & \multicolumn{1}{|c|}{X} \\ \hline\hline

\multicolumn{1}{|c}{\cite{6047135}} 
& \multicolumn{1}{|c}{} & \multicolumn{1}{|c}{} & \multicolumn{1}{|c}{} & \multicolumn{1}{|c}{} & \multicolumn{1}{|c}{} &
\multicolumn{1}{|c}{} & \multicolumn{1}{|c}{X} &
\multicolumn{1}{|c}{} & \multicolumn{1}{|c}{} & \multicolumn{1}{|c}{} &
\multicolumn{1}{|c}{X} & \multicolumn{1}{|c|}{} & \multicolumn{1}{|c|}{} \\ \hline
}

%21
\multicolumn{1}{|c}{\cite{sion2005query}} 
& \multicolumn{1}{|c}{X} & \multicolumn{1}{|c}{X} & \multicolumn{1}{|c}{X} & \multicolumn{1}{|c}{X} & \multicolumn{1}{|c}{X} &
\multicolumn{1}{|c}{X} & \multicolumn{1}{|c}{} &
\multicolumn{1}{|c}{} & \multicolumn{1}{|c}{} & \multicolumn{1}{|c}{} & \multicolumn{1}{|c}{} &
\multicolumn{1}{|c}{X} & \multicolumn{1}{|c|}{} & \multicolumn{1}{|c|}{} \\ \hline

%22\cite{di2014optimizing,de2016efficient}
\multicolumn{1}{|c}{\cite{di2014optimizing,de2016efficient}} 
& \multicolumn{1}{|c}{} & \multicolumn{1}{|c}{X} & \multicolumn{1}{|c}{} & \multicolumn{1}{|c}{X} & \multicolumn{1}{|c}{X} &
\multicolumn{1}{|c}{X} & \multicolumn{1}{|c}{} &
\multicolumn{1}{|c}{} & \multicolumn{1}{|c}{} & \multicolumn{1}{|c}{} & \multicolumn{1}{|c}{} &
\multicolumn{1}{|c}{X} & \multicolumn{1}{|c|}{} & \multicolumn{1}{|c|}{} \\ \hline

%23
\multicolumn{1}{|c}{\cite{xie2007integrity}} 
& \multicolumn{1}{|c}{X} & \multicolumn{1}{|c}{X} & \multicolumn{1}{|c}{} & \multicolumn{1}{|c}{X} & \multicolumn{1}{|c}{X} &
\multicolumn{1}{|c}{X} & \multicolumn{1}{|c}{} &
\multicolumn{1}{|c}{} & \multicolumn{1}{|c}{} & \multicolumn{1}{|c}{} & \multicolumn{1}{|c}{} &
\multicolumn{1}{|c}{X} & \multicolumn{1}{|c|}{} & \multicolumn{1}{|c|}{X} \\ \hline

\multicolumn{1}{|c}{\cite{vSQL}} 
& \multicolumn{1}{|c}{X} & \multicolumn{1}{|c}{X} & \multicolumn{1}{|c}{X} & 
\multicolumn{1}{|c}{X} & \multicolumn{1}{|c}{X} &
\multicolumn{1}{|c}{X} & \multicolumn{1}{|c}{} &
\multicolumn{1}{|c}{} & \multicolumn{1}{|c}{} & \multicolumn{1}{|c}{} & \multicolumn{1}{|c}{} &
\multicolumn{1}{|c}{} & \multicolumn{1}{|c|}{X} & \multicolumn{1}{|c|}{X} \\ \hline
\end{tabular}

\caption{Summary of existing approaches  ($\mathcal{A}$: authenticity; $\mathcal{S}$: soundness; $\mathcal{C}$: completeness; $\mathcal{F}$: freshness) 
%\Wendy{1. Under Deterministic approach, add "trusted hardware", and change [13] to this category. 2. [80] only handles join query. remove range query, and split it to a separate entry. 3. [60] verifies freshness too. 4. [45] verifies freshness too. move out of the 2nd entry, and create a separate entry. 5. Move [13] to be after [57] (i.e., together with tree-based approach) 6. Move [84] and [61] together, after [21]. 7. Move [79] after [21]. (i.e., together with sign.-based approach)}
}
\label{table:relatedwork2}
\vspace{-0.2in}
\end{table*}
In this survey, we categorize the existing result integrity authentication approaches into two types, based on the verification guarantee that these approaches can return: 
\begin{itemize}
\item The {\em deterministic} approaches that verify the query results with 100\% certainty. 
%Most of the deterministic approaches rely on the {\em proof} of the query results for verification. We call these approaches the {\em proof-based} authentication methods.
\item The {\em probabilistic} approaches that return a probabilistic verification guarantee of the query results. 
%Most of the probabilistic approaches use the {\em checkpoints}. We call these approaches the {\em checkpoint-based} authentication methods.
\end{itemize}

\subsubsection{Deterministic Authentication} 

\begin{figure}[!htb]
\vspace{-0.1in}
\begin{center}
\begin{tabular}{c}
 \includegraphics[width=0.5 \textwidth]{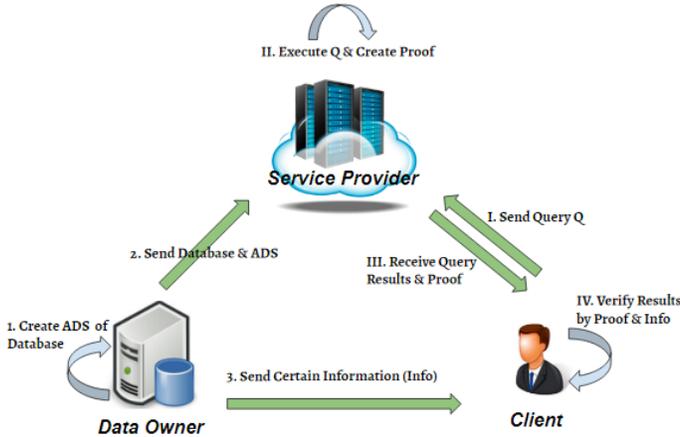}
\end{tabular}
\vspace{-0.1in}
\caption{\label{fig:vo-frame} An example of proof-based authentication method (Steps 1, 2 and 3: Verification preparation phase; Steps I, II, III and IV: Verification phase)}
\end{center}
\vspace{-0.15in}
\end{figure}

Consider the query $Q$ and its results $R$, all deterministic authentication solutions rely on {\em proofs} of the query results to return a {\em deterministic} integrity guarantee. Intuitively, besides the query results $R$, the SP constructs a short {\em proof} of $R$, and sends both the proof and $R$ to the client. The proof sometimes is referred as the {\em verification object} (VO). The client utilizes the proof to verify that $R$ satisfies certain requirements (e.g., soundness, completeness, and freshness).  
At a high level, most of the proof-based authentication methods follow the similar procedure below.
Before sending the dataset $D$ to the SP, the data owner computes some auxiliary information $A$ of $D$, which is known as an {\em authentication data structure} (ADS). 
Normally the ADS $A$ is much smaller than $D$. The data owner sends both $D$ and $A$ to the SP, while keeping $A$ locally. She may distribute certain information to the clients.  
When a client sends a query $Q$ to the SP, the SP executes $Q$ on $D$, and constructs a proof of the query results $R$, by using both $R$ and the auxiliary information $A$ obtained from the data owner. The SP sends both the query results and the proof to the client. The client verifies result integrity by using  
both the proof and the certain information that the data owner shared with her. 
Figure \ref{fig:vo-frame} illustrates the framework of the proof-based authentication methods. 

Based on how the proof is constructed, we categorize the proof-based methods into four types, namely {\em tree-based} solutions, {\em signature-based} solutions, {\em accumulation value} based solutions, and {\em trusted hardware} based solutions. 

In the tree-based solutions the ADS normally takes the tree format. As part of query execution,
the SP traverses the ADS tree and gathers the information of respective nodes to form the proof. From the query results and the proof, the client re-constructs the traversal path used in query execution and verifies that it is indeed authentic. 

By the signature-based methods, the data owner signs individual tuples in a chain fashion before uploading her dataset to the SP. During 
query execution, the SP gathers the signatures of 
the tuples in the query results, and compresses these signatures into an aggregated one. The aggregated signature constructed from the authenticated chain forms the proof of the query results. The client can authenticate the results by utilizing the signatures of the tuples in the results and the aggregated signature in the proof. 

The accumulation value based solutions are adapted from the cryptographic techniques that authenticate a number of set operations (e.g., intersection, set difference, and set summation). These techniques perform set authentication by constructing a proof of set operation results using accumulation values. The set authentication protocols are then used as building blocks for verification of particular SQL operations (e.g., multi-range selection, join, and aggregation). 

The trusted hardware based solutions rely on some tamper-proof, trusted hardware that is deployed on the untrusted server. The original query will be rewritten into sub-queries. All sub-queries that cannot be authenticated are processed inside the trusted hardware, while the results of the remaining sub-queries are verified by other authentication methods, e.g., the tree-based authentication methods. 

 \subsubsection{Probabilistic Authentication}

The probabilistic authentication methods return a {\em probabilistic} result integrity guarantee (i.e., how likely the returned results are correct given that they passed the verification). The existing probabilistic authentication methods can be classified into two types: (1) the {\em checkpoint-based} solutions, and (2) the {\em interactive-proof based} solutions.

The checkpoint-based solutions rely on {\em checkpoints} for verification. Unlike the proof-based authentication methods that require the SP to construct the proofs of the query results, the checkpoint-based authentication does not require the involvement of the SP. The key idea of the checkpoint-based authentication is to verify if the returned query results include some particular {\em checkpoints} (i.e., some specific records). If the results fail the verification based on those checkpoints, the client can determine that the results fail the authentication with 100\% certainty. Otherwise, the client believes that the results pass authentication with some probabilistic guarantee. Most of the checkpoint-based methods construct the checkpoints by inserting {\em counterfeit records}  $F$ into the dataset $D$ before outsourcing. The assumption is that SP cannot distinguish the real and counterfeit tuples easily. Then for any given query $Q$, a faithful SP should return $Q(F)$ as the part of query results. Thus, any query result that does not include $Q(F)$ can be caught as incorrect. 

The interactive proof based solutions extends the information-theoretic interactive proof system \cite{cormode2012practical} to SQL query authentication. The key idea is to translate SQL queries into arithmetic circuits, and verify if the outputs of these circuits are correct. Since the interactive-proof protocol can incur expensive communication and verification overhead, the verification methods takes a random input for the last step of interactive proof protocol, making the authentication method associated with a probabilistic integrity guarantee.

\vspace{-0.05in}
\section{Authenticated Data Structure (ADS)}
\label{sc:ads}
%Given a large set of tuples, an authenticated data structure (ADS) \cite{devanbu2002authentic, tamassia2003authenticated} enables efficient authenticity verification for any subset. 
In this section, we introduce two prevalent ADS structures, namely, Merkle hash tree (MHT) and Merkle B-tree (MB-tree), that are used by a number of existing ADS-based verification methods. 
\vspace{-0.05in}
\subsection{Merkle Hash Tree (MHT)}
\label{sc:mht}
\begin{figure}[!htb]
\vspace{-0.1in}
\begin{center}
\begin{tabular}{c}
\includegraphics[width=0.35\textwidth]{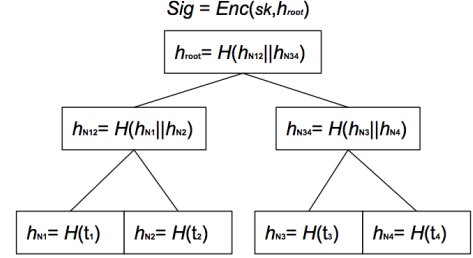}
\end{tabular}
\vspace{-0.1in}
\caption{\label{fig:MTree} An example of Merkle-tree}
\end{center}
\vspace{-0.15in}
\end{figure}
One of the widely-used ADS structures is {\em Merkle Hash tree} (MHT) \cite{merkle-1980}. 
A MHT $T$ is a tree in which each leaf node $N$ stores the digest of a tuple $t$: $h_{N} = H(t)$, where $H()$ is a one-way, collision-resistant hash function (e.g. SHA-1 \cite{eastlake2001us}). 
For each internal node $N$ of $T$, it is assigned the value $h_{N} = H(h_{N_{1}}||\dots||h_{N_{z}})$, where $N_{1}, \dots, N_{z}$ are the children of $N$, and $||$ is the concatenation operator. 
The hash value $h_{root}$ of the root node is used as the digest of the tree. To serve the authentication purpose, a trusted party (can be the data owner himself) generates the signature of the ADS as $Sig=Enc(sk, h_{root})$, where $Enc()$ is an asymmetric encryption function, and $sk$ is the private key. $Sig$ is shared with the clients for authentication.
Figure \ref{fig:MTree} illustrates an example of MHT.

\nop{
One of the commonly used authenticated data structure is based on Merkle tree \cite{Merkle1990} \Wendy{add reference to Merkle tree}. Authenticated Merkle Tree is a hash tree in which every internal node is labeled with the hash of the labels or values (in case of leaves) of its child nodes. Demonstrating that a leaf node is a part of the given hash tree requires processing an amount of data proportional to the logarithm of the number of nodes of the tree \ref{merkle-1980}. In particular, a Merkle hash tree $T$ is a tree in which each leaf node $N$ stores the digest of a tuple t: $h_N$ = h(t), where h() is a one-way, collision-resistant hash function (e.g. SHA-1 \cite{eastlake2001us}). For each non-leaf node $N$ of $T$, it is assigned the value $h_N =h(h_{N_1}||...||h_{N_z})$, where $N_1$,...,$N_z$ are the children of $N$, and $||$ is the concatenation operator. The hash value $h_root$ of the root node is used as the digest of the tree. 
\Wendy{Please copy the description of Merkle tree from our ICDE'17 submission to here. And also add the missing references.}.

\nop{And $H$ is a collision resistant hash functions that it is computationally infeasible to find two inputs $x_1 \neq x_2$, such that $H(x_1)=H(x_2)$, most of the existing work use SHA-1 \ref{eastlake2001u} as the hash function which has the advantage of being very fast to evaluate and takes variable-length inputs to a certain number of bytes output. Except the leaf node, each internal node is labeled with the hash of the labels of its child nodes, here labels are the hash value of its child nodes. Sometimes those labels also associate with some auxiliary information, for example the aggregation value or range value of its child node. In this example, the internal node is just hash value its child node hash values.
}
\Wendy{the description of hash of nodes should be given in the previous paragraph. Here using an example you should describe how the root signature is constructed instead.}

There are 4 leave nodes in the Merkle Tree. Each leaf node is associate with a value $v$. First, the hash value of leaf nodes $\{v_1,v_2,v_3,v_4\}$ are computed by the SHA-1 hash function $H$. Then, the hash value of internal node is the computed on the concatenation of the children's hash value by the same hash function, like $h_{12}=H(h_1||h_2)$. Finally, we can get the hash value of the root $h_1234$ by hashing $h_{12}||h_{34}$. In order to certify the digest of the ADS tree,

\Wendy{I don't understand this sentence. This sentence jumps out of nowhere is not connected with the next sentence.} the client also needs a public key digital signature scheme to sign the root of the Merkle Tree. A public key scheme, formally defined in \ref{Diffie:2006:NDC:2263321.2269104}, is a tool to authenticate the integrity and ownership of the signed message. In such a scheme, the signer generates a pair of keys (SK,PK), keeps the secret key SK secret and publishes the public key PK associated with her identity. For any message she sends, a signature $Sig$ is created by : $Sig=S(SK,m)$. The recipient of $Sig$ and m can verify $Sig$ by using the public key $V(PK,m,Sig)$. \Wendy{Section 4.1 has to be rewritten. First, use the description of Merkle tree in our ICDE'17 submission to discuss how Merkle tree and its root digest is constructed. Then use Figure 2 as an example to explain.}
}
\vspace{-0.05in}
\subsection{Merkle B-tree (MB-tree)}
\label{sc:mbtree}

Merkle B-tree
(MB-tree) \cite{li2006dynamic} enables efficient search on one-dimensional data, as it combines  MHT with $B^+$-tree. In MB-tree, the tuples are arranged in the same fashion as standard $B^+$-tree. The major difference is the incorporation of hash values in the tree nodes.
In particular, in the leaf node, each entry is associated with a hash value $h=H(t)$, where $t$ is the tuple in the entry. 
In the internal node, every pointer to a child node is combined with $h=H(h_1||\dots||h_k)$, where $h_1, \dots h_k$ are the hash values in the child node.

\nop{
\subsection{Merkle R-tree (MR-tree)}
R-tree \cite{guttman1984r} is a data structure for indexing multi-dimensional data. Given a set of multi-dimensional data points, the leaf entries of R-tree store the data points. The internal entries keep the minimum bounding hyper-rectangles (MBRs) that enclose the data points in the subtree. 
Figure \ref{fig:R-tree} shows an example of the 2-D space and the corresponding R-tree.
For any query $Q$, let $R_Q$ be its range filtering condition, the search starts from the root of the R-tree to the leaf nodes. For each internal node $N$, if its $MBR$ does not intersect with $R_Q$, the search stops at $N$. Otherwise, the search continues with the children of $N$, until it reaches the leaf nodes. The tuples of the leaf nodes that satisfy $R_Q$ are included in the query result. For example, in Figure \ref{fig:R-tree}, the subtree rooted at $R_6$ is not traversed as it does not intersect with $R_Q$.
% We say two MBRs $R_1$ and $R_2$ {\em overlap} if their ranges overlap at any dimension (i.e. attribute). Otherwise, we say $R_1$ and $R_2$ are disjoint. 

\begin{figure}[!htb]
\vspace{-0.1in}
\begin{center}
\begin{tabular}{cc}
\includegraphics[width=0.18\textwidth]{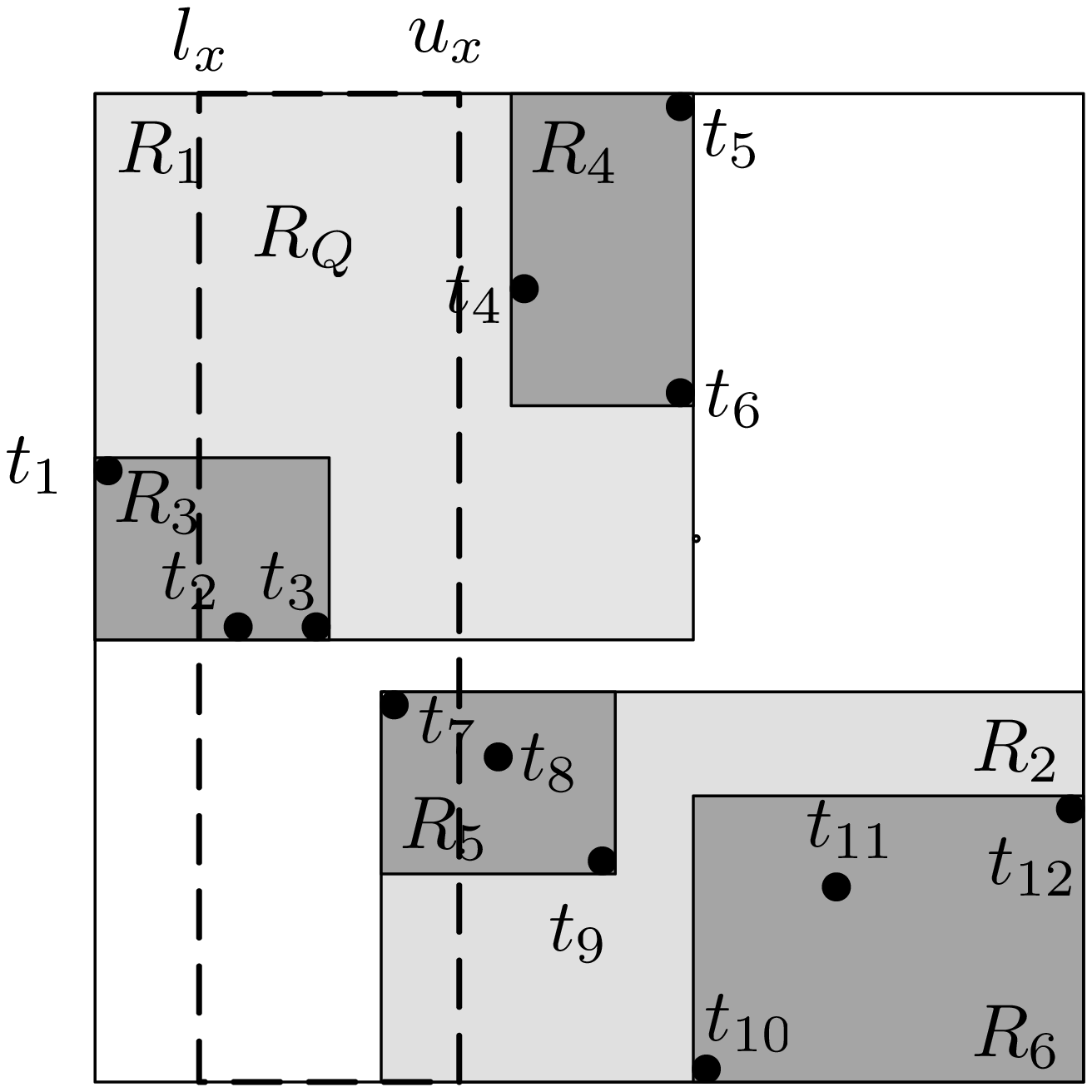}
&
\includegraphics[width=0.2\textwidth]{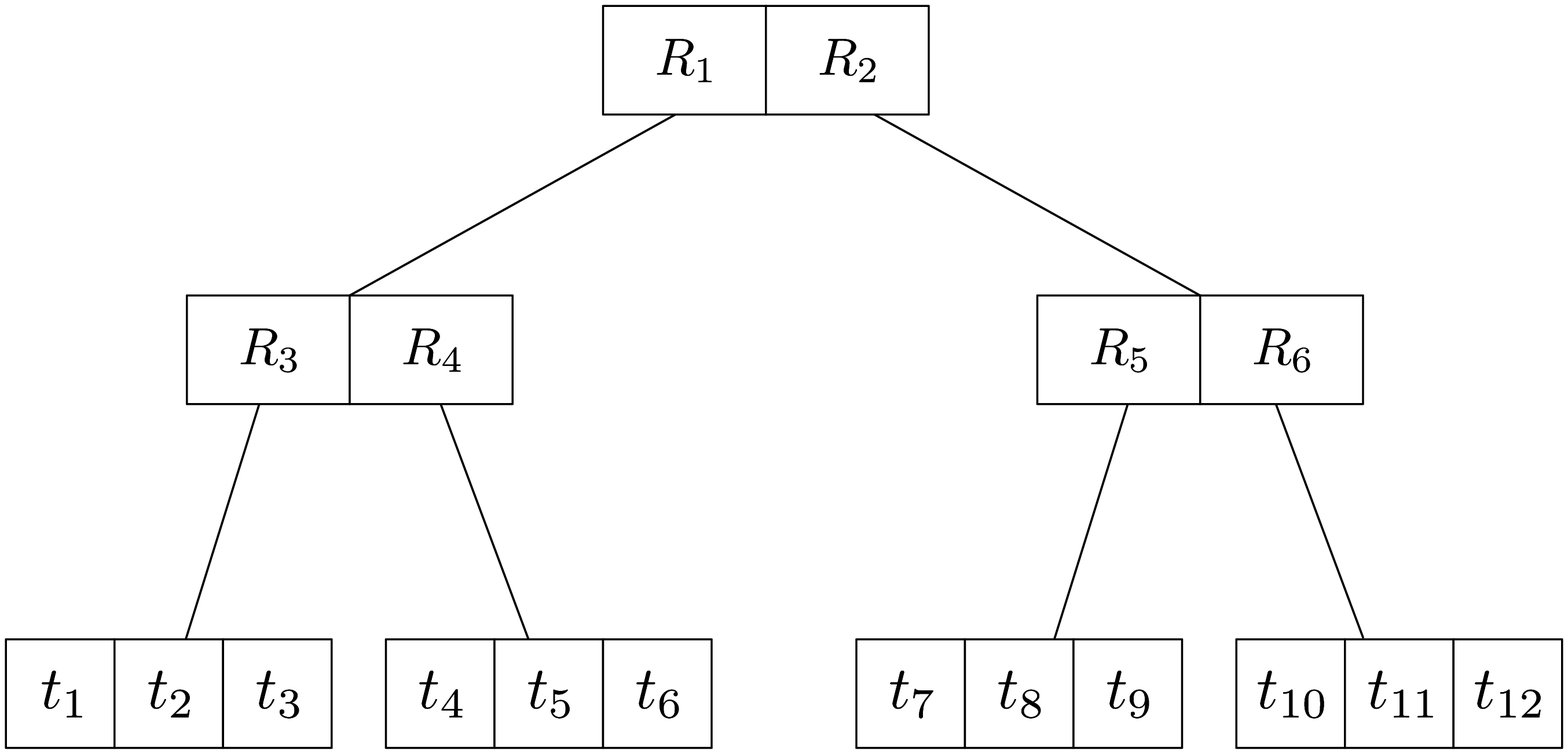}
\\
{\small (a) $MBR$s and filtering range $R_Q$}
&
{\small (b) R-tree structure}
\end{tabular}
\vspace{-0.1in}
\caption{\label{fig:R-tree} An example of R-tree}
\end{center}
\vspace{-0.15in}
\end{figure}

The Merkle R-tree (MR-tree) \cite{yang2008spatial} combines MHT \cite{merkle-1980} and R-tree. In MR-tree, each internal entry is augmented with a hash value. For any internal node that is parent of the leaf nodes, every entry contains a hash value $h=H(t_1 || \dots || t_z)$, where the tuples $t_1, \dots, t_z$ are enclosed in the child node. For other internal nodes (including the root), each entry contains the hash value $h=H((h_{N_1}, MBR_{N_1}) || \dots || (h_{N_z}, MBR_{N_z}))$, where the hash values and MBRs are in the child node. 
}
\vspace{-0.05in}
\section{Cryptographic Background}
\label{sc:crytobag}
In this section, we present the cryptographic background of the signature-based and accumulation value based verification methods.  

\nop{
\subsection{Bloom Filter}
\label{sc:bf}
\Wendy{Do we really need this subsection?}
Bloom filter \cite{bloom1970space} is a probabilistic data structure for testing whether an element is a member of a set. In general, a bloom filter is a bit array of $m$ bits, all set as 0 initially. There are $k$ different hash functions $H_1, \dots, H_k$, each of which hashes the given element to one of the $m$ bits in the bloom filter. Typically, $k$ is a constant and much smaller than $m$. In particular, given a set of elements $V=\{v_1, \dots, v_n\}$ and $k$ hash functions, for each value $v_i$, the bit at the $H_j(v_i)$-th position of the bloom filter is set to $1$, for each $j\leq k$. To query if an element $v\in V$, it can be performed by checking if the bit at the $H_j(v)$-th position in the Bloom filter is $1$, for every $1\leq j\leq k$. If any of the bits at these positions is 0, the element is definitely not in the set. If all are 1, then either the element is in the set, or the bits have by chance been set to 1 during the insertion of other elements, resulting in a false positive.
}

\nop{
\subsection{RSA Signature}
\label{sc:rsa}
\Wendy{RSA signature is like common sense. This part is not necessary.}
Rivest et al. \cite{rivest1978method} proposed the classic public key encryption scheme named {\em RSA}. Initially, an entity generates two random prime numbers $p$ and $q$, and computes $N$ as the product of them. The entity also finds a pair of integers $(e,d)$ such that $e,d\in \mathbb{Z}_N^*$, and $ed\equiv 1 \ mod \ \phi(N)$, where $\phi(N)=(p-1)(q-1)$. The entity publishes the public key $pk=(N, e)$, and keeps the secret key $sk=(d)$ confidential. 

One important application of the RSA encryption scheme is the RSA signature. Any message $v$ can be signed by the entity using its secret key as: 
\begin{equation}
\label{eq:rsa_1}
\sigma=H(v)^d\ (mod\ N),
\end{equation}
where $H$ is a full-domain cryptographic hash function that converts a message into an value in $\mathbb{Z}_N^*$. Any other entity who receives $(v, \sigma)$ can verify if $v$ corresponds to $\sigma$ by checking 
\begin{equation}
\label{eq:rsa_2}
\sigma^e \stackrel{?}{=} v\ (mod\ N).
\end{equation}
The signature is not forgeable. In other words, without the knowledge of the secret key, no party can produce a valid signature for any message.
}

\vspace{-0.05in}
\subsection{Bilinear Pairing}
\label{sc:pairing}
%\Wendy{Do we really need this subsection? Can we merge this section with where it mentioned bilinear pairing?}
%\Boxiang{Actually, I tried to merge this section to other sections. However, pairing is used in multiple places (aggregate signature and set operations), so I think it may be better to have it in a separate section.}
Bilinear pairing \cite{Menezes:1991:REC:103418.103434} uses a pairing between two cryptographic groups to a third group with a mapping. 
In particular, let $G$ and $G_T$ be two groups of order $p$ (with $p$ a $\lambda$-bit prime), a {\em bilinear pairing} on $(G, G_T)$ is a map 
$e:  G \times G \rightarrow G_T$ that satisfies the following conditions: 
\begin{itemize}
\item Bilinearity: $\forall a, b\in \mathbb{Z_p}$, $P, Q\in G$, $e(P^a, Q^b) = e(P, Q)^{ab}$; 
\item Non-degeneracy: $e(g, g)\neq 1$, where $g \in G$ is a generator of group $G$; 
\item Computability: There exists an efficient algorithm to compute $e$.
\end{itemize}
\vspace{-0.05in}
\subsection{Aggregate Signature}
\label{sc:bgls}

An aggregate signature scheme is a digital signature that supports aggregation. Informally, given $n$ 
signatures on $n$ distinct messages, all these signatures are aggregated into a single short signature \cite{boneh2003aggregate}. 

\subsubsection{BGLS Scheme} 

The BGLS \cite{boneh2003aggregate} aggregate signature scheme (named after its four authors) relies on bilinear pairing for signature aggregation. 
\nop{
In particular, given a set of messages $\{m_1, \dots, m_n\}$, the prover \Wendy{who is prover?} constructs an aggregated signature $\sigma_{1,n}$ that corresponds to $\prod_{i=1}^n m_i$ from the individual signatures. In the verification step, the verifier can efficiently evaluate the authenticity of the messages via $\sigma_{1,n}$. 
}
In particular, given a bilinear map $e:G\times G\rightarrow G_T$ over two groups of prime order $p$, let $s\in \mathbb{Z_p}$ be a secret key,  the public key $v$ is generated as
%each user picks a random secret key $s\in \mathbb{Z_p}$, and publishes his/her public key 
\begin{equation}
\label{eq:bgls_0}
v=g^s\in G,
\end{equation}
where $g$ is the generator of group $G$. 
For any message $m_i$, let $H:\{0,1\}^* \rightarrow G$ be a full-domain hash function. The signature $\sigma_i$ of $m_i$ is generated as 
\begin{equation}
\label{eq:bgls_1}
\sigma_i=H(m_i)^s.
\end{equation}
Clearly, $\sigma_i\in G$. 

Given $n$ signatures $\{\sigma_1, \dots, \sigma_n\}$ where $\sigma_i = H(m_i)^{s_i}$, their aggregate signature is generated as 
\begin{equation}
\label{eq:bgls_2}
\sigma_{1,n}=\Pi_{i=1}^n \sigma_i. 
\end{equation}
Again, $\sigma_{1,n}$ is a member in group $G$.

Upon receiving an aggregate signature $\sigma_{1,n}$, the set of messages $\{m_i\}$, and the public keys $\{v_i\}$, whether $\sigma_{1,n}$ corresponds to the messages $\{m_i\}$ can be verified by first computing $h_i=H(m_i)$ for $1\leq i\leq n$. The messages are accepted if
\begin{equation}
\label{eq:bgls_3}
e(\sigma_{1,n}, g) = \Pi_{i=1}^n e(h_i, v_i).
\end{equation}

%The advantage of BGLS is that with a single aggregate signature $\sigma$, anyone can verify the authenticity of an arbitrary number of messages.

\subsubsection{Homomorphic Linear Authentication (HLA) Scheme}
\label{sc:hlt}
Homomorphic Linear Authentication (HLA) scheme \cite{ateniese2009proofs} allows to check the authenticity of the messages against the generated aggregate signature without the knowledge of individual messages. 
\nop{
In the HLA scheme, given a vector of random coefficients $\vec{c}=\{c_1, \dots, c_n\}$, the prover \Wendy{who is prover?} computes a single aggregate signature $\sigma_{1,n}$ that corresponds to $\sum_{i=1}^n c_im_i$. 
The verifier can determine the authenticity of the messages $\{m_1, \dots, m_n\}$ from $\sigma_{1,n}$.
}
The HLA scheme consists of the following algorithms:

\begin{itemize}
	\item $(sk, pk)\leftarrow KeyGen(1^{\ell})$: given a security parameter $\ell$, this algorithm outputs a pair of public and private keys $(sk, pk)$.
	\item $\sigma_i \leftarrow TagGen(sk, m_i)$: for message $m_i$, this algorithm generates a signature $\sigma_i$ based on the private key $sk$.
	\item $\sigma_{1,n} \leftarrow HLAAgg(\vec{c}, \vec{Tag})$: this linear aggregation algorithm takes the input of a vector of coefficients $\vec{c}=\{c_1, \dots, c_n\}$ and a vector of tags $\vec{Tag} = \{\sigma_1, \dots, \sigma_n\}$ with respect to a vector of messages $\vec{m} = \{m_1$, \dots, $m_n\}$. This algorithm produces the aggregate signature $\sigma_{1,n} = \sum_{i=1}^{n}c_im_i$.
	\item $\{0,1\}\leftarrow Vrfy(pk, m', \sigma_{1,n})$: given the input as the public key $pk$, a candidate message $m'$ and an aggregate signature $\sigma_{1,n}$, this algorithm returns 1 if $\sigma_{1,n}$ is valid with respect to $m'$, and 0 otherwise.
\end{itemize}

%The algorithm $Vrfy(pk, \vec{t}', \sigma)$ does not need to obtain $\vec{t}'$. Instead, the verifier only needs to access $\varphi(pk, \vec{t}')$, where $\varphi$ is a irreversible function. In this way, even without the knowledge of individual tuples in $\vec{t}'$, the verifier is able to check the validity.
\vspace{-0.05in}
\subsection{Accumulation Value}
\label{sc:acc}
An accumulator scheme \cite{benaloh1993one,baric1997collision} allows aggregation of a large set of inputs
into one constant-size value. 
%For a given element, there is a witness that the element was included into the accumulated value whereas it is not possible to compute a witness for an element that is not accumulated.
Papamanthou et al. \cite{papamanthou2011optimal} proposed an accumulation value scheme for set operation verification.  In particular, given a set of elements $S=\{x_1, \dots, x_n\}$, the accumulation value of $S$ is constructed as: 
\begin{equation}
\label{eq:acc}
acc(S) = g^{\prod_{i=1}^n (x_i+s)}, 
\end{equation}
where $g$ is the generator of a bilinear group, and $s$ is a randomly chosen secret value. 

\nop{
Bilinear pairing \cite{Menezes:1991:REC:103418.103434} is the use of a pairing between two cryptographic groups to a third group with a mapping. 
In particular, let $G$ and $G_T$ be two groups of order $p$ (with $p$ a $\lambda$-bit prime), a {\em bilinear pairing} on $(G, G_T)$ is a map 
$e:  G \times G \rightarrow G_T$ that satisfies the following conditions: (1) $\forall a, b\in \mathbb{Z_p}$, $P, Q\in G$, $e(P^a, Q^b) = e(P, Q)^{ab}$; and (2) $e(g, g)\neq 1$, where $g \in G$ is a generator of group $G$.
}

For any pair of subsets $S_1$ and $S_2$ of $S$ such that: (1) $S_1 \cup S_2 = S$; and (2) $|S_1|+|S_2|=|S|$, it must be true that $e(acc(S_1), acc(S_2)) = e(acc(S), g)$, where $e$ is the bilinear pairing function. 
Based on the discrete log assumption \cite{stinson2005cryptography}, it is NP-hard for any polynomial-time adversary without the knowledge of the secret key $s$ to find a different set $S_3\neq S_2$ such that $e(acc(S_1), acc(S_3)) = e(acc(S), g)$.

\vspace{-0.05in}
\subsection{Verification Protocols for Set Operations}
\label{sc:set_protocol}

\subsubsection{Set Intersection}
\label{sc:set_intersection}
Given a collection of sets $\cal S$ = $\{S_1, \dots, S_t\}$, $I = S_1 \cap
S_2 \cap \cdots \cap S_t$ is the {\em correct} intersection of $\cal S$ if and only if:
\begin{itemize}
  \item $I \subseteq S_1 \wedge \cdots \wedge \subseteq S_t$ (subset condition);
  \item $(S_1 - I) \cap \cdots \cap (S_t - I) = \emptyset$ (completeness condition).
\end{itemize}
%Here the completeness condition is on the set intersection. To avoid confusion with our completeness verification of query result, in the remainder of the paper, we use the term {\em intersection completeness} for the completeness condition on the set intersection. 

\nop{
For every set $S_j\in\cal S$, we define polynomial 
%$P_j(s) = \prod_{x \in S_j - I} (x + s)$.
   \begin{eqnarray}
    \label{eqn:pj}
      P_j(s) = \prod_{x \in S_j - I} (x + s),
    \end{eqnarray}
where $s$ is a randomly chosen value by the client and kept as secret. 
}

Papamanthou et al. proposed a set intersection verification protocol to verify that $I$ is a correct intersection of $\cal S$ \cite{papamanthou2011optimal}. 
For each set $S_i$, the SP demonstrates $I\subseteq S_i$ by providing $acc(W_i)$, where $W_i=S_i-I$, and $acc()$ is  the function that calculates the accumulation value (Equation \ref{eq:acc}). If 
$e(acc(I), acc(W_i))=e(acc(S_i), g)$ for every set $S_i$, where $e$ is the bilinear pairing, the client is assured that the subset condition is met. The intersection completeness is proved by finding a set of polynomials of $s$, namely $\{q_1, \dots, q_t\}$, such that $\sum_{i=1}^t P(W_i)q_i=1$, where $P(W_i)=\prod_{x_j\in W_i}(x_j+s)$.
 
\nop{
\noindent{\bf Proof construction by the server.} Given an intersection result $I=\{i_1, \dots, i_\delta\}$, to prove that $I= S_1 \cap \dots \cap S_t$, the set intersection verification protocol requires the server to construct the proof $\Pi(I)$ of the intersection $I$. The proof $\Pi(I)$ consists of four parts:
(1) An encoding of the result (as a polynomial) as the coefficients $\cal B$ =$\{b_{\delta}, b_{\delta-1}, \cdots, b_0\}$ of the
    polynomial $(s + i_1)(s + i_2) \cdots (s + i_{\delta})$, where 
    $s$ is a randomly chosen value by the client and kept as secret; 
		
(2) A set of {\em accumulation values} $\cal A$ =$\{acc(S_j)|\forall S_j\in\cal S\}$, where $acc(S_j) = g^{\prod_{x\in S_j}(s+x)}$, along with their respective proofs on the Merkle tree $\cal T$. The proof shows that the digest value $acc(S_j)$ is indeed computed from the original dataset $D$; 
%In particular, let $path(i)$ be a list of nodes on the path from leaf $i$ to the root and $sib(j)$ denote a sibling of node $j$ in $\cal T$. Then $\mathsf{MTproof}()$ algorithm outputs an ordered list containing the hashes of the siblings $sib(j)$ of the nodes $j$ in $path(i)$; 
%For example, consider the Merkle tree $\cal T$ in Figure \ref{fig:index} (c), and the value $h_1$. To prove $h_1$ comes from $\cal T$, $\mathsf{MTproof()}$ returns $\{h_2, h_{34}\}$. 
%How to use the proofs to verify the authenticity of $v$ will be discussed in Section \ref{sc:ver}.

(3) The {\em subset witness} $\cal W$ = $\{W_{j}|\forall S_j\in\cal S\}$ as the proof of the subset condition. In particular, for each set $S_j\in\cal S$, the server computes its {\em subset witness} $W_j = g^{P_j(s)}$, where $g$ is a generator of a group $G$ from an instance of bilinear pairing parameters, and $P_j(s)$ is computed by Equation (\ref{eqn:pj}); and 

(4) The {\em intersection completeness witness} $\cal C$ = $\{C_{j}|\forall S_j\in\cal S\}$ as the proof of the  intersection completeness condition. 
In particular, for each set $S_j\in\cal S$, the server computes its completeness witness $C_{j} = g^{q_j(s)}$, such that $q_1(s) P_1(s) + q_2(s) P_2(s) + \cdots + q_t(s) P_t(s) = 1$, where $P_j(s)$ is computed by Equation (\ref{eqn:pj}).
%The complexity of constructing $\cal W$ is $O(\delta log\delta)$, where $\delta=|I|$. The complexity of constructing $\cal A$ based on a Merkle tree of level $\lceil 1/\epsilon \rceil$ levels and $m$ leaves is $O(m^{\epsilon}log m)$, where $\epsilon\in(0, 1)$ is a user-specified constant. The complexity of constructing $\cal W$ is $O(NlogN)$, where $N = \sum_{j=1}^{t}|S_j|$.  And the complexity of constructing $\cal C$ is $O(N log^2N log logN)$. 

The total complexity of proof construction is $O(Nlog^3N + m^{\epsilon}logm)$, where $N = \sum_{j=1}^{t}|S_j|$, $m$ is the number of leaves of the Merkle tree, and $\epsilon\in(0, 1)$ is a user-specified constant that is used to specify the number of levels of the Merkle tree. After constructing the proof $\Pi(I)=\{\cal B, \cal A, \cal W, \cal C\}$, the server sends $\Pi(I)$ together with $I$ to the client. 

\noindent{\bf Correctness verification by the client.} After receiving $\Pi(I)$ together with $I$, the client verifies the following:

(1) whether the coefficients 
$\cal B$ =$\{b_1, \dots, b_\delta\}$ are computed correctly by the server; 

(2) whether any given accumulation value $v$ is indeed calculated from the original dataset; 

(3) whether $I$ satisfies the subset condition by checking whether 
%\begin{eqnarray}
%    \label{eqn:verifycorrectness}
\[e (\prod_{k=0}^{|I|} (g^{s^k})^{b_k}, W_{j}) \stackrel{?} =  e(acc(S_j),g), \text{ for } j = 1, \cdots, t, \]
%\end{eqnarray}
where $b_k\ (0\leq k\leq\delta)$ is a coefficient in $\cal B$, $e$ is a bilinear pairing, and $W_{j}\in\cal W$ is  the subset witness of $S_j$; 
 
(4) whether $I$ satisfies the intersection completeness condition by checking whether
%\begin{eqnarray}
%    \label{eqn:verifycompleteness}
\[     \prod_{j=1}^t e (W_{j}, C_{j})
      \stackrel{?}= e(g, g), \text{ for } j = 1, \cdots, t,\]
   %\end{eqnarray}
where $W_{j}\in\cal W$ and $C_{j}\in\cal C$ are the subset witness and intersection completeness witness in $I$ for the set $S_j$. 
}

\subsubsection{Set Difference} 
\label{sc:set_difference}

Papadopoulos et al. \cite{papadopoulos2014taking} designed a method for efficient authentication of set difference results based on the accumulation values. 
In particular, given two sets $X_1$ and $X_2$ such that $X_2\subseteq X_1$, to demonstrate $X=X_1\setminus X_2$, the prover constructs $acc(X)$ (Equation \ref{eq:acc}). The verifier simply checks $e(acc(X_2), acc(X) \stackrel{?}{=} e(acc(X_1), g)$ for verification, where $e$ is the bilinear pairing, and  $acc()$ is  the function that calculates the accumulation value (Equation \ref{eq:acc}).

\nop{
In particular, let $X_1$ and $X_2$ be two sets under the constraint that $X_2\subseteq X_1$. In order to prove that the returned set $X$ is indeed the set difference between $X_1$ and $X_2$, i.e., $X\stackrel{?}{=} X_1\setminus X_2$, the SP constructs $acc(X)$ (Equation \ref{eq:acc}) and transmits it to the client. The client verifies whether $X$ is the set difference by checking: 
\begin{equation}
\label{eq:papadopoulos2014taking_1}
e(acc(X_2), acc(X) \stackrel{?}{=} e(acc(X_1), g),
\end{equation}
where $e: G_1\times G_1 \rightarrow G_2$ is the bilinear pairing over two cyclic multiplicative groups $G_1$ and $G_2$ (Section \ref{sc:pairing}), and $g$ is the generator of $G_1$. 
%Equation (\ref{eq:papadopoulos2014taking_1}) The client verifies the set difference relationship with $O(1)$ bilinear pairings.
}

\subsubsection{Summation over Sets}
\label{sc:set_sum}

Given a set $S=\{x_1, \dots, x_n\}$, Zhang et al. \cite{zhang2015integridb} proposed the {sum verification protocol} based on accumulation values to check if $sum \stackrel{?}{=} \sum_{i=1}^n x_i$. 
Let $M(S)=\prod_{i=1}^n (x_i^{-1}+s)=a_ns^n + \dots + a_1s + a_0$. Obviously, $\sum_{i=1}^n x_i=a_1/a_0$. Based on this reasoning, the prover provides $a_0$, $a_1$, as well as proof that they are the coefficients of the smallest exponent in $M(S)$, for the purpose of verification. 
\nop{
In the setup phase, the client constructs an accumulation value 
\begin{equation}
\label{eq:zhang2015integridb_1}
acc(S)=g^{\prod_{i=1}^n (x_i^{-1}+s)},
\end{equation}
where $g$ is the generator of the Elliptic group \cite{Menezes:1991:REC:103418.103434}, $acc()$ is  the function that calculates the accumulation value (Equation \ref{eq:acc}), and $s$ is a random number and kept secret by the client. Note that Equation (\ref{eq:zhang2015integridb_1}) is different from Equation (\ref{eq:acc}) in that we use the inverse of $x_i$.
To prove the correctness of $sum$, the SP constructs a proof that includes: (1) $a_0=\prod_{i=1}^n x_i^{-1}$; (2) $a_1=\sum_{i=1}^n (\prod_{j\neq i} x_j^{-1})$; (3) $W_1 = g^{s^{n-1}+\dots+a_2s+a_1}$; and (4) $W_2=g^{s^{n-2}+\dots+a_3s+a_2}$. 
After receiving $sum$ as well as the proof, the client first verifies the correctness of $a_0$ by checking 
\begin{equation}
\label{eq:zhang2015integridb_2}
e(g^s, W_1) \stackrel{?}{=} e(acc(S)/g^{a_0}, g),
\end{equation}
where $e$ is the bilinear pairing function, and $acc(S)$ is obtained in the setup phase. 
Next, the client verifies the correctness of $a_1$ by checking 
\begin{equation}
\label{eq:zhang2015integridb_3}
e(g^s,W_2) \stackrel{?}{=} e(W_1/g^{a_1}, g).
\end{equation}
In the last step, the correctness of $sum$ can be verified by checking
\begin{equation}
\label{eq:zhang2015integridb_4}
sum \stackrel{?}{=} a_1/a_0.
\end{equation}
By calculating 4 pairings and 1 group division, the client can efficiently verify the sum of $n$ values.
}

\vspace{-0.05in}
\section{Deterministic Authentication Approaches}
\label{sc:vo}

In this section, we overview the authentication approaches that use verification object (VO) for authentication. There are three different methods to construct VO: (1) use the authenticated data structure; (2) use signature aggregation and chaining; and (3) use cryptographic accumulation values. All these methods return a deterministic integrity guarantee (i.e., with 100\% certainty). Besides these three methods, we also discuss the verification method that relies on the trusted hardware to return deterministic guarantee. Next, we discuss each  of these four different methods respectively. 

\vspace{-0.05in}
\subsection{Tree-based Authentication}

Quite a few existing authentication approaches use the  authenticated data structure (ADS) to construct the verification objects. Most of these works use different variants of Merkle hash tree (MHT) for specific query types.
%Most of these work follow the same strategy: a trusted party (possibly the data owner) constructs an ADS from the dataset to be outsourced, and maintain the auxiliary information of the ADS (typically much smaller than the size of outsourced data) locally. The auxiliary information of the ADS is shared with the data requester for query authentication. Both dataset and its ADS are sent to SP during outsourcing. When SP executes the query, it prepares a verification object (VO) of its results by facilitate the ADS. SP sends both the query results and the VO to the data requester. The data requester leverages the VO and the auxiliary  information of ADS to authenticate the returned results. 

%Queries supported and ADS
\begin{figure}[!htb]
\vspace{-0.1in}
\begin{center}
\begin{tabular}{c}
\includegraphics[width=0.4 \textwidth]{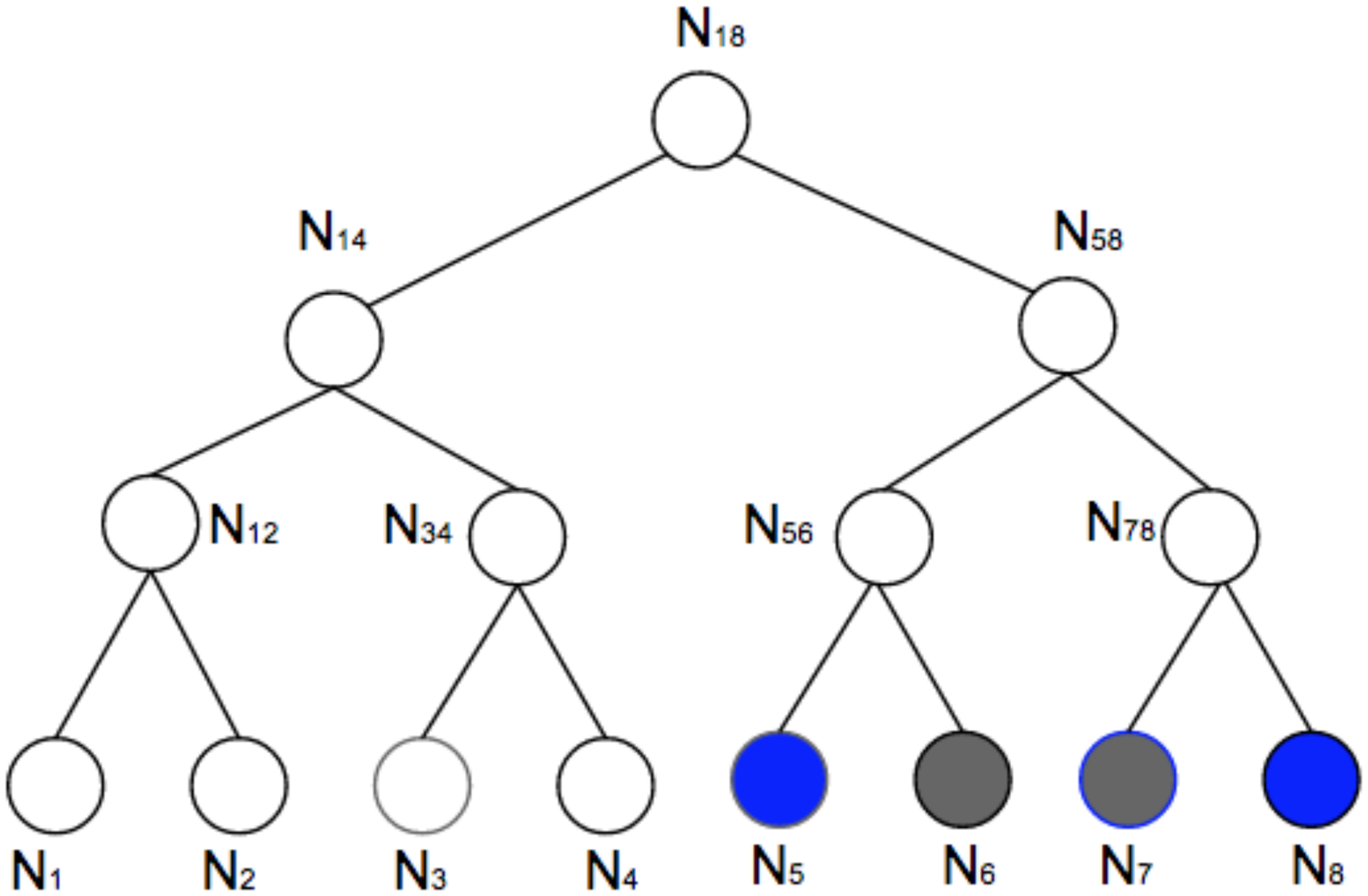}
\end{tabular}
\vspace{-0.1in}
\caption{\label{fig:AnotherMT} Illustration of MHT-based Verification \cite{devanbu2003authentic}}
\end{center}
\vspace{-0.15in}
\end{figure}

Devanbu et al. \cite{devanbu2003authentic} is one of the first works that use MHT to verify inclusiveness (i.e., soundness) and completeness of SQL query evaluation. 
\nop{Before outsourcing the dataset $D$ to SP, the data owner constructs a Merkle hash tree of $D$ for verification preparation. The leaves of the Merkle hash tree correspond to individual records of $D$. After the Merkle hash tree is constructed, the data owner sends both $D$ and the Merkle tree to SP, while keeping the root signature of the Merkle hash tree locally. }
They use Merkle hash tree (MHT) to construct the ADS. During query execution, the SP traverses the MHT and gathers the respective nodes in the MHT to construct the VO. For instance, given a single-dimensional range query $Q$, the VO includes the lowest common ancestor in MHT that cover the tuples in $Q$ and their nearest lower/upper bound nodes, as well as the nodes in the {\em proximity subtrees} that contain the boundary nodes occurring consecutively to the smallest (largest) element of $Q$.  Consider the Merkle tree in Figure \ref{fig:AnotherMT} as an example.  Assume the leaf nodes (colored gray) $N_6$ and $N_7$ falls in the query range. The nearest lower (upper, resp.) bound node is $N_5$ ($N_8$, resp). For VO construction, the SP identifies: (1) node $N_{58}$, which is the lowest common ancestor of nodes $N_6$ and $N_7$; and (2) the proximity trees that include $N_{58}$ and the two paths from $N_{58}$ to $N_5$ and to $N_8$. The SP includes the hash value of $N_{58}$ and the proximity trees in VO.  
%The VO consists of the hash values needs to reconstruct the root digest. i.e. the hash value of node $N_5$, $h_{N_5}$. The VO is sent to the client together with the query result $q$. 
The authenticity of $Q$ can be verified by re-constructing the root hash of $N_{58}$ using the hash values of $N_5$ - $N_8$, which are included in VO. 
The completeness of $Q$ can be verified via the proximity sub-trees in VO.  
Devanbu et al. \cite{devanbu2003authentic} also considered other types of SQL queries, including projection, join, set union and intersection. The key idea is similar to the verification of selection queries. We omit the details for these types of queries. 
Given a dataset $D$ that contains $n$ tuples and a query $Q$ whose result contains $t$ data points, the VO size  of $Q$ by \cite{devanbu2003authentic} is $O(t + log_2(n))$.

\nop{which corresponds to a non-empty contiguous sequence $q = \{
t_i,  \dots, t_j\}$ in the Merkle hash tree $\cal T$. The greatest lower bound (GLB) and lowest upper bound (LUB) of $q$ in $\cal T$ are defined as:

$GLB(q):= \{t|t\in r \wedge t.A < t_i.A$ 
$\wedge(\neg \exists t' \in r : t'.A > t.A \wedge t'.A < t_i.A)\}$

$LUB(q):=\{t|t\in r \wedge t.A > t_j.A$ 
$\wedge (\neg \exists t' \in r : t'.A < t.A \wedge t'.A > t_j.A)\}$

Given the sequence $q$, the {\em lowest common ancestor} $LCA(q)$ of $q$ in $\cal T$ is defined as the root of the minimal subtree in $\cal T$ that has all tuples in $q$ as leaf nodes. Consider a consecutive pair of tuples (leaf nodes) $s$ and $t$ in $\cal T$, the lowest common ancestor $LCA(s, t)$ along with the two paths from the root to $LCA(s, t)$ to $s$ and $t$ constitute the {\em proximity subtree} of $s$ and $t$, denoted by $ptree(s, t)$.  The proximity tree can show that $s$ $(t, resp)$ occurs as the smallest (largest, resp.) element of the range. 

For the returned query result $q$,  SP constructs the verification object of $q$ by the following steps.
First, it identifies $l := LCA(q \cup GLB(q) \cup LUB(q))$ in $\cal T$, as well as the path from $l$ to the root of $\cal T$. 
Next, it identifies the proximity subtrees which shows that $GLB(q)$ $(LUB(q), resp)$ occurs 
consecutively to the smallest (largest, resp.) element of $q$. Then the VO includes the root hash on $l$ and the hash values of all leaf nodes in the 
entire subtree with $l$ as the root. 

Consider the Merkle tree in Figure \ref{fig:AnotherMT} as an example.  Assume the leaf nodes (colored gray) $t_6$ and $t_7$ falls in the query range, apparently $GLB(q) = t_5$ and $LUB = t_8$. Following the algorithm, the node $l$ is identified as $N_6$.  The proximity trees include two paths from $N_6$ to $t_5$ and to $t_8$ respectively. The VO consists of the hash values needs to reconstruct the root digest. i.e. the hash value of node $N_5$, $h_{N_5}$. The VO is sent to the client together with the query result $q$. 

To verify $q$ is sound and complete regarding the range $[l,u]$, besides $q$, the client retrieves two additional tuples, $GLB(q)$ (respectively,
$LUB(q)$) which is immediately smaller (larger) than the smallest (largest) tuple in r with regard to the answer set $q$. Then the client 
uses the hash of the values in $q$ plus the the paths to $GLB(q)$ and $LUB(q)$ to construct a root hash that includes $q$, $LUB(q)$ and $GLB(q)$ as the leaf nodes. If the constructed root hash is identical to the root hash in VO, $q$ is verified as both sound and complete.
}
%Continue the example in figure \ref{fig:AnotherMT}, the client (1) calculates the hash values of the tuples in the query result, the GLB and the LUB : $h(t_5)$, $h(t_6)$,$h(t_4)$ and $t_7$, then (2) combines the VOs $H(t_3)$ , $H(n_1)$ and $H(t_8)$ to reconstruct the root digest of the Merkle hash tree. If the root digest matches the original one, then it verifies the authenticity of the query result. Then client (3) checks that the GLB and the LUB are indeed out of the query range $[l,u]$ and they are consecutive tuples to the edge of the query result. In this case, GLU $t_4$ is close to the left edge $t_5$ and LUB $t_7$ is close to the right edge $t_7$. If any  satisfied value between GLB and LUB are missing in the query result, the client cannot reconstruct the root digest of the Merkle hash tree. Since the tuples are sorted, there are no satisfied value below the GLB and beyond the LUB the. Thus it verifies the completeness of the query results.

\nop{
\textbf{VO Construction for Projection Query} In general, given a projection query, the service provider returns the hash values that need to reconstruct the root digest of the Merkle hash Tree same as the selection query. But the client has to eliminate the duplicated tuples after projection on the attributes, i.e the client does the projection locally. In special case, after the projection, large amount of tuples has the same value. Assume the projection query result has sequence of leaf nodes like $a_1a_2a_3b_1b_2b_3c_1c_2c_3c_4$, the same letter indicates the tuples has same values after projection. Then service provider provides authentication path VOs for $[a_1]$, $[a_3,b_1]$, $[b_3,c_1]$, and $[c_4]$, which are the border tuples of two different projection value.

\textbf{VO Construction for Join Query} For the join query, the data owner has to apply the join operation for tables locally and creates the ADS for the joined tables at the preparation step. For example, in order to support equi-join, the data owner needs to apply the full outer join of two scheme $R$ and $S$, the tuples appear in the join table has three types: (1) tuples from $R$ matches tuples from $S$ (2) tuples from $R$ do not match tuples from $S$, (3) tuples from $S$ do not match tuples from $R$. After the data owner creates the full outer join table, then creates the ADS based on the join table. When the SP receives any equi-join query from the client, like join $R$ and $S$ where the tuples match. The SP needs to return the boundary values and the authentication path as VOs. The VO construction process of join query is the same as we discussed in selection query.

\textbf{VO construction for Set Union and Intersection} Assume there are two sets of tuples $U$ and $V$ from two query evaluation respectively. For the two intermediate results, the SP will apply set operation on it. In order to verify the correctness and completeness of the set operation. The client needs to verify each tuple from the final result one by one. Take set union for example, each tuples t in $U \cup V$ , the service provider applies selection query for $t$, the SP provides the $VO$ to prove that $t$ exists in either $U$ or $V$. And the VO provides the authentication paths same as the VOs we discussed in the selection query. In order to prove the completeness, the client needs to go a single pass over the union to make sure no tuple is missing.

\textbf{Verification of Other Queries } 
In order to support multi-dimensional selection query, the multi-dimensional range tree (MDRT) are also  proposed in this paper. As shows in figure \ref{fig:MDRT},the key idea of MDRT is that each internal node contains a link to another Merkle tree whose leaf nodes are sorted by a different attribute. This second level Merkle tree includes the tuples that the first level internal node contains. Follow the same reasoning, the third level Merkle tree's leaf nodes are sorted by a different attribute. All the digests of Merkle hash tree in the MDRT are sent to the client.

The verification of projection query and join query follows the similar procedure as the selection query. For projection query, the client first needs to find the witness of each element $e\in q$ indeed exists in relation scheme. To do that, the SP needs to provide VO of the authentication path for the element, so that the client can reconstruct the root digest of the ADS. 
In special case, selection result is poorly distributed, for example, the final result $R$ only contains a few unique elements. Then checking each element in the intermediate result (before projection) will be very expensive because the final result is very small compared to the intermediate result. Thus, the SP can provide VOs of the authentication path for each border projection value. For example, if we have a sequence of tuples $\{1_1,1_2,1_3,1_4,1_5,2_1,2_2,2_3,2_4,3_1,3_2,3_3,3_4\}$, the intermediate result contains 12 elements, but the final result only contains 3 elements. The subscript indicates position of the value in the ADS. The border projection values are $1_1$, $1_5$, $2_1$, $2_4$, $3_1$ and $3_4$. The SP provides VOs for each consecutive border projection values rather than all elements. The client can get unique projected tuples from those border values and also verifies the authenticity when the digest matches the original one. Similarly, the VO of join query includes boundary values and the authentication path, thus verification of the join query is to reconstruct root digest of a bigger Merkle hash tree.

%ADS
\begin{figure}[!htb]
\vspace{-0.1in}
\begin{center}
\begin{tabular}{c}
\includegraphics[width=0.4 \textwidth]{./figs/mdrt.eps}
\end{tabular}
\vspace{-0.1in}
\caption{\label{fig:MDRT} An example of MDRT \cite{devanbu2003authentic}}
\end{center}
\vspace{-0.15in}
\end{figure}

\textbf{Verification of Set Union and Intersection} For set union, the client needs to check each element in the union set belongs to both sets. To do that, the client can use authentication path VOs provided by the SP to verify an element belongs to a certain scheme. For example, give two sets $R=\{1,2,3\}$ and $S=\{2,4,5\}$, the union is $U=\{1,2,3,4,5,6\}$, for each element $u\in U$, the SP provides VOs that $u$ belongs to scheme $R$ and $S$. The client verifies the authenticity by reconstructing the root digest. Then The client applies a single scan of $U$ to make sure no missing value in $R$ or $S$ that does not included in $U$. This step verifies the completeness of the query result. For set intersection, the authenticity verification is similar, check any tuple in the intersection result $I$ both belongs to $R$ and $S$. In order to check the completeness of the result, the client needs to check that each tuple in $R-I$ does not belongs to $S$. 
}

%{\bf Weakness of the approach} Overall, the verification method is not flexible because all precomputed the Merkle trees only support queries based on the sorted attribute. If the client build a Merkle hash tree for each attribute, then it will increase the storage space at the SP side. For the projection query the client needs to do computation locally, such as eliminating duplicated tuples, it involves high verification cost at the client side. For the set union and set intersection operation, the client needs to check the authenticity for each element in the final result which could involve large VO space. The author didn't consider the authenticity of the digest from the data owner. The client receives all the digest of the trees from the data owner, these digests may be changed during the transmission. Therefore, the data owner should sign the root digest with his/her private key, so that the client can verify the authenticity of the digest.

One weakness of  \cite{devanbu2003authentic} is that the VO size depends on the size of the query answers and the outsourced dataset. To eliminate this dependence on data and query size, 
Pang et al.  \cite{pang2004authenticating} designed  a new ADS named {\em verifiable B-tree} 
(VB-tree). VB-tree is similar to MB-tree (Section \ref{sc:mbtree}). It constructs a Merkle hash tree on top of the $B+$-tree by adding the digests on every B-tree node. The digests are computed using a cumulative and commutative hash function. 
%VB-tree leaf nodes.  
For the VO construction, the SP identifies the smallest subtree (called {\em enveloping tree}) in VB-tree that covers all the tuples in the query results. For a given range selection query, the VO of the selection results includes: (1) the signed digest for the node at the root of the smallest subtree (called {\em enveloping tree}) in VB-tree that covers all the result tuples of the query; and (2) the signed digest for each node in the enveloping subtree that represents those branches that do not overlap the result.  
The result verification is similar to \cite{devanbu2003authentic}; the client tries to re-construct the signed digest of the root of the VB-tree, and matches it with the local copy that is shared by the data owner. 
Unlike \cite{devanbu2003authentic} whose 
 VO contains the digests all the way to the root of
the tree index, the VO of \cite{pang2004authenticating} only needs to
contain proofs for the smallest subtree that envelops the
query result. Therefore, the VO size grows linearly with the size of the query results, but is independent of the data size. 
However, VB-tree cannot be used for the verification of result completeness.

\nop{For the selection queries, the VO includes: (1) the signed digest for the node at the root of the smallest subtree (called {\em enveloping tree}) in VB-tree that covers all the result tuples of the query; and (2) the signed digest for each node in the enveloping subtree that represents those branches that do not overlap the result. The result verification is also similar to \cite{devanbu2003authentic}. 
}

\begin{figure}[!htb]
\vspace{-0.1in}
\begin{center}
\begin{tabular}{c}
\includegraphics[width=0.45 \textwidth]{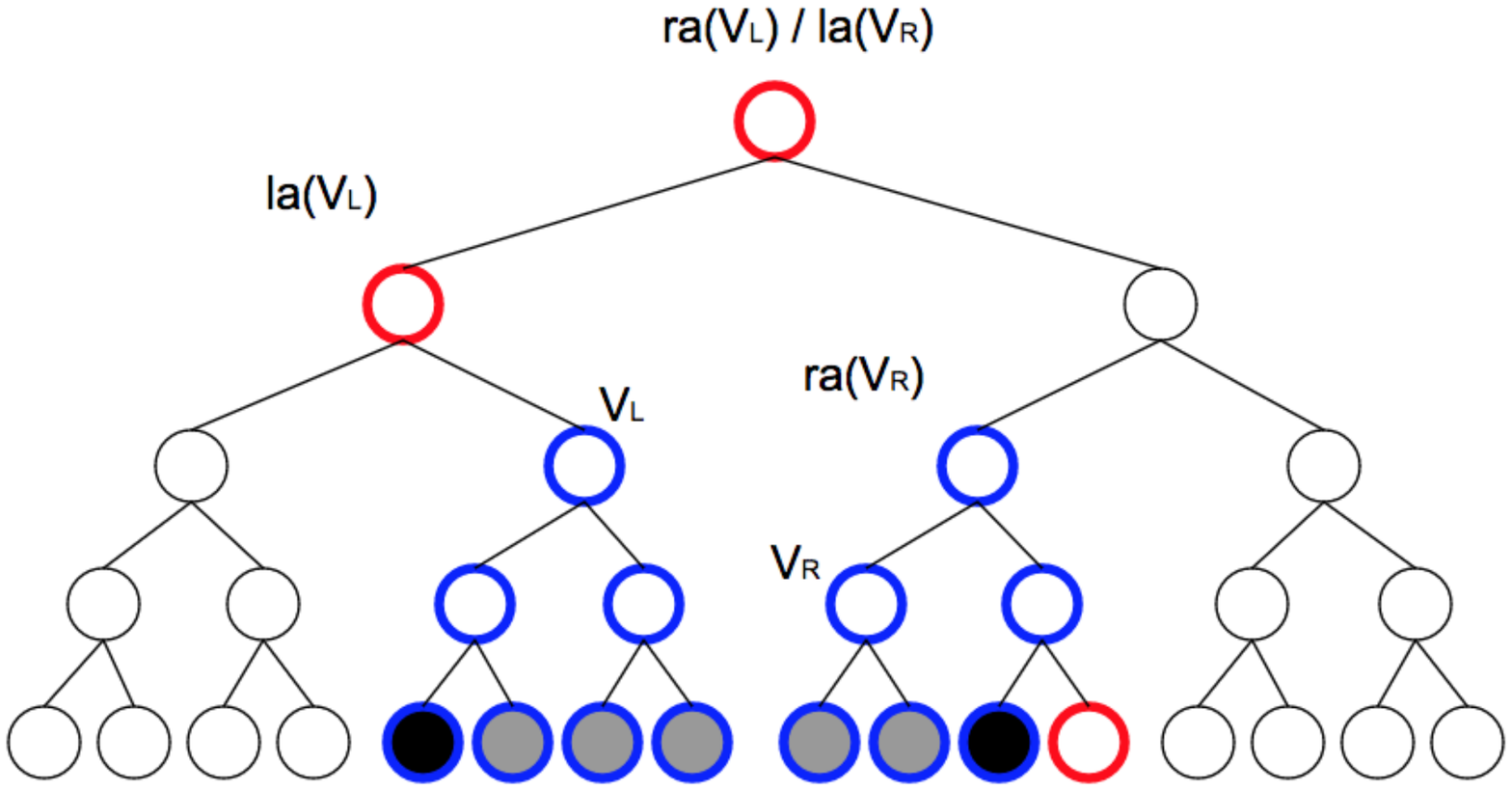}
\end{tabular}
\vspace{-0.1in}
\caption{\label{fig:hybrid-tree} An example of HAT ($la(v)$ (resp. $ra(v)$) denotes the least command ancestor of left (resp. right) adjacent nodes of $v$)}
\end{center}
\vspace{-0.15in}
\end{figure}

Although VB-tree \cite{pang2004authenticating} eliminates the complexity dependency on data size, it may suffer from the strong assumption that the SP must be trusted to some degree \cite{Nuckolls2005}. 
Furthermore, VB-tree uses a one-way function based 
on modular exponentiation, which was shown infeasible to handle large answer sizes \cite{Nuckolls2005}. 
%VB-tree may incur  excessive overhead to both VO size and verification time due to the use of expensive primitives for hash construction in the tree. 
Therefore, \cite{Nuckolls2005} proposed a new ADS named {\em Hybrid Authentication Tree (HAT)}, which incorporates fast hash functions with an efficient RSA based one-way quasi-commutative accumulator \cite{benaloh1993one}. 
%$H_n$ that with a given $n$, it takes the input $x, y$, and outputs $H_n(x, y) = x^y\ mod\ n$. To ensure $H_n$ is one-way, the modulus $n$ is chosen to be a rigid integer, meaning $n = p\bullet q$ where $p$ and $q$ are safe primes of the same bit size. 
A HAT is a binary search tree with data stored at the leaves, along with
a digest procedure that incorporates a fast collision intractable hash function and an accumulator.
 In HAT, the hash values of the leaf nodes are generated by applying the standard hash function on the data values, while the hash values of the internal nodes incorporate the hash values from the lowest common ancestors of all left and right adjacent nodes respectively. 
The final digest value of the HAT tree is computed as the accumulator of the hash of all the nodes along with the root hash. 
Given a range query $[a, b]$, SP returns all data items in the range and two boundary leaves in HAT. The verification is based on the concept of {\em covering nodes}, which is a set of nodes with disjoint HAT subtrees whose leaves are the exact answer to the range query. The verification method utilizes the fact that if the two covering nodes $v_R$ and $v_L$ are adjacent, then their leaf nodes are continuous and no values are missing between the two boundary values of the range $[a, b]$. This is verified by computing the hash values of internal nodes as well as the accumulator proof of the covering nodes and their immediate adjacent left and right nodes returned by SP. 
Finally, the client checks if the digest of root of the HAT matches the local copy shared by the data owner. 
Take Figure \ref{fig:hybrid-tree} as an example. The leaf nodes in gray color denote the query results. $V_L$ and $V_R$ denote the two covering nodes. The SP generates the regular hash value by using the hash function $f_1()$ and the accumulate hash value by using the hash function $f_2()$. Given a query $Q$, besides the query results, the SP returns the VO that consists of the following information: 
 (1) the nearest lower/upper bound nodes (denoted in black color); 
(2) the two accumulation values $f_2(root)$ excludes the digest of $f_2(V_L)$ and $f_2(V_R)$ respectively, 
(3) the digest values that are needed to reconstruct $f_1(V_L)$ and $f_1(V_R)$; and 
(4) the digest values that are needed to reconstruct $f_2(V_L)$ and $f_2(V_R)$. The digest values of the nodes in red (blue, resp.) color circle outline are returned (computed, resp.) by the SP (the client, resp.). With those digest values, the client first verifies that $V_L$ and $V_R$ are adjacent by checking if the digest of $ra(V_L)$ equals to the digest of $la(V_R)$. Then it checks if the two accumulation values together with $f_1(V_L)$ and $f_1(V_R)$ can reconstruct $f_2(root)$. 
%\Wendy{This example looks incorrect. Why $v_L$ and $v_R$ are also colored black? The query results and internal nodes should be colored differently. Bo, can you re-read paper [54], and use the notations in [54]? I don't think the paper uses the notation of $(A-d_{V_L})$ and $(A-d_{V_L})+d_{V_L}'$. \Bo{Revised}}

\nop{
Take the HAT in figure \ref{fig:hybrid-tree} as an example. Given a range query $[a, b]$, SP returns black leaf nodes in the range as query result, and (1) the digest information of gray nodes (2) $H(A-d_{V_R})$ and $H(A-d_{V_L})$ \Wendy{What does $A-d_{V_R}$ mean?} as VOs to help client to verify the HAT digest $\Sigma$ and the accumulation value $A$ respectively.
The accumulation value $A$ is computed using the accumulator H with some initial value x: $A=H(x,d_1,d_2,...,d_n)$, where the $d_i$ is the digest for each node in the tree. The HAT digest is computed by $h(A,d_{root})$. $V_R$ and $V_L$ are the two covering nodes of the query result.
The verification of soundness is to recover the HAT digest value, while the completeness of the query result is to check if $h(ra(v_L))=h(la(v_R))$, if so the two covering nodes $v_R$ and $v_L$ are adjacent, then their leaf nodes are continuous and no values are missing between the two boundary values of the range $[a, b]$ \cite{Nuckolls2005}. In order to verify if two covering nodes are adjacent or not, the digest of each node $v$ consists of the lowest common ancestor of its left and right adjacent nodes respectively. For example, $d_{V_R}=h(h(V_R),h(la(V_R)),h(ra(V_R)))$. The client needs to recover $d_{V_R}$, $d_{V_L}$ and $\Sigma$ by herself, then compares that if $H(A-d_{V_R}$ (from server) + $d_{V_R}) = H(A-d_{V_L}$ (from server) $=d_{V_L})$. At last the client compares if $\Sigma$ matches the original HAT digest.
}
\nop{
For each data item $r$, the data owner computes $key(r)$ and the hash output $h(r)$, where $key(r)$ is simply a unique identifier for $r$, e.g. the primary key for a relational tuple.

The digest of the authentication structure is constructed in the following way:

\noindent 1. $f_1$ is the standard Merkle hash using $h$, $d$ is the value associated with leaf node.
\begin{equation*}
f_1(v) =
    \begin{cases}
      h(f_1(lc(v)),f_1(rc(v))) & v\ is\ internal\\
      h(key(d),h(d)) & v\ is\ leaf
    \end{cases}              
\end{equation*}
where $lc(v)$ and $rc(v)$ are the left and right nodes to $v$ in HAT. 

\noindent 2. $f_2$ incorporates the hash values from adjacency lca nodes.
\begin{equation*}
	f_2(v) = h(f_1(v),f_1(la(v)),f_1(ra(v))),
\end{equation*}
where $la(v)$ and $ra(v)$ denote the lowest common ancestor of $v$
and all left and right adjacent nodes respectively. 

\noindent 3. A value $A$ is computed using the accumulator $H$ with some initial value $x$. If $v_1,v_2,...,v_m$ are the nodes of the tree,
\begin{equation*}
	A=H(x,f_2(v_1),f_2(v_2),...,f_2(y_m))
\end{equation*}
$H$ is quasi-commutative so the node order will not affect the result.

\noindent 4. Let $root$ denote the root of the tree. The final digest value $\Sigma$ is computed as
\begin{equation*}
	\Sigma = h(A,f_2(root))
\end{equation*}
}

\nop{
The leaf nodes of the HAT are tuples with key and values, which are sorted by the $key()$ value. For the internal nodes, except standard hash values, the data owner also computes the hash value of adjacency lowest common ancestor (lca) for each node which defined as $f_2(v)=h(f_1(v),f_1(la(v)),f_1(ra(v))$. The $f_1$ is the standard hash function, and the $la(v)$ is the lca of v and its left adjacent nodes in the tree. Note that, the node v have more than one adjacent nodes, but their lca is same, thus la(v) denotes one internal node. Similarly, $ra(v)$ is lca of node v and its right adjacent nodes.
}

%After creating those hash values, the data owner accumulates the hash values by an efficient RSA based one-way accumulator which is a one way hash function with quasi-communicative property. A hash function is one-way if given $x\in X$ and $y,y'\in Y$, it is hard to find $x'\in X$ such that $f(x,y)=f(x',y')$. The function f is quasi-communicative if $f(f(x,y_1),y_2)=f(f(x,y_2),y_1), \forall x\in X_k, y_i \in Y$. So the oder of the accumulated value will not change the final accumulation value. In particular, Benaloh and de Mare propose a one-way quasi-commutative accumulator based on an RSA modulus and prove it's security in \cite{Benaloh1994}. Given n, the authors define $H_n(x,y)=x^y \ mod\ n.$ $H_n$ is quasi-commutative by the laws of exponents: $(x^{y_1})^{y_2}=(x^{y_2})^{y_1}$. In ensure $H_n$ is one-way, the modulus $n$ is chosen to be a rigid integer. The data owner calculates the accumulation value $A=H(x,f_2(v_1),f_2(v_2),...,f_2(v_m))$, and then constructs the digest $\Sigma$ of the tree as $h(A,f_2(root))$, where $x$ is an initial value. The data owner sends the digest of HAT to the client and sends the database and HAT to the service provider.

\nop{
{\bf VO Construction for Selection Query} Given a range selection query, the SP needs to returns (1) the boundary nodes of the query range, (2) the accumulation value proof for the two cover nodes $v_R$ and $v_L$. The cover nodes $v_R$ and $v_L$ are the nodes with smallest subtree that cover the query range. (3) the hash values in the authentication path to the cover node $v_R$ and $v_L$, and (4) $f_2(root)$. Take the figure \ref{fig:hybrid-tree} for example, assume the black nodes are the query result, $v_L$ and $v_R$ are two covering nodes with smallest subtree whose leaf nodes contains all the query nodes. The VOs include (1) the boundary value $t_3$ and $t_8$. (2) the accumulation proof $z_{v_L}$ and $z_{v_R}$, where $z_{v_L}$ is accumulation value $\{A-{f_2(v_L)}\}$, and $z_{v_R}$ is $\{A-{f_2(v_R)}\}$. (3) no such VOs in this case (4) $f_2(v_5)$.
}

\nop{
Given a range query $[a, b]$, SP returns all data items in the range and two boundary leaves in HAT. The client can compute $key(d)$ 
and $h(d)$ for each data item $d$ in the answer except for the boundary leaves. The verification is based on the concept of {\em covering nodes}. 
Given a range query, the covering nodes is a set of nodes with disjoint subtrees whose leaves are the exact answer to the range query. 
Based on the covering nodes  $v_L$ and $v_R$, \cite{Nuckolls2005} designs the following verification method. The verification method utilizes the fact that if the two covering nodes $v_R$ and $v_L$ are adjacent, then their leaf nodes are continuous and no values are missing between the two boundary values of the range $[a, b]$.
The details of the Verification are as follows:

\noindent 1. Compute $f_1(v_L)$ and $f_1(v_R)$. The client can calculate $h(d)$ directly, then reconstruct the digest of internal node incrementally.  

\noindent 2. Compute $f_2(v_L)$ and $f_2(v_R)$ values and verify $v_L$ and $v_R$ are adjacent by checking if $f_1(ra(v_L))=f_1(la(v_R))$.

\noindent 3. Check that $H(z_{v_L},f_2(v_L))\stackrel{?}{=}H(z_{v_R},f_2(v_R))\stackrel{?}{=}A$, where $z_{v_L}$ and $z_{v_R}$ are the accumulator proofs returned by SP. 

\noindent 4. Check that $h(f_2(root), A)=\Sigma$. 
}

%The client computes $f_2(v_L)$ and $f_2(v_R)$ values and check that $v_L$ and $v_R$ are indeed adjacent by checking if $f_1(ra(v_L))=f_1(la(v_R))$ and both $H(z_{v_L}, f_2(v_L))$ and $H(z_{v_R}, f_2(v_R))$ equal to $A$.

\nop{
After reconstructing the hash values $f_2(v_L)$ and $f_2(v_R)$, the client stops reconstructing the hash value above the covering nodes $v_L$ and $v_R$. Instead, the accumulation value $A$ is computed. Finally, the client reconstructs the digest $h(A, h_2(v_5))$ and compares with the original digest which verifies the authenticity of the query result if the digests matches.

\Wendy{This part is different from Section 3.4 of paper [35]. Please copy from [35], with minor change of wording. }

\begin{figure}[!htb]
\vspace{-0.1in}
\begin{center}
\begin{tabular}{c}
\includegraphics[width=0.3 \textwidth]{./figs/hybrid-tree.eps}
\end{tabular}
\vspace{-0.1in}
\caption{\label{fig:hybrid-tree} An example of Hybrid Merkle Tree \Wendy{what is the difference of this figure from Figure 4?} \Bo{figure 4 we introduced the LGB and GLB with blue color. It is confusing to use it again.}}
\end{center}
\vspace{-0.15in}
\end{figure}

{\bf Weakness of the approach}
First of all, this method only supports selection query which is quite limited for other SQL queries like projection, join and aggregation queries. Secondly, this method is only efficient when the query result $T$ is small. Otherwise, the size of verification objects is still large. One challenge for the client is that the client cannot verify that $v_R$ and $v_L$ are the covering nodes with smallest subtrees. Otherwise, the SP can always return the root as the covering nodes, the verification of the query results is still depend on the height of the ADS which contradicts to the tuition of the paper. In addition, the client needs to accumulate large amount of node's hash values, and the computation cost at the client side is high because the accumulation function involves modular exponentiation which is very expensive. Thirdly, the author didn't consider the integrity of the ADS digest. If the data owner sends the ADS digest without signature, then the digest maybe incorrect at the client side.
}

\nop{
\begin{figure}[!htb]
\vspace{-0.1in}
\begin{center}
\begin{tabular}{c}
\includegraphics[width=0.4 \textwidth]{../figs/sig-tree.eps}
\end{tabular}
\vspace{-0.1in}
\caption{\label{fig:sigbased-tree} An example of Signature Based Tree}
\end{center}
\vspace{-0.15in}
\end{figure}

{\bf Verification preparation by data owner} The data owner prepares the ADS like the figure \ref{fig:sigbased-tree}. It is a signature based tree (SB-tree), each leaf node has a signature of the hash value of the current record concatenates with next record, and the first and last one can concatenate with a special marker. The data owner creates the signatures for leaf nodes and then create a Merkle $B^+$-tree based on those values which leads to a ABS-tree. The aggregated signature B-tree (ASB-tree), it is a scheme to aggregate the signature of hash values at leaf nodes. The scheme called Condensed-RSA \cite{Mykletun2004} can aggregate the signatures signed by the same signer, with the aggregated signature has the same size as individual signature which computes as fast as the individual signature too. }

\nop{
An EMB-tree consists of regular $B^+$-tree entries, which are augmented with an embedded MB-tree. At each node of the Merkle tree, it contains a triplets $(k_i,p_i,h_i)$, where $k_i$ is key of the node, $p_i$ is the pointer to the embedded Merkle tree at this node, and $h_i$ is the hash value of the embedded tree. The leaf nodes of the embedded Merkle tree at a EMB-tree node point to the children nodes of the EMB-tree node. Thus, given a query range, we start from the root and find the next children nodes we need to search at the embedded Merle tree until we reach the leaf nodes of the Merkle tree. 
}

\begin{figure}[!htb]
\vspace{-0.1in}
\begin{center}
\begin{tabular}{c}
\includegraphics[width=0.4 \textwidth]{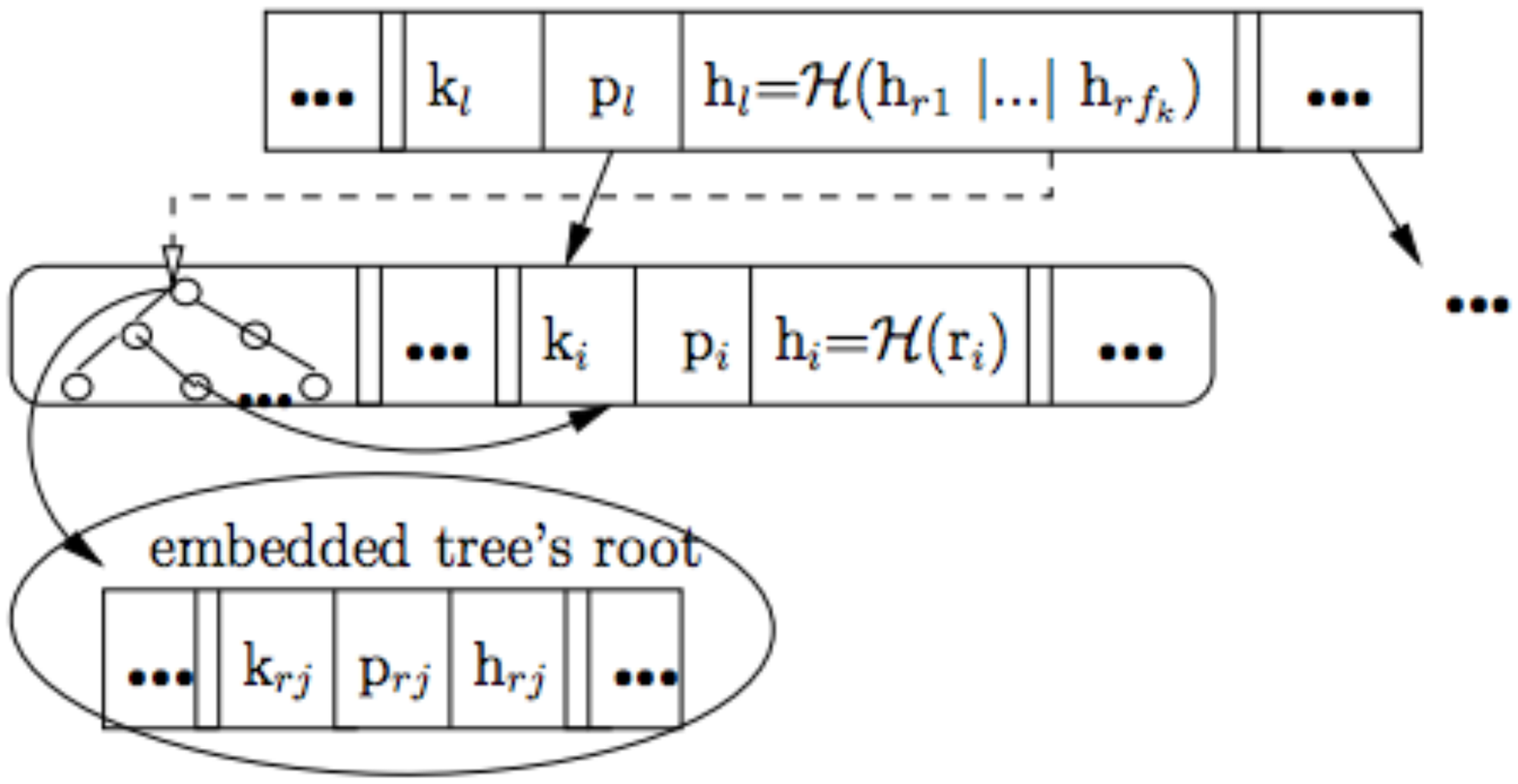}
\end{tabular}
\vspace{-0.1in}
\caption{\label{fig:EMB-tree} An example of EMB-Tree Node \cite{li2006dynamic} }
\end{center}
\vspace{-0.15in}
\end{figure}

So far the existing works mainly consider static scenarios, where owners never issue data updates. 
Li et al. \cite{li2006dynamic} are one of the first work that designs dynamic ADS for query authentication. 
They design three ADS structures, namely, 
the Aggregated Signatures with B+-trees (ASB-tree), 
the Merkle B-tree (MB-tree), and the embedded MB-tree (EMB-tree). The ASB-tree is a B+-tree whose leaf nodes contain hash values that are constructed from the consecutive pairs of tuples. For example, for the leaf node whose entry corresponds to the tuple $r_i$, assume the consecutive tuple of $r_i$ is $r_j$, based on some sorted order on a given attribute. Then $r_i$ corresponds to a leaf node in the ASB-tree, which is associated with a pair $(r_i, s_i)$, where $s_i$ = $S(r_i|r_j)$ ($|$ denotes some canonical pairing of strings that can be uniquely parsed back into its two components).
Chaining tuples in this way enables to verify that there are no tuples missing in-between the lower/upper bound node. 
%All the hash values  are signed by the data owner, and the internal node of the ASB-tree contain a signed accumulated hash value \cite{10.1007/978-3-540-30108-0_10} based on the signed hash value of its children node. Thus the root of the ASB-tree has aggregated signature for all the tuples. Given a query range, the client can verity the result by checking the aggregated signature rather than checking the individual signature for each tuple which significantly reduces the verification cost.
%\Wendy{Bo, you only have discussed EMB-tree. Please add the description of ASB-tree.} 
The MB-tree has been discussed in Section \ref{sc:mbtree}. An EMB-tree consists of regular $B^+$-tree entries augmented with an embedded MB-tree. Each EMB-tree node consists of a triplet $(k_i,p_i,h_i)$, where $k_i$ is key of the node, $p_i$ is the pointer to the embedded Merkle tree at this node, and $h_i$ is the hash value of the embedded tree. An example of EMB-trees is shown  in Figure \ref{fig:EMB-tree}. When there are updates on the data, only the path from the affected leaf (i.e., the updated data tuple) to the root is updated. The VO construction and query authentication procedure based on the EMB-tree is similar to \cite{devanbu2003authentic}. 
%To construct the VO, the SP follows a path from the root to the leaves of the external tree as in the normal $B^+$-tree. For every node visited, the algorithm scans all the triplets on the data level of the embedded  tree to find the key that needs to be followed to the next level. When the right node is found the SP also initiates a point query on the embedded tree of the node using this key. The point query will return all the  hash values that are needed for computing the concatenated hash of the node. To authenticate the query results the client uses the normal Merkle hash tree authentication algorithm to construct the hash value of the root node of each embedded tree (assuming that proper boundary information has been included in the VO for separating groups of hash values into different nodes) and then to compute the final hash value of the root of the EMB-tree.
Besides the design of the EMB-tree structure, \cite{li2006dynamic} 
conducted the analysis of some existing approaches, including the aggregated signatures for $B^+$-trees, the Merkle B-tree, and the EMB-trees, over all six metrics: (1) the computation overhead
for the owner, (2) the owner-SP communication cost, (3) 
the storage overhead for the SP, (4) the computation
overhead for the SP, (5) the client-SP communication
cost, and (6) the computation cost for the client. 
The experimental results show that, the ASB-tree has the highest  construction cost, the most expensive VO construction cost and verification cost. But it has the smallest VO size. The Merkle B-tree has the largest VO size, but the cheapest verification cost and update cost. The EMB-tree  achieves a good compromise between ADS construction overhead and verification cost. 

\nop{
{\bf VO construction for Selection Query} 
For ASB-tree, given a query range $[l,u]$. Instead of returning all signatures of the query results, the SP returns (1) the aggregated signature $s_{\pi}$ based on the query results, (2) the boundary values of the query result as VOs. Take the figure \ref{fig:EMB-tree} for example, if the leaf nodes $r_2$ and $r_3$ 
fall in the query range $[l,u]$, (1) the aggregation $s^{\pi}=Acc(S(r_1||r_2),S(r_2||r_3))$ and (2) the left boundary $r_1$ and the right boundary $r_4$ are returned as VOs.

For EMB-tree. The SP needs to return all the hash values needed for reconstructing the root signature of the MB-tree, as well as the boundary values for the query. However, the boundary values include the boundary values of the embedded Merkle tree as well as the boundary values at the leaf nodes of the Merkle tree.

\textbf{VO construction for Join Query} The join query verification is transferred to the selection query verification. Given two scheme $R$ and $S$, the SP selects the smaller scheme min(R,S), then for each record in the smaller scheme, the SP constructs the VO for each of the following selection queries: for each $r_k$ in $R$, $q_k$ = SELECT * FROM S WHERE $r.{A_j}=r_k.A_j$. The condition of the selection query is based on the conditions of join query.

{\bf Verification of Selection Query} 
For the ABS-tree, the client needs to (1) calculate the hash values of the returned records that fall in the query range, (2) calculate the aggregate signature of query results. (3) check the boundary values are next to the edge of the query result and also out of the query range.
Specifically, given query result $r_2$ and $r_3$, the client (1) calculates the hash values $h_2=S(r_2||r_3)$ and $h_3=S(r_3||r_4)$, (2) the aggregate signature of the query result $m^{\pi} = \prod_{\forall q}h_q(mod \ n)$ $=h_2*h_3 (mod) \ n$ and get the accumulated signature $m^{\pi}$, where n is the RSA modulus from the public key of the owner. 
If $s^{\pi}$ equals to $m^{\pi}$, the client verifies authenticity of the query result. Since the SP can only calculate $s_{\pi}$ by the signed hash values from the data owner. (3) The client checks that the boundary values $r_1$ and $r_4$ do not belong to $[l,u]$, and they are next to the left-most satisfied value and right-most satisfied value. This step verifies the completeness of the query result because the hash values combine the consecutive values together. If the SP does not return all the satisfied values between the boundaries, the client cannot get the original hash value of the consecutive records, then the second step will fail. Since there are no satisfied record beyond the boundary values, there are no missing record in the query results.

For the EMB-tree, the client checks the boundary values and reconstructs the root signature of the tree follow the authentication path. It is the similar process as we discussed in the paper \cite{devanbu2003authentic}. Due to the duplicity of the description, we omit this discussion here.

{\bf Weakness of this Approach}
The author mentioned that the signature based ADS incurs large storage space at the SP side. Except, the aggregated RSA signature still has large VO size and very high update cost compare to other ADS because one insertion or deletion will trigger a lot of expensive signature update. In addition, the authors didn't mention where does the update operations happen. Because the SP is not trusted, if the updates of the ADSs happens at the data owner side, the client needs to keep the ADSs locally which is against goals of the outsourcing scheme. Thus, some other schemes need to be introduced to allow the update happens at the SP side while keep the freshness of the database.
}

\nop{
\begin{figure}[!htb]
\vspace{-0.1in}
\begin{center}
\begin{tabular}{c}
\includegraphics[width=0.5 \textwidth]{./figs/AISM.eps}
\end{tabular}
\vspace{-0.1in}
\caption{\label{fig:AISM} An example of AISM \cite{yang2009authenticated} }
\end{center}
\vspace{-0.15in}
\end{figure}
}
Most of the existing works discussed so far mainly focus on range queries. Yang et al.\cite{yang2009authenticated} initiated the investigation of  result authentication of join operations. They proposed 
three authenticated join algorithms depending on
the ADS availability: (1) Authenticated Indexed Sort Merge Join
(AISM), which utilizes a single ADS in one of the base relations,
(2) Authenticated Index Merge Join (AIM) that requires an ADS
for both relations, and (3) Authenticated Sort Merge Join (ASM),
which does not rely on any ADS. 
\nop{
The AISM algorithm relies on the MB-tree and the ranked-list for authentication. 
%An example that illustrates the AISM algorithm is shown in Figure \ref{fig:AISM}. 
It treats the verification of join query of two relations as the verification of executing range queries for one relation based on the join attribute. Consider two relations $R$ and $S$, for each record $r \in R$, the SP executes a range query where the join attribute value of records in $S$ equals to the join attribute value of $r$. The SP returns digests for branches all the way to the root node of $S$ as VO. the verification process is similar to \cite{devanbu2003authentic}. The client needs to verify the signature of $S$ and $R$ and get the join table locally. }
Consider the join of two tables $R$ and $S$ on the attribute(s) $J$.  The AISM algorithm relies on the MB-tree and the ranked-list for authentication of join results of $R$ and $S$. In the pre-processing step, the outer table $R$ is sorted on the join attribute $J$, with the output as a rank list. The rank list outputs the {\em verifiable order} by which the client follows to verify the signature of records in the results. Given a set of records $R'\subseteq R$, the query of finding the join records of $R'$ in $S$ is equivalent to a single range query $Q = \sigma_{c_l≤J≤c_u}S$, where $c_l$ and $c_u$ are the lower- and upper- bound value of $R'$ on the join attribute $J$. The result of the range query evaluation (and thus the join results) can be authenticated by utilizing  the MB-tree \cite{li2006dynamic} of $S$ as the ADS. The query authentication method is similar to  \cite{devanbu2003authentic}.
In order to save communication cost, the AIM algorithm constructs two MB-trees for $R$ and $S$ respectively.  Based on the two MB-trees $T_R$ and $T_S$, the SP alternatively searches the matching tuples on each ADS and prepares the VO.
ASM method does not require any ADS at all. It sorts the data values of the join attributes $J$ of $R$ and $S$ as two rank list respectively. Then the two sorted tables $R'$ and $S'$ are merged as a single rank list, with the records that can be joined marked. Then it creates a bitmap $B_R$, in which the bit of the records of $R$ that have join partners in $S$ is set as 1, otherwise 0. Both signatures of $R$ and $|R|$, and $B_R$ are added to VO. The same process repeats for $S$. The client verifies the result by checking the bitmap values of the received results with VO. Compared with AISM and AIM, ASM is  less efficient as it does not utilize any ADS. However, this also brings flexibility for authentication of complex
queries such as multi-way joins. 
%So the client no longer needs all records of $R$ to recover the signature of $R$. She can performs index-traversal and leaf-scan on both tree to retrieve matching and boundary records by treating the join operation as a range query (similar to AISM) for VO construction.

\nop{
Without losing generality, let us consider the join of two tables $R$ and $S$ on the attribute(s) $J$.  In the pre-processing step, the outer table $R$ is sorted on the join attribute $J$, with the output as a rank list. The rank list outputs the {\em verifiable order} by which the client follows to verify the signature of records in the results. Given a set of records $R'\in R$, the query of finding the join records of $R'$ in $S$ is equivalent to a single range $Q = \sigma_{c_l≤J≤c_u}S$, where $c_l$ and $c_u$ are the lower- and upper- bound value of $R'$ on the join attribute $J$. The AISM algorithm utilizes the MB-tree \cite{li2006dynamic} to authenticate the results of $Q$.
}

\nop{
Authenticated Index Merge join (AIM) utilizes ADSs on the join
attribute in both input relations. Based on the two MB-trees $T_R$ and $T_S$ 
on $R$ and $S$ respectively, the SP alternatively searches the matching tuples on each ADS and prepares the VO. In particular, the SP starts from the first tuple $r_1$ in $R$, performs index-traversal and leaf-scan on $T_S$ to retrieve matching and boundary records by treating the join operation as a range query (similar to AISM) for VO construction. After it finishes, it switches the roles of $T_R$ and $T_S$, and performs
index-traversal and leaf-scan on $T_R$. It repeats until all records in the join results are traversed. Compared with AISM, AISM inserts all $R$ records into VO, while AIM only adds those with join tuples, and boundary records/digests for the remaining ones. Thus AIM reduces the communication cost with smaller VO size.
}

\nop{
{\bf Verification by data requester} 

AISM, AMI and ASM bind two authentication structure together and aim to save the VO size for join query. But the basic idea is to verify that one record exists in the scheme and let the client to build the join results locally according the VOs. 
The verification process is the following steps: (1) for each element in the rank-list of $R$, the client creates the join tuples if there is a matched tuple in $S$. Except that, the clients need to recomputes the hash value for those matched tuples and boundary values to reconstruct the root digest of $S$. (2) the client needs to check the boundary values for each element to make sure no missing values tuples between the boundary values. In addition, no hash values should be included between the boundary values. (3) the client uses the hash values to reconstruct the root digest of $S$ incrementally. (4) the client finally checks if the root digest of $S$ matches the original one or not. The verification steps verify the correctness of the join results because the client verifies the authenticity by recovering the root digest and also gets the join result by himself, and verify the completeness of the join results because the client can make sure no values between the boundary values are omitted by the SP. Otherwise, any changes to the original values or values omitted in the VOs will cause mismatch of the digest of $S$. 

{\bf Weakness of the methods}
The order of the ranked-list records are required for signature verification which could be very expensive for the client. If the join query involves multiple schemes, the data structure will become very complicated, it will definitely increase overhead at the SP side.
}

Li et al. \cite{li2010authenticated} extend the study to aggregation queries. They consider both static and dynamic cases for query authentication. 
For the static case, first, they show that authenticating SUM queries is equivalent to authenticating prefix sums \cite{ho1997range}. 
To authenticate prefix sum, Li et al. designed a new ADS structure named the {\em authenticated prefix sums tree} (APS-tree), which takes the format of of a multi-way Merkle hash tree, in which each leaf entry is converted into a base-f number. 
To construct the VO of a given query $Q$, the SP returns the $2^d$ corner prefix sum values as the part of VO, where $d$ is the number of dimensions of the dataset. Besides, the VO contains the hash values of the APS-tree nodes that are needed to authenticate the sum results, as well as the encoding of the path for each node. 
Based on the VO, the client can authenticate the query result by authenticating  each of the $2^d$ elements of the answer set by computing the hash of the root for each path and then comparing it with the local signature generated by the data owner.

\nop{ Without lost of generality, assume $d=2$ and given a query $Q=SUM(A_q)|S_1 = [a_1,b_1], S_2=[a_2,b_2]$. The query can be converted to four corner points in the prefix sum array from the rectangle area defined by the query range. In particular, $PS[b_1,b_2]-PS[b_1,a_2-1]-PS[a_1,b_2]+PS[a_1-1,a_2-1]$. }
\nop{
In particular,  given a $d$-dimension database $D$ and a query $Q = SUM(A_q)|S_1 = [a_1, b_1], \dots, S_d = [a_d, b_d]$ on $D$, 
$Q$ can be considered as an array $C$ in which each coordinate that contains one or more database tuples contains the SUM of attribute $A_q$ of these tuples. The prefix sum array prefix sum ($PS$) of $C$ can be represented as: $\forall x_j \in D_j,j\in[1,d]$:
$$PS[x_1,...,x_d]=\sum_{i_1=0}^{x_1}...\sum_{i_d=0}^{x_d}C[i_1,i_2,...,i_d].$$  In other words, an entry $PS[x_1, \dots, x_d]$ in the prefix sum array $PS$ stores the total sum of all entries before $C[x_1,...,x_d]$, including $C[x_1,...x_d]$ itself. 
To authenticate prefix sum, Li et al. designed a new ADS structure named the {\em authenticated prefix sums tree} (APS-tree). 
First,  the data owner coverts PS into a one dimensional array $PS'[k]$, where $k = \sum_{j=1}^{d-1}(i_j\times \prod_{n=j+1}^{d}M_n)+i_d$. 
Second, the data owner constructs the APS-tree in the format of a f-way Merkle hash tree. The root node has no label. The first level nodes
have base-$f$ representations 1, 2, $\dots$, $f$, from left to right respectively. The second level nodes have labels 11, $\dots$, 1f, 21, $\dots$, 2f, $\dots$, f1, $\dots$, ff, and so on all the way to the
leaves. A leaf entry in the APS-tree with $PS'$ offset $k$ is converted into a base-f number $\lambda_1 ...\lambda_h$ with $h$ digits. The APS-tree can be considered as a Merkle hash tree; the data owner constructs the root signature of the APS-tree. 

To construct the VO of a given query $Q$, the SP returns the $2^d$ corner prefix sum values as the part of VO. Besides, the VO contains the hash values of the MHT that are needed to authenticate each such element $k$, as well as the encoding of the path for each element. The SP returns VO to the client with the query result.

Based on the VO, the client can authenticate the query result by authenticating  each of the $2^d$ elements of the answer set. First, the client verifies each element by computing the hash of the root for each path and then comparing it with the local signature generated by the data owner. Next, using the query ranges $[a_i, b_i]$ and the domain sizes, the client maps each element’s position value $k$ back to
the coordinate in the $d$ dimensional prefix sum array. The client can authenticate whether all required elements are returned if all the elements are verified correctly.
}

\begin{figure}[!htb]
\vspace{-0.1in}
\begin{center}
\begin{tabular}{c}
\includegraphics[width=0.4 \textwidth]{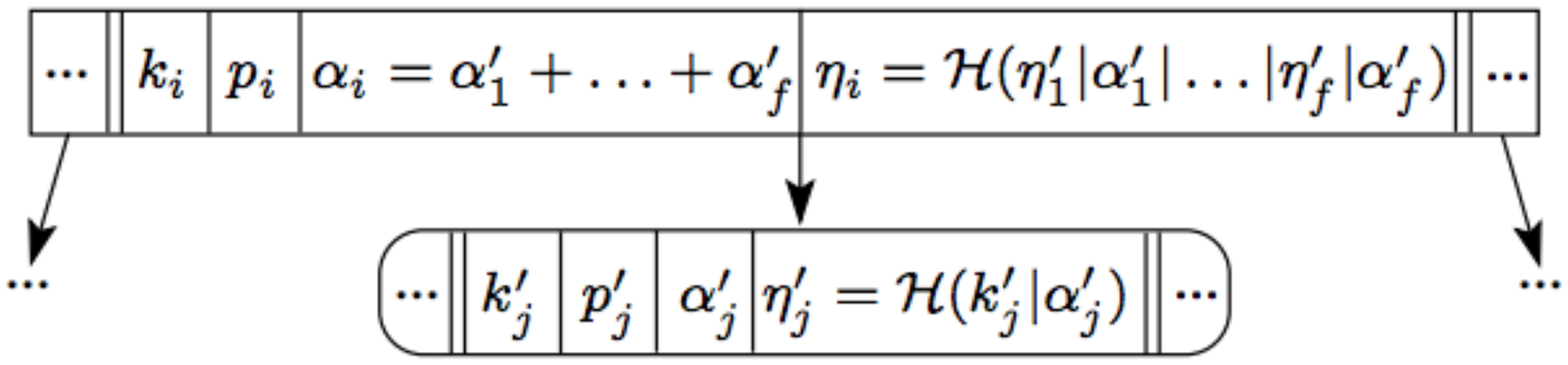}
\end{tabular}
\vspace{-0.1in}
\caption{An example of AAB-tree node \cite{li2010authenticated} ($k_i$: value range; $p_i$: pointer to the children node;  $\alpha_{i}$: aggregation of all children nodes; $\eta_{i}$: the node digest.}
\label{fig:AAB-tree}
\end{center}
\vspace{-0.15in}
\end{figure}

\nop{
The d-dimensional space can be reduced to a $|D_1| \times \cdot \cdot \cdot \times |D_d|$ array $C$. Every coordinate of the array that contains one or more database tuples stores the SUM of attribute $A_q$ of these tuples. 
In another word, $PS[x_1,...,x_d]$ stores the total some of all entries . IN d-dimensions, one only needs to consider the $2^d$ corner points in the prefix sum array from the d-dimensional hyper-cube defined by the query range. For example, the sum range query $SUM(A_q)|[a_1,b_1],[a_2,b_2]$ can be calculated by prefix sum as follows:
$PS[b_1,b_2]-PS[b_1,a_2-1]-PS[a_1-1,b_2]+PS[a_1-1,a_2-1]$. 
The range sum query is transferred to the corner points of the prefix sum, it will take less accumulate operation to verify the sum query which leads to low verification cost. The data owner coverts PS into a one dimensional array PS', and then constructs a f-way Merkle hash tree based on PS', i.e the authenticated prefix sums tree (APS-tree). The element $PS[i_1,...i_d]$ corresponds to element $PS'[k],k=\sum_{j=1}^{d-1}(i_j\times \prod_{n=j+1}^{d}M_n)+i_d$, where $M_n$ indicates the domain of attribute $A_n$. A leaf entry in the APS-tree with $PS'$ offset $k$ is converted into a base-f number $\lambda_1 ...\lambda_h$ with h digits. Retrieving $PS'[k]$ is possible by following the node with label that is the prefix of the label for $k$.
}
%AAB-tree
The APS-tree only can be used for static databases. It cannot work well for dynamic settings as it may lead to high querying cost. Therefore, Li et al. \cite{li2010authenticated} designed another ADS named {\em authenticated aggregation B-tree} (AAB-tree) for the dynamic case.  The AAB-tree is an MB-tree with each node associated with an aggregate value as the sum of the aggregate values of its children, and a hash value over the concatenation of both the hash values and the aggregate values of the children. An example of AAB-tree node shows in Figure \ref{fig:AAB-tree}. The AAB-tree can be used for authentication of one-dimensional aggregate queries in a dynamic setting since the owner can easily issue deletions, insertions and updates
to the tree, which handles them similarly to a normal B+-tree. It can be easily extended to deal with the multi-dimension aggregate queries in the similar way as the APS-tree. Other than COUNT and AVG, AAB-tree supports authentication of MIN and MAX as well, by replacing the SUM aggregate in each entry with the MIN/MAX aggregate. 
%Consider a query $Q$ that returns the sum on the attribute $A_q$ of those tuples that satisfy the range query $S_1 = [a, b]$. In the AAB-tree, each leaf node corresponds to a tuple $t$ with key $k = t.S_1$, aggregate value $\alpha = t.A_q$, and an associated hash value $\eta = H(k|\alpha)$. Each internal node of the AAB-tree has keys $k$ computed in the same way as in the normal $B^+$-tree. Each key is associated with an aggregate value $\alpha$ = $\alpha_1 + \dots + \alpha_f$ (i.e., the sum of the aggregate values of its children), and a hash value $H(\eta_1|\alpha_1|\dots|\eta_f|\alpha_f)$, which is to compute the hash value over the concatenation of both the hash values and the aggregate values of the children. An example of AAB-tree node shows in Figure \ref{fig:AAB-tree}. VO construction and verification can be performed in the similar way as MB-tree. 
\nop{
The AAB-tree is a MB-tree with aggregation sum value stored in the node. The aggregation sum value in a internal node is the summation of the values stored in its children node. Take figure \ref{fig:sigbased-tree} for example, the node in the left second level, except the hash values $h_{12}$, it also contains a summation value $v_1+v_2$ of its children node's values $v_1$ and $v_2$. In order to save the verification cost, the SP will return the minimum cover set (MCS) which covers the entries in the range completely. The VO of the query range includes those tuples and hash values used to reconstruct the root signature as well as the summation value used to verify the aggregation result. An example of AAB-tree node shows in Figure \ref{fig:AAB-tree}, where $k_i$ denotes the value range, $p_i$ denotes pointer to the children node, the $\alpha_{i}$ is the accumulation value of all leaf node values included in this subtree, and the $\eta_{i}$ is the digest of the node which is the hash value on the concatenation of the aggregation values and hash values of children nodes.
}

\nop{
%AAR-tree
The AAR-tree is MR-tree which is similar to the AAB-tree, but because it can support multiple range query, so each node in the tree is associated with multiple aggregation value from different attribute.

For all the ADSs, the data owner has to sign the digest of the root and send the root signature to the client.

{\bf VO construction by SP} 

\textbf{Selection Query} Even though the authors focus on the verification of aggregation query, but AAB-tree and AAR-tree can support selection without changing the data structure. The SP provides the verification path for each selection query like the Merkle tree we discussed the paper \cite{devanbu2003authentic}. For brevity, we omit the details of the VO construction here.

\textbf{Aggregation Query} Take the SUM query for example. For APS-tree, the SP returns the $2^d$ corner prefix sum values as query results, so that the client can calculate the final result based on the prefix sum values locally. In order to let the client verify the correctness of the prefix sum values, for each prefix sum value, the SP needs to return the authentication path that needed to reconstruct the root digest of APS-tree as VOs.

For AAB-tree, the SP finds the minimum covering set $MCS$ for the tuples that satisfies the query range. $MSC$ is a set of leaf nodes or internal nodes that cover exactly the same tuples that satisfy the range query. Then SP returns (1) the left and right boundary values (two leaf nodes), and (2) the hash values in the authentication path as VOs. For example,
selection query results are the black leaf nodes shows in the figure \ref{fig:AAB-example}. The MSC is tuples $t_4$ and internal node $N_6$. The SP returns (1) the boundary value $t_3$ (2) hash value $h_{12}$, $h_{56}$, $h_{78}$ and the accumulation value $S_{58}$ as VOs.

\textbf{Other aggregation query} The client can change the aggregation function to MIN/MAX in the ADSs, and the VO construction process is the same. As for COUNT query, it is a special case of SUM and is thus handled similarly. As for the MEDIAN and QUANTILE queries, they are transfered to the special selection query first, then the VO construction does not have any difference. For brevity, we omit the details here.

\begin{figure}[!htb]
\vspace{-0.1in}
\begin{center}
\begin{tabular}{c}
\includegraphics[width=0.4 \textwidth]{./figs/AAB-example.eps}
\end{tabular}
\vspace{-0.1in}
\caption{\label{fig:AAB-example} An example of AAB-tree}
\end{center}
\vspace{-0.15in}
\end{figure}

{\bf Verification by data requester}
For APS-tree, given the domain size for attribute and the fan-out $f$ of the APS-tree, the client needs to authenticate each
of the $2^d$ elements of the answer set. First, the client verifies each element by computing the hash of the root for each path and then comparing it with the digital signature. If the client follows VO of all the element can finally leads to the correct match of the digital signature, then it verifies that the SP return the correct and complete prefix sum values, the client can get the final sum value according to the prefix sum values.

For AAB-tree, continues the example in figure \ref{fig:AAB-example}. The client (1) calculates the hash values for each element in the MSC: the hash value of $t_4$ which is $h_3$ and hash value of $N_6$ which is $h_58$.  (2) checks the boundary values are out of the query range and next to the edge of the satisfied tuples in the leaf nodes and calculates its hash value $h_4$ (3) reconstructs the hash values of the internal nodes and it's accumulation values returned in the VO incrementally. 
If the client can verify the root signature of the query result is correct, then it proves result of the selection query is correct and complete. The accumulation values in MSC is the final result, thus the client add $t_4$ and $s_{58}$ together and get final result of the sum query. 

{\bf discuss about the update}
The insertion or deletion tuples to the ADSs will change the accumulation stored in the tree. So beside updating the data structure, all the accumulation value of the affected nodes have to be updated.

{\bf weakness of the methods}
The size of the APS tree is depend on the domain of each attribute and the dimension of the database, which could lead to large storage cost. The updating a single tuple in the database might necessitate updating the whole tree.
}

\begin{figure}[!htb]
\vspace{-0.1in}
\begin{center}
\begin{tabular}{c}
\includegraphics[width=0.5 \textwidth]{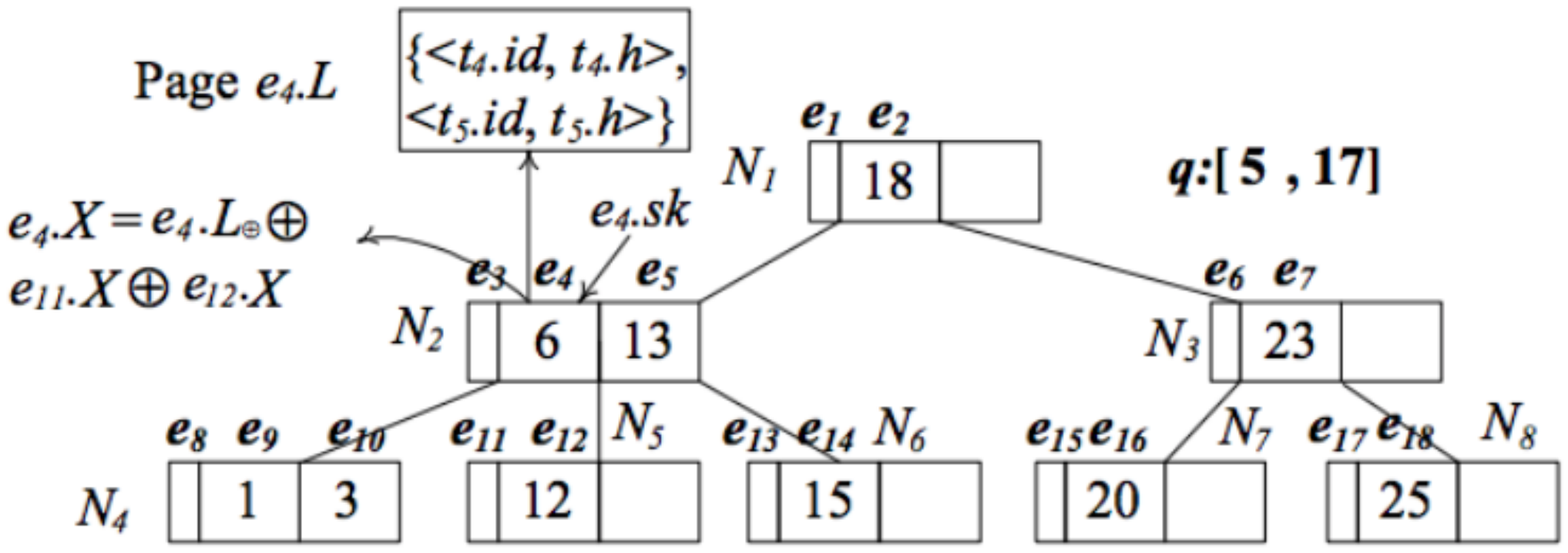}
\end{tabular}
\vspace{-0.1in}
\caption{\label{fig:XOBTree} An example of XB-tree  \cite{4812487}}
\end{center}
\vspace{-0.15in}
\end{figure}

Most of the existing ADS-based verification methods involve the data owner into the verification setup for ADS construction, which may not be feasible for the data owner who may have limited computational resources. Furthermore, the VO is typically large, which may bring significant communication overhead. To address these two drawbacks, Stavros Papadopoulos et al. \cite{4812487} separate authentication from query execution by exploiting a {\em trusted entity} (TE). 
The high-level idea is that the data owner sends her dataset to TE. TE generates the digest of each record by using a one-way, collusion-resistant hash function.
When the client receives the result of a range query from the SP,  the client sends the query to TE. TE evaluates the query on the dataset received from the data owner,  and produces a {\em verification token}, which is constructed by applying the exclusive-OR (XOR) operator on the digests of tuples in the query result. TE transmits the verification token to the client. The client computes the XOR of the digests of the records in the returned results by the SP, and matches it against the verification token by TE. 
%The high-level idea is explained as follows. First, the data owner sends the dataset $D$ to TE. For each record $r_i\in D$, TE generates a triple $r_i=<r_i.id,r_i.a,r_i.h>$, where $r_i.id$ is the ID of the record, $r_i.a$ is the value of $_i$ on attribute $a$, and $t_i.h$ is the hash value $r_i$ generated by a one-way, collusion-resistant hash function. When the client receives the result $R^S$ for a range query $Q$ from the SP,  the client sends $Q$ to TE. TE evaluates $Q$ on its collection of triples, and determine $TS = \{t_i, \dots, t_j\}$ that satisfies $Q$. It then produces a {\em verification token} $VT =  t_i.h \oplus \dots \oplus t_j.h$, where $\oplus$  is the exclusive-OR (XOR) operator of the digests of tuples in a set $S$. TE transmits $VT$ to the client. The client computes $VT_{R}$ (i.e., the XOR of the digests of the records in the returned results) locally, and matches it against $VT$. The match of $VT$ and $VT_{R}$ proves the authenticity, soundness, and correctness of the query results. 
To facilitate the authentication process,  \cite{4812487} includes the design of a new ADS named XOR B-Tree (XB-Tree) that integrates the B-tree with XOR values. The XB-tree is very similar to the B-tree; the difference is that each entry of the XB-tree 
is associated with a bit string that represents the results of XOR operator.
%$e$ of an internal node takes the form of $<e.sk,e.L,e.X,e.c>$, where: (1) $e.sk$ is a search key, (2) $e.L$ is a pointer to a disk page containing the IDs and digests of the tuples in $T$ whose values on the attribute $a$ equal $e.sk$, (3) $e.X$ is, and (4) $e.c$ is a pointer to the child node e. Given two consecutive entries in the same internal node $e_{i-1},e_i$, the sub-tree rooted at node $e_i.c (e_{i-1}.c)$ contains entries whose search keys are larger (smaller) than $e_i.sk$. 
An example of XB-tree is shown in Figure \ref{fig:XOBTree}.
Search on the XB-tree and VO construction are very similar to those on the MB-tree.

\nop{
The figure \ref{fig:XOBTree} illustrates an $XB$-Tree indexing tuples $t_1-t_{14}$ with search keys $\{1,3,4,6,6,12,$ $13,15,18,18,20,23,23,25\}$. Take $e_4$ in node $N_2$ for example: (1) $e_4.sk=6$, (2) $e_4.L$ points to the page that accommodates tuples $\{<t_4.id,t_4.h>,<t_4.id,t_5.h>\}$, (3) $e_4.c$ points to node $N_5$, (4) $e_4.X$ is equal to $e_4.L_{\oplus} \oplus e_{11}.X \oplus e_{12}.X=t_4.h \oplus t_5.h \oplus t_6.h$. The first entry $e_3$ of node $N_2:e_3.X=e_8.X \oplus e_9.X \oplus e_{10}.X$ and $e_3.c=N_4$, but there are no $e_3.sk$ and $e_3.L$ attributes. 
}

\nop{
The client will send the query request to both SP and ET, the $ET$ returns digest $VT$ and the ST returns result $RT$. The client computes $RS_{\oplus}^{SP}$, i.e. the $XOR$ of the digest of the records in $RS^{SP}$ locally, and checks if $RS_{\oplus}^{SP}$ matches $VT$. If the SP returns the client a corrupt result $RS_{\oplus}^{SP} = (RS-DS) \cup IS$, where $DS$ is a subset of the actual results an $IS$ is a set of fake tuples. The SP can escape detection iff. the $VT$ produced for $RS$ is equal to $RS_{\oplus}^{SP}$. This happens iff. $RS_{\oplus}^{SP} = ((RS-DS) \cup IS)_{\oplus} \Leftrightarrow RS_{\oplus}^{SP} = RS_{\oplus}^{SP} \oplus DS_{\oplus} \oplus IS_{\oplus} \Leftrightarrow DS_{\oplus} = IS_{\oplus}$. The authors prove that it is computationally infeasible for the $SP$ to find sets of records DS and IS such $DS_{\oplus}=IS_{\oplus}$.
}
\nop{
\Bo{I copied part of the sentence from the paper with minor word changing}
There are so many different ADSs, some of them must have some properties in common, thus Charles Martel et al. \cite{Martel2004} proposed a more general model called search $DAGs$ for authentication data structure. The existing work on simple structures such as binary trees, multi-dimensional range trees, tries, and skip lists can be placed easily in the Search DAG framework. However, unlike the paper \cite{839932} which describes  a method for generating authenticated versions of arbitrary data structures. Their approach is to authenticate each part of the data structure accessed. Verifying the result of a membership query in a dictionary $D$ incurs a cost of $O(T_qlog|D|)$ where $T_q$ is the query time in the unauthenticated dictionary. In the paper \cite{Martel2004}, the authors' approach eliminates the log factor overhead by incorporating the internal structure of the unauthenticated data structure in a very general way. 

In particular, a search $DAG$ consists of a directed acyclic graph (DAG) $G=(V,E)$ and an associated deterministic search procedure $P$. DAG has a unique source node $s$ and nodes of the graph contain information relevant to the search procedure, like the successor. The digest of $DAG$ is constructed in the following way:

\begin{equation*}
f(v) =
    \begin{cases}
      h(a(v)), & v\ sink\ node,\\
      h(a(v),f(v_1),f(v_2),...,f(v_K)) & otherwise
    \end{cases} 
\end{equation*}

where $v_i,...,v_k$ are the successor s of $v$ in order. The $DO$ sends the digest value $f(s)$, the hash function and the search procedure $P$ to the client for verification.

For each query, let $v_1(=s),v_2,...,v_r$ be the nodes visited by $P$ and let $u)1^i,...,u_{k_i}^i$ be the successors of $v_i$. The SP needs provides following values as VOs:
$a(s),f(u_1^1),...,f(u_{k_1}^1)$; $a(v_2),f(u_1^2),...,f(u_{k_2}^2)$;...;
$a(v_r),f(u_1^r),...,f(u_{k_r}^r)$; Each vector is ended by a ";" and is called a step of the $VO$. The client can verifies the result by hashing the individual items $a(s),f(u_i^1)$ and checks if it matches $f(s)$ or not. Then the client finds the next visited node $v_2$ in the successors of $s$. If $v_2$ is the $k$th successor of $s$, the client first hashing the node $v_2$ and checks if it matches $f(u_k^1)$. When it matches, the client can input $a(v_2)$ into $P$, and continue to find the next visited node in the successors of $s$ or $v_2$. After inputing $a(v_r)$, $P$ will produce the answer and halt.
}

So far most of the techniques incorporate the Merkle hash tree with a data index to construct ADS. As part of query execution, the SP traverses the ADS and gathers the respective nodes to form the VO. 
Mouratidis et al. \cite{Mouratidis:2009:PMD:1541533.1541543} designed a novel scheme called {\em Partially Materialized 
Digest scheme} (PMD) that separates MHT from the data index. 
Instead, it uses a {\em main index} (MI) for querying, and a separate {\em digest index} (DI) for query authentication. 
The MI is a standard $B^+$-tree on key. Range queries are evaluated on MI in the same way as on the $B^+$-tree. The DI is constructed as follows. First, for each leaf MI node $E_i$, a lower MHT is built on top of it. Then 
the upper MHT is built over the roots of the lower trees. The
root of the upper tree is signed with the owner's private
key. Similar to \cite{devanbu2003authentic}, for a given range query $Q$, 
 besides the records that satisfy the range of $Q$, the SP identifies 
two boundary records, $p^−$ and $p^+$, falling immediately to
the left and to the right of the query range. The VO is constructed by using the DI only. 
It contains (1) the signed DI root, (2) all left sibling hashes to the path of $p^−$, and (3) all right sibling hashes to the path of $p^+$. To verify if the received result is correct and complete, the client  combines $R^S$ with the components (2) and (3) of VO to reconstruct the root of DI. If the signature of 
the root of DI (i.e., component (1) of the VO) matches the
locally computed root hash, the result is correct and complete.  
By separating these two indice, PMD significantly outperforms
the performance of MB-tree \cite{li2006dynamic} in terms of query response time, storage overhead and index construction cost. 

\nop{
. In this case, it will facilitate both the query speed on the data as well as retrieving the VOs for that query. 
The main idea is about creating a set of smaller Merkle hash trees for the external nodes for the data values index and build another Merkle hash tree (MHT) based on the root of these smaller Merkle hash tree. For example, the index of the data values could be a $B^+$-tree, divide the last level of the $B^+$-tree, i.e. the external nodes into several groups and then build a MHT for each group. The verification process is similar to the process discussed in \cite{devanbu2003authentic}, the SP returns the query result which can feed into a subset of these smaller MHTs and other hash values which are necessary to reconstruct the root signature of the bigger MHT. Since the index of the digest information is suited to disk-resident data, the authors proposed effective index compression method which leads to low space consumption and fast VO computation. 
}

\nop{
Most of the published work tree-based verification method requires the SP to store at least $O(nlogn)$ verification objects and the user needs $O(logn)$ of those VOs to verify the query result. However, for 2-dimension grid such as GIS image data, Mikhail J. et al. \cite{4497478} proposed more efficient ADS based on grid of the data which can reduce the size of verification of object stored at SP to $O(n)$. Besides, the SP needs constant time to find constant hash values that needed for the client to verify the query result. 

For one dimensional data corresponds to the while interval $[1,n]$, the data owner creates an authenticated tree from the root which contains $\sqrt[]{n}$ children, those $\sqrt[]{n}$ children nodes evenly divide the interval from their parent . Continue this process until the internal node of the tree only contains two cells of the grid. Thus, the height of the tree will be $O(nlognlog)$. Some auxiliary information about its interval for each node has to be created. Let $I_{v,1},...I_{v,\sqrt[]{l}}$ be the sub-interval of $I_v$ associated with the $\sqrt[]{l}$ children of v, the following auxiliary structures are stored at each node $v$:
\begin{itemize}
\item The VO of each individual cell in the interval $I_v$
\item A  $\sqrt[]{l} \times \sqrt[]{l}$ table $\tau_v[i,i']$, 1$\leq i \leq i' \leq \sqrt[]{l}$, contains the integrity VO of the region of cells numbered from $i\sqrt[]{l}$ to $i'\sqrt[]{l}$.
\item For every cell $i$ that falls in $I_{v,j}$, an array $L_v[i]$ that contains the VO of the cells of $I_{v,j}$ that are $\leq i$.
\item For every cell $i$ that falls in $I_{v,j}$, an array $R_v[i]$ that contains the VO of the cells of $I_{v,j}$ that are $\geq i$.
\item The information needed to support nearest common ancestor queries in constant time (see, e.g., \cite{}).
\end{itemize}

When processing a query $q\in[a,b]$, the SP can determine the VOs in constant time as follows:
\begin{itemize}
\item If a=b then returns VO of $a$, otherwise continue with the next steps.
\item Compute the node $v$ in the authenticated tree that is the nearest common ancestor of leaves $a$ and $b$. Let $L_{v,i}$ (resp., $I_{v,i'}$) be the sub-interval of $I_v$ to which $a$ ($b$) belongs.
\item Returns $R_v[a]$,$L_v[b]$, $\tau[a',b']$, where $a'$ ($b'$) is the rightmost (leftmost) cell of $I_{v,i}$ (resp., $I_{v,i'}$).
\end{itemize}
In such settings, the client cannot know the hash values except the data falls in the query range, the client verifies the result by matching the hash values of the query result with the VOs provided by the SP.

\Bo{For two dimensional grid, the scheme is so complicated, I cannot fully understand yet, what I understand is it defines the boundary values in the grid as VOs}

\Wendy{the following papers are missing:
1. Separating authentication from query execution in outsourced databases, 
2. A general model for authenticated data structures
3. Partially materialized digest scheme: an
efficient verification method for outsourced databases
4.  Ensuring correctness over untrusted private database
5. Efficient data authentication in an environment
of untrusted third-party distributors.}
}

\nop{
Wei Wei dt al. \cite{Wei2014} studies how to make the verification of the outsourcing query result more practical at the SP side. The main idea is to transfer the MB-tree into a table, and let the SP to retrieve the verification object more efficient. In specific, the author defined radix-path identifier (rpid) for each node in the MB-tree. Let $i$ denotes an index of an pointer, $f$ denotes the fanout of the MB-tree and $r_b$ denotes the radix base, the rpid is defined as 
\begin{equation}
rpid =
    \begin{cases}
      i & if \ l==0,\\
      rpid_{p} & if \ l>0.
    \end{cases}              
\end{equation}
where $l$ is the level of the tree and $rpid_p$ is the parent of $rpid$. The $rpid$ has the following properties: (1) pointer identifiers is continuous in a node, but it is not continuous between nodes; (2) identifier of parent node can be calculated directly; (3) the min and max identifier can be easily derived. 
The authors proposed two methods to transfer the rpid into relational scheme. The first one is the single authentication table (SAT) which contains 4 attributes : id, rpid, hash, level. Those attributes describe the node in the MB-tree. Since the SAT suffers large update cost, so the authors proposed the second method: level-based authentication table (LBTA) which consists of several scheme for each level of the MB-tree. Since the level information is presented by the scheme itself, thus it only contains the $id$, $rpid$ and the hash values as attributes.

\nop{
There is an example of the LBTA shows in figure \ref{fig:LBTA}.

\begin{figure*}[!htb]
\vspace{-0.1in}
\begin{center}
\begin{tabular}{c}
\includegraphics[width=0.7 \textwidth]{./figs/LBAT.eps}
\end{tabular}
\vspace{-0.1in}
\caption{\label{fig:LBTA} An example of AAB-tree \cite{Wei2014}}
\end{center}
\vspace{-0.15in}
\end{figure*}
}
In this way, the authentication table is outsourced along with the original database, the authentication table can provide efficient data retrieval for integrity verification. 
The experiment results show that verification objects for each query result can be efficiently retrieved in the authentication table.

The VO constructions for SP and the verification process for the client is exactly the same as we discussed in the paper \cite{devanbu2003authentic}.
}

\vspace{-0.05in}
\subsection{Signature-based Authentication}

An alternative verification approach is to generate the verification object in the format of a cryptographic signature. Intuitively, using the signature of individual tuples only can verify the authenticity of the returned tuples, i.e., if the tuples indeed exist in the original dataset. To verify the soundness and completeness of query results, a new integrity authentication method  that relies on {\em signature aggregation} for VO construction is designed. 
The key idea is to generate an authenticated chain from the signatures of tuples, and outputs the aggregated signature of the authenticated chain as the  VO. 
In this section, we overview the works based on signature aggregation. 

% Authentication and Integrity in Outsourced Databases
Mykletun et al. \cite{mykletun2006authentication} (appeared in 2004, published in 2006) proposed the first signature-based method for range query authentication. By their approach, the data owner constructs a standard RSA signature  \cite{rivest1978method} for each record, and sends these signatures together with the dataset. The SP uses the {\em Condensed RSA} to compress the RSA signatures \cite{rivest1978method} that are associated with all records in the query results into a single signature. The single signature then is sent to the client together with the query results as its VO. The client can verify the correctness and authenticity of the returned query results by reconstructing the single signature from the signatures of returned result tuples. 
%\Bo{I think aggregate the tuples values to verify the aggregated single signature, rather aggregate each signature}
The major advantage of this method is that the VO size is significantly reduced by combining multiple tuple signatures into a single signed message.
%By this approach, the preparation process at the data owner side is the same as the standard RSA approach; the data owner generates the public key $pk=(N, e)$ and secret key $sk=(d)$, where $ed \equiv 1\ mod\ \phi(N)$, the order of the group $\mathcal{Z}_N^*$ is $\phi(N)=(p-1)(q-1)$, and $p$ and $q$ are two random prime numbers. Moreover, for each tuple $t_i\in D$, the data owner produces its signature $\sigma(t_i)=(H(t_i))^d\ (mod\ N)$, where $H$ is a full-domain hash function from $\{0,1\}^*$ to $\mathcal{Z}_N$. 
%The data owner transmits $D$ as well as the signatures $\sigma(t_i)$ to the SP. Suppose that the SP executes a query $Q$ issued by the client, and retrieves the set of tuples $R^S=\{t_a, \dots, t_b\}$ as the result. To prove the authenticity of every tuple in $R^S$, instead of sending the signatures $\sigma(t_i)$ for individual tuples in $R^S$, the SP calculates a single aggregate signature as:

\nop{
\begin{figure}[!htb]
\vspace{-0.1in}
\begin{center}
\begin{tabular}{c}
\includegraphics[width=0.3 \textwidth]{./figs/Narasimha2006.eps}
\end{tabular}
\vspace{-0.1in}
\caption{\label{fig:Narasimha2006} An example of range query for \cite{Narasimha2006}}
\end{center}
\vspace{-0.15in}
\end{figure}
}
Although effective, \cite{mykletun2006authentication} cannot verify the completeness of the query results. 
Narasimha et al. \cite{Narasimha2006} extended \cite{mykletun2006authentication} by combining signature aggregation with signature chaining for the authentication of range queries. Unlike \cite{mykletun2006authentication} that aggregates the signatures of individual tuples, \cite{Narasimha2006} constructs the VO from the signature of each individual tuple along
with its immediate predecessor, i.e., $sign(H(t_i||t_{i-1})$, where $t_{i-1}$ is the immediate predecessor of $t_i$ on the search attribute. By including the predecessor in the signature, it forms a
chain of tuples ordered on the search attribute. 
By using the VO, the client verifies that the set of tuples received in 
the result indeed form a valid chain. 
%The scheme is also applicable to multiple search attributes.

Simultaneously, Pang et al. \cite{pang2005verifying}  devised a new signature-based scheme for range and join queries. 
%\Bo{We omit discussion how to verify join query directly ??} 
It constructs the VO from the signature of each individual tuple along
with both of immediate predecessor and  successor, i.e., $sign(H(g(t_{i-1})||g(t_i)||g(t_{i+1})))$, where $t_{i-1}$ and $t_{i+1}$ are the immediate predecessor and successor of $t_i$ on the search attribute respectively,  $sign$ is the digital signature signing function (e.g., RSA or DSA \cite{rivest1978method}), $H$ is the hash function, $||$ is the concatenation operator, and $g(t_i)$ is the digest of $t_i$. The proposed scheme can support joins on primary key and foreign key, by using signatures on the key attribute of a relation to generate
proof of the completeness of query results from that
relation. 
\nop{
\begin{figure}[!htb]
\vspace{-0.1in}
\begin{center}
\begin{tabular}{c}
\includegraphics[width=0.25 \textwidth]{./figs/R-tree3.eps}
\end{tabular}
\vspace{-0.1in}
\caption{\label{fig:cheng2006authenticating} An example of multi-dimensional range query for authentication methods of \cite{cheng2006authenticating}}
\end{center}
\vspace{-0.15in}
\end{figure}
}

One weakness of the signature aggregation approach \cite{pang2005verifying} is that it only supports single-dimension range queries on the key attribute, since it requires the tuples to be ordered based on their key attribute values. To address the limitations of the signature chaining approaches in \cite{pang2005verifying}, Cheng et al. \cite{cheng2006authenticating} combined signature aggregation with a data partitioning structure (e.g., R-tree \cite{guttman1984r} and KD-tree \cite{bentley1975multidimensional}) to support authentication of multi-dimensional range query evaluation. 
Before outsourcing, the data owner constructs an R-tree or KD-tree of the dataset to get data partitions. Each partition $P$ will be associated with a signature, which is aggregated from the signatures of two auxiliary bounding tuples of $P$. By including the signature of those partitions that do not overlap with the query range, instead of the records in these partitions, it  reduces the VO size significantly. 

\nop{
\begin{figure}[!htb]
\vspace{-0.1in}
\begin{center}
\begin{tabular}{c}
\includegraphics[width=0.4 \textwidth]{./figs/zheng2012efficient.eps}
\end{tabular}
\vspace{-0.1in}
\caption{\label{fig:zheng2012efficient} An example of range query for \cite{zheng2012efficient}}
\end{center}
\vspace{-0.15in}
\end{figure}
}

Zheng et al. \cite{zheng2012efficient} designed the signature-based authentication for various types of queries, including aggregation queries.  
It utilizes the Homomorphic Linear Authentication (HLA) (Section \ref{sc:hlt}) to construct the aggregate signature for authentication of SUM queries. To support authentication of other types of queries including range  selection queries, the authentication method combines HLA aggregation signature with MB-tree \cite{li2006dynamic}, by associating the MB-tree entries with the HLA aggregation signatures. 
%The advantage of the HLA aggregate signature is that in order to authenticate the SUM query, the client only needs to set the challenge coefficient vector $\vec{c}$ as a sequence of $1$'s.

\nop{Given $n$ tuples, the data owner first computes the signature $\sigma_i$ for each tuple $t_i$ following the $TagGen()$ function of HLA (Section \ref{sc:hlt}). Then the data owner constructs a Merkle B-tree $T$ \cite{li2006dynamic} from all the tuple-signature pairs. 
Given a range query $[l,u]$, the SP finds the two boundary tuples that just fall out of the filtering range, and includes them as well as the set of siblings of the nodes on the path from the boundary tuples to root in $T$. Moreover, the SP also generates an aggregate signature for all the matching tuples. Different from \cite{Narasimha2006}, the aggregate signature is generated via the $HLAAgg(\vec{c},\vec{\sigma})$ routine (Section \ref{sc:hlt}), where $\vec{c}$ is a set of challenge coefficients randomly generated by the client, and $\vec{\sigma}$ is a vector of individual signatures of the matching tuples. 
Take the query in Figure \ref{fig:zheng2012efficient} as an example. 
The matching tuples are $t_3, \dots, t_6$. Therefore, the VO includes the aggregate signature $\sigma_{3,6}$, the boundary tuples $t_2$ and $t_7$, and the paths for the boundary tuples, which are $path(t_2)=\{t_1, N_{34}, N_{5,8}\}$ and $path(t_7)=\{t_8, N_{56}, N_{14}\}$.
The client verifies the authenticity by calling the $Vrfy(pk,R^S,\sigma_{3,6})$ routine, and reconstructing the root hash of $T$ from $t_2$ and $path(t_2)$ as well as $t_7$ and $path(t_7)$. The completeness is inspected via $t_2<l$ and $t_7>u$. The advantage of the HLA aggregate signature is that in order to authenticate the SUM query, the client only needs to set the challenge coefficient vector $\vec{c}$ as a sequence of $1$'s.}

\nop{
Given a relation $R=\{A_1, \dots, A_k\}$, let $A_1$ be the search key, and $L$ and $U$ be the lower- and upper-bound of attribute values on $A_1$. Firstly, the data owner calls the $KeyGen(1^{\lambda})$ routine of HLT (Section \ref{sc:hlt}) to produce a pair of public and private key. Next, the data owner inserts two artificial tuples $t_0$ and $t_{n+1}$ such that $t_0.A_1=L$ and $t_{n+1}.A_1=U$.
Then for each tuple $t_i\in R$, its individual signature $\sigma_i$ is calculated by following the $TagGen(sk, t_i)$ function of HLT. Furthermore, the data owner creates a pair $(t_i.A_1, \sigma_i)$ for each tuple $t_i$, and constructs a Merkle B-tree $T$ \cite{li2006dynamic} from all the pairs. In the last step, the client publishes the root signature $sig(T)$ of the Merkle B-tree, and transmits $R$ as well as $T$ to the SP.
Given a selection query $\sigma_{A_1\in[l, u]}(R)$, the SP searches for the set of tuples $R^S=\{t_a, \dots, t_b\}$. In addition to that, the SP constructs a proof that consists of:
(1) two boundary nodes $t_{a-1}$ and $t_{b+1}$ such that $t_{a-1}.A_1$ is the largest value that is smaller than $l$, and $t_{b+1}.A_1$ is the smallest value that is larger than $u$. Besides, the SP includes the $path(t_{a-1})$ and $path(t_{b+1})$, where $path(t_i)$ is a sequence of values for nodes from $t_i$ to the root of the Merkle B-tree, as well as the values of the nodes' siblings;
(2) the aggregate signature $\sigma$ generated via the $HLTAgg(\vec{c}, \vec{\sigma})$ function (Section \ref{sc:hlt}), where $\vec{\sigma}=\{\sigma_a, \dots, \sigma_b\}$, and $\vec{c}=\{c_a, \dots, c_b\}$ is a challenge coefficients randomly generated by the client. The authenticity of $R^S$ can be verified by the client based on the $Vrfy(pk, R^S, \sigma)$ function of HLT scheme (Section \ref{sc:hlt}). The client can easily verify the soundness of $R^S$ by checking if for every tuple $t_i\in R^S$, $t_i.A_1\in [l, u]$. 
The completeness verification relies on the Merkle B-tree $T$. First, the client checks the authenticity of $t_{a-1}$ and $t_{b+1}$ by reconstructing the root hash values of $T$ from $path(t_{a-1})$ and $path(t_{b+1})$, and comparing them against the root signature that is published by the data owner. If they match, the client is assured that $t_{a-1}$ and $t_{b+1}$ are from the original database $R$. Next, the client simply checks if $t_{a-1}.A_1<l$ and $t_{b+1}.A_1>u$ to validate the completeness of $R^S$.

Zheng et al. \cite{zheng2012efficient} also dealt with authentication of projection, join and aggregation queries. Here we briefly present the authentication method for aggregation queries. \cite{zheng2012efficient} only consider SUM. For a SUM query $\mathcal{G}_{sum(A_x)}(\sigma_{A_1\in[l, u]}(R))$ for $1\leq x \leq m$, the proof construction procedure is the identical to that of the selection query, except that the challenge coefficient vector $\vec{c}$ is a sequence of $1$'s. 
In specific, let $\{t_a, \dots, t_b\}$ be the set of tuples whose $A_1$ attribute value is the range $[l, u]$, and $sum$ be the aggregation result returned by the server. In addition, the server returns $t.A_j=\sum_{i=a}^b t_i.A_j$ for $j\neq x$. The server constructs proof from the Merkle B-tree to demonstrate the correctness of $\{t_a, \dots, t_b\}$, and computes the aggregate signature $\sigma$ from $HLTAgg(\vec{c}, \vec{\sigma})$, where $\vec{c}$ is a vector of $1$'s, and $\vec{\sigma}=\{\sigma_a, \dots, \sigma_b\}$.
The verification process of the aggregation is similar to that of the selection query. The client validates the correctness of $sum$ by checking if $sum+\sum_{j\neq x}t.A_j$ corresponds to the aggregate signature $\sigma$. 
}

\nop{
{\bf Projection query.}
Upon receiving a projection query $\pi_{A_1, \dots, A_k}(R)$, the SP returns $R^S=\{(t_i.A_1, \dots, t_i.A_k)|1\leq i \leq n\}$ as the result, and constructs a proof that consists of: 
(1) the proof for the selection query $\sigma_{A_1\in(L,U)}(R)$, where $L$ and $U$ are the lower- and upper-bound values for the search attribute $A_1$;
(2) the set of search attribute and signature pairs for all tuples $\{(t_1.A_1, \sigma_1), \dots, (t_n.A_1, \sigma_n)\}$;
and (3) $R.A_j=\sum_{i=1}^{n} c_i*t_i.A_j$ for every attribute $k+1 \leq j \leq m$ that is out of the projection field, where $\vec{c}=\{c_1, \dots, c_n\}$ is a sequence of random coefficients generated by the client.

{\bf Join query.}
Given a equi-join query $R\Join_{R.A_x=S.B_y} S$ on two databases, where $1\leq x,y\leq m$, let $A_2$ and $B_2$ be the primary key of $R$ and $S$ respectively. The SP returns the set of matching tuples $R^S=(R^*,S^*)$ such that for every tuple $t_i\in R$, there is at least a matching tuple $s_j \in S^*$. 
To prove the correctness of the join result, the SP executes two projection queries $\pi_{A_2, A_x}(R)$ and $\pi_{B_2, B_y}(S)$, and constructs the proofs respectively. Additionally, the SP includes two set of triples $\{(t_i.A_2, t_i.A_x, \sigma_i)|t_i\in R\}$ and $\{(s_j.B_2, s_j.B_y, \sigma_j)|s_j\in S\}$ in the proof.

{\bf Aggregation query.}
The approach in \cite{zheng2012efficient} only provides support for sum as the aggregation function.
For a sun query $\mathcal{G}_{sum(A_x)}(R)$ for $1\leq x \leq m$, the proof construction procedure is the identical to that of the projection query, except that the challenge coefficient vector $\vec{c}$ is a sequence of $1$'s.  
The verification process of the aggregation is similar to that of the projection query. \Wendy{Can you add details for Verification of SUM query? I don't see how it is similar to projection queries.}

{\bf Projection query.}
After receiving the projection result $R^S=\{(t_i.A_1, \dots, t_i.A_k)|1\leq i \leq n\}$, the client calculates $R.A_j=\sum_{i=1}^n c_i*t_i.A_j$ for $1\leq j\leq k$, where $\vec{c}=\{c_1, \dots, c_n\}$ is the same list of coefficients that are sent to the SP for proof construction. Next, the client computes the aggregated signature $\sigma$ from $HLTAgg(\vec{c}, \vec{\sigma})$, where $\vec{\sigma}=\{\sigma_1, \dots, \sigma_n\}$ are individual signatures included in the proof. The authenticity of $R^S$ is inspected via $Vrfy(pk, \vec{M}, \sigma)$, where $Vrfy$ is the verification routine of HLT scheme (Section \ref{sc:hlt}), and $\vec{M}=\{R.A_1, \dots R.A_m\}$. 
The result can pass the authenticity verification only if $R^S$ and $\{R.A_{k+1}, \dots, R.A_{m}\}$ are valid with regard to the aggregate signature $\sigma$.
The completeness verification relies on the proof of the selection query $\sigma_{A_1\in(L,U)}(R)$.

{\bf Join query.}
Upon receiving the join result $R^S$ and the proof, the client first investigates the authenticity of $\{(t_i.A_2, t_i.A_x, \sigma_i)|t_i\in R\}$ and $\{(s_j.B_2, s_j.B_y, \sigma_j)|s_j\in S\}$ according to the projection proofs of $\pi_{A_2, A_x}(R)$ and $\pi_{B_2, B_y}(S)$. 
Next, the client indentifies two sets of indices $\alpha\subseteq \{1, \dots, n\}$ and $\beta\subseteq \{1, \dots, n'\}$ such that for every $i\in \alpha$, there exists $j\in \beta$ to satisfy $t_i.A_x=s_j.B_y$. 
The completeness of the join result $R^S=\{R^*, S^*\}$ is validated by checking if $|R^*|=|\alpha|$ and $|S^*|=|\beta|$.
In the last step, the client verifies the authenticity of $R^*$ by generating a vector of random coefficiently $\vec{c}=\{c_1, \dots, c_{|\alpha|}\}$, computing the aggregate signature $\sigma$ from $\{\sigma_i|i\in \alpha\}$ and executing $Vrfy(pk, R^*, \sigma)$. The client applies the same operation for $S^*$.

{\bf How to deal with data updates.}
If the data owner intends to insert a new tuple $t$ into $R$, firstly, the data owner prepares the pair $(t.A_1, \sigma)$, where $\sigma$ is the HLT signature of $t$. Then the data owner sends $(t.A_1, \sigma)$ to the SP, and lets the SP find the appropriate position for $t$. The SP returns $t_i$ and $t_{i+1}$ such that $t_i.A_1<t.A_1<t_{i+1}.A_1$, as well as $path(t_i)$ and $path(t_{i+1})$. The data owner verifies the authenticity of $t_i$ and $t_{i+1}$ based on $path(t_i)$ and $path(t_{i+1})$, and updates the signature of the Merkle B-tree with the new tuple $t$. The new signature is sent to all the clients.

{\bf Weakness of the approach.}
Similar to \cite{Pang:2009:SVO:1687627.1687718} , the approach in \cite{zheng2012efficient} only allows selection query on the predefined single search attribute.
Besides, the client needs to do a lot of operations to authenticate projection query.
In the join verification, it is the client who joins the two relational instances, which incurs high verification complexity. The support for authenticated aggregation query is limited in SUM operations only.
}

\vspace{-0.05in}
\subsection{Accumulation Value based Verification}
% Taking authenticated range queries to arbitrary dimensions

Most of the ADS-based authentication approaches share one weakness for authentication of multi-dimensional range queries: they require to construct one ADS for each possible combination of dimensions in the database.  
This makes the number of required ADS structures scales exponentially with
the number of dimensions. To address this weakness, 
Papadopoulos et al. \cite{papadopoulos2014taking} integrate the accumulation values with Merkle hash tree (MHT) for the authentication of multidimensional range query evaluation on outsourced databases. They use the authentication protocol of set intersection and difference (Section \ref{sc:set_difference}) as the building block. 
The authentication method consists of two steps. Let $R$ be the query result. The first step is that the SP computes the set of hash values $R_i$ of the tuples that satisfy the query on dimension $i$. It also computes the proof $\pi_{R_i}$ that can be used to verify that $R_i$ is the set of tuples that match the range constraint on dimension $i$. The second step is that the SP computes $R{=}\bigcap R_j$ and generates the proof $\pi$ of $R$. The client uses the set intersection verification protocol on $\pi$ (Section \ref{sc:set_difference}) to verify the correctness of $R$. 
\nop{
In particular, for the first step, given a set of hash values of tuples $\{h_1,h_2,h_3,h_4,h_5,h_6\}$ which is ordered based on attribute $i$. Assume $R_i = \{h_3,h_4\}$ is the hash values of tuples satisfy constrain on attribute $i$, SP prepares a prefix set of hash values $P_i=\{h_1,h_2,h_3,h_4\}$ and a suffix set of hash values $S_i=\{h_3,h_4,h_5,h_6\}$, then the correctness of $R_i=P_i \cap S_i$ can be verified by using the set difference protocol (Section \ref{sc:set_difference}). 
Similarly, for the second step, the correctness of the final query results on multiple dimensions can be verified by using the set intersection protocol. We omit the details here.
}
\nop{
\begin{itemize}
  \item Step 1 (1-D verification): For each range constraint $A_i\in[l_i, u_i]$, the SP computes the set $R_i = \{h_j= H(t_j) | l_i\leq t_j.A_i\leq u_i\}$. It also computes the proof $\pi_{R_i}$ that can be used to verify if $R_i$ is the set of tuples that match the range constraint. 
  %by facilitating the set difference verification protocol.
   \item Step 2 (Verification of multi dimensions): The SP computes $R^S{=}\bigcap R_j$ and generates the proof $\pi$ of $R^S$. The client uses the set intersection verification protocol on $\pi$ to verify the correctness of $R^S$. 
\end{itemize} 

For Step 1, the data owner orders the hash values $h_j$ (of tuples $t_j$ ) according to the $t_j.A_i$ values, where $A_i$ is an attribute in the query with constraint $A_i\in[l_i, u_i]$. The {\em prefix set} $P_{i,j}$ is defined to be the set of all hash values appearing at positions $1, \dots, j$ in the ordering. Similarly, the {\em suffix set} $S_{i,j}$ is defined as the set of all
hash values appearing at positions $n-j+1, \dots, n$ in the ordering. Let $R_i$ be the set of tuples that satisfy  $A_i\in[l_i, u_i]$. Now assume that $k_i' + 1$, $k_i$ are the two positions in this ordering corresponding to $R_i$ 
It must hold that $R_i = P_{i,k_i} \ P_{i, k_i}'$. Based on this, the correctness of $R_i$ can be verified by using the set difference protocol (Section \ref{sc:set_difference}). Similarly, for Step 2, the correctness of the final query results on multiple dimensions can be verified by using the set intersection protocol. We omit the details here. 
}

\nop{
computes the digest of each tuple as 
\begin{equation}
\label{eq:papadopoulos2014taking_2}
h(t_i)=H(t_i.A_1||\dots||t_i.A_m),
\end{equation}
where $H: \{0, 1\}^* \Rightarrow G_1$ is a cryptographic hash function. 
Next, for each attribute $A_j$ ($1\leq j \leq m$), the data owner creates a permutation $\psi_j$ over $\{1, \dots, n\}$ such that $t_{\psi_j(i)}.A_j \geq t_{\psi_j(i+1)}.A_j$. In other words, the permutation $\psi_j$ is a  descending order of tuples based on attribute values of $A_j$. Then for each integer $1\leq i\leq n$, the data owner creates a triplet 
\begin{equation}
\label{eq:papadopoulos2014taking_3}
\tau_{i,j}=(v_{i,j},v_{i+1,j},acc(P_{i,j})),
\end{equation}
where $v_{i,j}=t_{\psi_j(i)}.A_j$ is the $i$-th largest value on attribute $A_j$, $P_{i,j}=\{h(t_{\psi_j(1)}), \dots, h(t_{\psi_j(i)})\}$ is the {\em prefix set} that consists of the hash values of tuples appearing in positions $1, \dots, \psi_j(i)$ in the ordering of $A_j$, and $acc(P_{i,j})$ is the accumulation value of $P_{i,j}$ (Equation \ref{eq:acc}).
For each attribute $A_j$, the data owner constructs a Merkle hash tree $T_j$ from the triplets $\{\tau_{i,j}|1\leq i\leq n\}$. In the last step, the data owner builds a Merkle hash tree $T$ over the set of pairs $\{(j, \delta_j)|1\leq j\leq m\}$, where $\delta_j$ is the root hash value of $T_j$, and $m$ is the number of attributes in $D$.
The data owner sends the $m+1$ MHTs to the SP and publishes $\delta$, where $\delta$ is the root digest of $T$.

\begin{figure}[!htbp]
\begin{center}
\includegraphics[width=0.4\textwidth]{./figs/taking2014.eps}
\vspace{-0.3in}
\caption{\label{fig:papadopoulos2014taking_1}Framework of the authentication protocol in \cite{papadopoulos2014taking}}
\end{center}
\end{figure}

{\bf VO construction by SP.}
Given a multidimensional range query with range $R^Q=\{[l_1, u_1], \dots, [l_d, u_d]\}$, where $d$ is an arbitrary number between $1$ and $m$, the SP searches for the query result $R^S=\{t_i\in R^Q|1\leq i \leq n\}$. The proof constructed by the SP consists of the following: 
(1) an accumulation value $acc(R^S)$ for the result; 
(2) for each query dimension $1\leq j\leq d$, the SP computes $R^S_j=\{t_i\in [l_j, u_j]|1\leq i \leq n\}$, and identifies the two indices $k$ and $k'$ such that $R^S_j=P_{k,j}\setminus P_{k',j}$. Then the SP locates the triplets $\tau_{k,j}$ and $\tau_{k',j}$, and retrieves the paths $path(\tau_{k,j})$ and $path(\tau_{k',j})$ from the MHT $T_j$, where $path(\tau_{i,j})$ is a sequence of values for nodes from $\tau_{i,j}$ to the root of $T_j$, as well as the values of the nodes' siblings. Next, the SP locates the path $path((j, \delta_j))$ from $T$. Furthermore, the SP prepares the set difference proof $\pi_{R^S_j}=acc(R^S_j)$ to demonstrate that $R^S_j=P_{k,j}\setminus P_{k',j}$. 
For each query dimension $j$, $\tau_{k,j}$, $\tau_{k',j}$, $path(\tau_{k,j})$, $path(\tau_{k',j})$, $path((j, \delta_j))$, and $\pi_{R^S_j}$ are included in the proof;
(3) the SP constructs a set intersection proof $\pi(R^S)$ (Section \ref{sc:set_intersection}) to demonstrate that $R^S=\bigcap_{j=1}^d R^S_j$.

{\bf Verification by data requester.}
Given the query result $R^S$ and the proof, for each query dimension $1\leq j\leq d$, the client first verifies the authenticity of $(j, \delta_j)$ by reconstructing the root based on $path((j, \delta_j))$ and comparing it with $\delta$, where $\delta$ is the root digest of $T$ published by the data owner. Next, the client checks the authenticity of the triplet $\tau_{k,j}$ ($\tau_{k',j}$ resp.) based on $path(\tau_{k,j})$ ($path(\tau_{k',j})$ resp.) and $\delta_j$. The soundness and completeness of $R^S_j$ is verified by evaluating (1) if $v_{k,j}\leq l_j\leq v_{k+1,j}$ and $v_{k',j}\leq u_j \leq v_{k'+1,j}$; and (2) if $R^S_j=P_{k,j}\setminus P_{k',j}$ by using the set difference verification protocol (Equation (\ref{eq:papadopoulos2014taking_3})). 
After finishing the verification for all the query dimensions, the client verifies the authenticity, soundness and completeness of $R^S$ by checking if $R^S=\bigcap_{j=1}^d R^S_j$ by applying the set intersection verification protocol (Section \ref{sc:set_protocol}).

{\bf How to deal with data updates.}
Upon inserting a new tuple $t$, the data owner first calculates the digest $h(t)$ according to Equation (\ref{eq:papadopoulos2014taking_2}). Next, for each dimension $1\leq j \leq m$, the data owner finds the appropriate position $k$ for $t$ such that $t_k.A_j\geq t.A_j \geq t_{k+1}.A_j$. The data owner update the element $v_{i+1,j}$ in the triplet $\tau_{k,j}$. Starting from $i=k+1$ to $n$, the data owner updates the triplet $\tau_{i,j}$ by adding $t$ into $P_{i,j}$. In other words, the data owner raises $acc(P_{i,j})$ by the power of $(h(t)+s)$, where $s$ is the secret key (Section \ref{sc:acc}). The data owner updates the internal nodes of $T_j$ accordingly and obtains the new root digest $\delta_j$ of $T_j$. Finally, the data owner updates the MHT $T$ and publishes the new root digest.
The authors further reduces the update complexity at the data owner side by organizing the hash values along each dimensions into buckets.
}

The verification scheme of \cite{papadopoulos2014taking} only supports the verification of range queries. To address the limitation of \cite{papadopoulos2014taking}, Zhang et al. \cite{zhang2015integridb} proposed a verification system named {\em IntegriDB} to accommodate a wide range of SQL queries, including multi-dimensional range queries, JOIN, SUM, MAX/MIN, COUNT, and AVG. The authors designed a {\em sum verification protocol} (Section \ref{sc:set_sum}) and a new ADS named 
Authenticated Interval Tree (AIT-tree), and facilitates them as the key components of {\em IntegriDB}. 
The AIT-tree is designed for key-value pairs. Given a set of key-value pairs $S=\{(k,v)\}$, the AIT-tree is a binary-structured tree constructed from $S$.  For each key-value pair $(k,v) \in S$, it corresponds to a leaf node in the AIT-tree, which stores a triplet $(k,v,h)$ with the hash value $h=H(k||v)$. The leaf nodes are arranged in the order of the keys. The value
stored at an internal node $N$ corresponds to the accumulation value (Section \ref{sc:set_sum}) of the leaves in the subtree rooted at $N$. 
\nop{In particular, 
\begin{equation}
\label{eq:zhang2015integridb_5}
f(N)=Enc_{sk} \big(\prod_{u\in B(N)} (u^{-1}+s)\big) || g^{\prod_{u\in B(N)} (u^{-1}+s)},
\end{equation}
where $Enc_{sk}$ is the encryption routine of a CPA scheme with the secret key $sk$, and $B(N)$ is the set of values stored in $N$'s subtree.}
%For the setup, the data owner constructs an AIT-tree for each table, and each pair of columns in that table. Then data owner gets the root digest for each tree as well as the hash values for each tuple in each table. 
The verification of one-dimensional range query result is similar to \cite{devanbu2003authentic}. The verification of multi-dimensional range queries and join queries is performed by the authentication protocol of set intersection (Section \ref{sc:set_intersection}), by considering the results of each individual dimension/table as a set. Regarding the aggregation queries, they are authenticated by different methods. The SUM queries are verified by using the set summation protocol (Section \ref{sc:set_sum}). 
In order to facilitate the verification of COUNT operations, the data owner creates an additional attribute $A_j'$ for each attribute $A_j$. The value of $A_j'$ of the record $t_i$ is set as  $t_i[A_j']=t_i[A_j]+1$. The data owner constructs an {\em AIT} for the new attribute $A_{j'}$. To prove the correctness of $COUNT(A_j)$, the SP prepares the proof for $SUM(A_j)$ and $SUM(A_{j'})$, as $COUNT(A_j)=SUM(A_{j'})-SUM(A_j)$.
MAX/MIN queries can be reduced to single-dimensional range queries. In particular, let $j_{max}$ be the maximum value for $A_j$ that is returned by the SP. The SP prepares the proof for the range query $\sigma_{A_j\geq j_{max}}(D)$. When there are data updates (i.e., insertion/deletion) on the dataset, the data owner updates the AIT-tree by updating the corresponding leaf node $N$ and all the nodes on the path from $N$ to the root of the AIT-tree, as well as the accumulation values of all internal nodes on this path. 

%for different AIT-tree to help the client verify the correctness of the intersection. As for the join query, for example $R\Join_{R.A_j=S.A_k} S$, the SP first carries range query on $R$ based on constrain on join attribute and prepares the VO for the range query result, similarly SP executes the same range query for relation $S$ and returns VO to client. The client verifies the correctness of the range query respectively and get the join result by combine tuples from different range query result.

\nop{
Next, we first introduce the AIT-tree, then present the design of {\em IntegriDB}.
}
\nop{
Given a set $S=\{x_1, \dots, x_n\}$, let $sum$ be the result returned by an untrusted SP upon the request of sum operation over $S$. in order for the client to efficiently check if $sum \stackrel{?}{=} \sum_{i=1}^n x_i$, Zhang et al. proposed the {sum verification protocol} based on accumulation values (Section \ref{sc:acc}).  
In the setup phase, the client constructs an accumulation value 
\begin{equation}
\label{eq:zhang2015integridb_1}
acc(S)=g^{\prod_{i=1}^n (x_i^{-1}+s)},
\end{equation}
where $g$ is the generator of the Elliptic group \cite{Menezes:1991:REC:103418.103434}, and $s$ is a random number and kept secret by the client. Note that Equation (\ref{eq:zhang2015integridb_1}) is different from Equation (\ref{eq:acc}) in that we use the inverse of $x_i$.
To prove the correctness of $sum$, the SP constructs a proof that includes: (1) $a_0=\prod_{i=1}^n x_i^{-1}$; (2) $a_1=\sum_{i=1}^n (\prod_{j\neq i} x_j^{-1})$; (3) $W_1 = g^{s^{n-1}+\dots+a_2s+a_1}$; and (4) $W_2=g^{s^{n-2}+\dots+a_3s+a_2}$. 
After receiving $sum$ as well as the proof, the client first verifies the correctness of $a_0$ by checking 
\begin{equation}
\label{eq:zhang2015integridb_2}
e(g^s, W_1) \stackrel{?}{=} e(acc(S)/g^{a_0}, g),
\end{equation}
where $e$ is the bilinear pairing function, and $acc(S)$ is obtained in the setup phase. 
Next, the client verifies the correctness of $a_1$ by checking 
\begin{equation}
\label{eq:zhang2015integridb_3}
e(g^s,W_2) \stackrel{?}{=} e(W_1/g^{a_1}, g).
\end{equation}
In the last step, the correctness of $sum$ can be verified by checking
\begin{equation}
\label{eq:zhang2015integridb_4}
sum \stackrel{?}{=} a_1/a_0.
\end{equation}
By calculating 4 pairings and 1 group division, the client can efficiently verify the sum of $n$ values.
}

\nop{
for a given table, assume it has $m$ attributes. For each pair columns $(j, k)$ ($1\leq j,k \leq m$) of the table, let $S_{j,k}=\{(t_{1,j}, t_{1,k}), \dots, (t_{n,j}, t_{n,k})\}$ be the projection over the attribute $A_j$ and $A_k$. The data owner constructs an {\em AIT} $T_{j,k}$ from $S_{j,k}$, where for each internal node $N$, the value $v$ in the triplet of $N$ is computed as
\begin{equation}
\label{eq:zhang2015integridb_5}
f(N)=Enc_{sk} \big(\prod_{u\in B(N)} (u^{-1}+s)\big) || g^{\prod_{u\in B(N)} (u^{-1}+s)},
\end{equation}
where $Enc_{sk}$ is the encryption routine of a CPA scheme with the secret key $sk$, and $B(N)$ is the set of values stored in $N$'s subtree.
}

\nop{
\begin{itemize}
	\item For each $(k,v) \in S$, it corresponds to a leaf node in the AIT-tree, which stores a triplet $(k,v,h)$, and the hash value $h=H(k||v)$. The leaf nodes are arranged in the order of the keys. 
    \item Each internal node also stores one triplet $(k,v,h)$, where $k$ is the maximum key in the left subtree, $v$ is calculated by applying a randomized function $f$ (the choice of $f$ will be discussed later) over all the values stored in the subtree, and $h$ is obtained by hashing the set of triplets in its children.
\end{itemize}
Using the keys the AIT-tree nodes the indexing information, the proofs for range queries over keys can be constructed in the same way as the Merkle B-tree \cite{li2006dynamic}. 

Now we are ready to discuss the design of {\em IntegriDB} in detail. For the setup, the data owner constructs an AIT-tree for each table, and each pair of columns
in that table. 
%this requires storage quadratic in the number of columns, this is fine for practical purposes since the number of columns is typically much smaller than the number of rows.) 

Given a dataset $D$ with $n$ tuples and $m$ attributes, the data owner first randomly generates a pair of secret keys $s$ and $sk$ according to the security parameter $1^{\lambda}$, where $s$ is used for generating the accumulation values (Equation \ref{eq:zhang2015integridb_1}), and $sk$ is the secret key of a CPA encryption scheme. 
After that, for each pair of columns $(j, k)$ ($1\leq j,k \leq m$), let $S_{j,k}=\{(t_{1,j}, t_{1,k}), \dots, (t_{n,j}, t_{n,k})\}$ be the projection over the attribute $A_j$ and $A_k$. The data owner constructs an {\em AIT} $T_{j,k}$ from $S_{j,k}$, where for each internal node $N$, the value $v$ in the triplet of $N$ is computed as
\begin{equation}
\label{eq:zhang2015integridb_5}
f(N)=Enc_{sk} \big(\prod_{u\in B(N)} (u^{-1}+s)\big) || g^{\prod_{u\in B(N)} (u^{-1}+s)},
\end{equation}
where $Enc_{sk}$ is the encryption routine of a CPA scheme with the secret key $sk$, and $B(N)$ is the set of values stored in $N$'s subtree.
}

\nop{
Now we are ready to discuss the design of {\em IntegriDB} in detail. For the setup, the data owner constructs an AIT-tree for each table, and each pair of columns in that table.  
After constructing the tree $T_{j,k}$, the data owner gets the root digest $\delta_{j,k}$. 
Additionally, the data owner computes the hash value $h_i=H(t_i)$ for each tuple $t_i\in D$.
Then, the data owner sends $\{T_{j,k}|1\leq j,k \leq m\}$ to the SP, and publishes $\{\delta_{j,k}|1\leq j,k \leq m\}$ and $\{h_i|1\leq i\leq n\}$.

In order to facilitate the verification of multidimensional range query, in the verification preparation phase, the data owner creates a {\em reference column} in which every element is
guaranteed to be distinct. Let $A_0$ denote the reference column. For each attribute $A_j$ ($1\leq j\leq m$), the data owner creates an auxiliary {\em AIT} $T_{j,0}$ from the set $S_{j,0}=\{(t_{1,j}, 1), \dots, (t_{n,j}, n)\}$, and publishes its root digest $\delta_{j,0}$.  Without loss of generality, assume that the SP receives a 2-dimensional selection query with the range $R^Q=\{[l_j, u_j], [l_k, u_k]\}$. The SP searches for the query result $R^S=\{t_i|l_j\leq t_{i,j}\leq u_j,\  and\  l_k\leq t_{i,k}\leq u_k\}$. 
Let $R_j$ ($R_k$ resp.) be the index of the tuples that satisfy the range requirement on $A_j$ ($A_k$ resp.), i.e., $R_j=\{i|l_j\leq t_{i,j}\leq u_j\}$ ($R_k=\{i|l_k\leq t_{i,k}\leq u_k\}$ resp.). 
Let $R^*$ be the tuple index of $R^S$. 
It is straightforward that $R^*=R_j\cap R_k$. 
The SP prepares the proof in two steps. 
In the first step, the SP constructs the proof of $R_j$ and $R_k$ by calling the range query routine of $T_{j,0}$ and $T_{k,0}$ with the range $[l_j, u_j]$ and $[l_k, u_k]$ respectively. Let $\pi_j$ and $\pi_k$ be the proofs. 
In the second step, the SP constructs the set intersection proof $\pi^*$ (Section \ref{sc:set_intersection}) to demonstrate that $R^*=R_j\cap R_k$. 
The SP sends $R^S$, $\pi_j$, $\pi_k$ and $\pi^*$ for verification.
}

\nop{
Regarding the join query $R\Join_{R.A_j=S.A_k} S$, the SP constructs the proof in three steps.
In the first step, on the dataset $R$, the SP executes the selection query with the range $R^Q=\{[l_j, u_j]\}$, and prepares the proof $\pi_j$ in the way of selection query, except that $\pi_j$ is constructed from the AIT $T_{j,j}$. Let $C_j$ be the set of unique values in $R.A_j$. The SP obtains $acc(C_j)$ and $\pi_j$. Similarly, the SP collects $acc(C_k)$ and $\pi_k$ by traversing $T_{k,k}$ of $S$.
Secondly, the SP identifies the common values of $C_j$ and $C_k$ by getting $C^*=C_j\cap C_k$, and constructs the set intersection proof $\pi^*$ of $C^*$.
In the last step, for each unique element $x\in C^*$, the SP executes the selection query $[x,x]$ on $R.A_j$, obtains the query result $R^x_j$, and prepares the proof $\pi_j^x$ by visiting the AIT $T_{j,j}$ of $R$. Likewise, the SP gets $S^x_k$ and $\pi_k^x$ from $S$. Apparently, every pair of tuple $(t_i, t_{i'})$ must exist in the returned result $R^S$, where $t_i\in R_j^x$, and $t_{i'}\in R_k^x$.
Finally, the SP returns $R^S$, and the proof, which includes $acc(C_j), acc(C_k), \pi_j, \pi_k, C^*, \pi^*$, and $(\pi_j^x,\pi_k^x)$ for each $x\in C^*$. When the client receives the join result $R^S$ and the proof, the client first verifies $\pi_j, \pi_k, \pi^*$ to confirm the correctness of $C^*$. Then for each $x\in C^*$, the client checks the correctness of $R_j^x$ and $R_k^x$ by inspecting $\pi_j^x$ and $\pi_k^x$ respectively. Then for each tuple $t_i\in R_j$ ($t_i\in R_k^x$, resp.), the client compares $H(t_i)$ against $h_i$ of $R$ ($S$, resp.), which is published by the data owner, to validate the authenticity of $t_i$.
}

\nop{
{\bf Verification by data requester.}

{\bf Selection query.}
The client verifies the correctness of $R^S$ in three steps. 
First, the client checks the correctness of $R_j$ and $R_k$ by verifying $\pi_j$ and $\pi_k$ against $\delta_{j,0}$ and $\delta_{k,0}$. Note that it is not necessary for the client to learn the intermediate result $R_j$ and $R_k$. Instead, only the accumulation values $acc(R_j)$ and $acc(R_k)$ (Equation (\ref{eq:zhang2015integridb_1})) are sufficient. 
Secondly, the client validates $R^*$ by using the set intersection proof $\pi^*$.
Lastly, having obtained the indices of the matching tuples, for each index $i\in R^*$, the client computes the hash $H(t_i)$, where $t_i$ is included in the result $R^S$, and compares it with $h_i$, which is published by the data owner.
If $R^S$ passes the verification, the client is assured of the authenticity, soundness and completeness of $R^S$.

{\bf Join query.}

{\bf Aggregation query.}
The verification of SUM operation is straightforward. The client only needs to follow Equation (\ref{eq:zhang2015integridb_2}-\ref{eq:zhang2015integridb_4}) to validate the result.
In order to check the correctness of COUNT operations, the client first inspects $SUM(A_j)$ and $SUM(A_{j'})$ via the proof. After that, the client learns the correct COUNT from their difference.
The verification of MAX/MIN results is the same as that of selection query. As long as the result of selection query $\sigma_{A_j\geq j_{max}}(D)$ only includes one element, the client is assured of the correctness of $j_{max}$.
}

%To update the values in the triplet $(k,v,h)$, the data owner can recover $acc(B(N))$ or $\prod_{u\in B(N)} (u^{-1}+s)$ by decrypting $Enc_{sk} \big(\prod_{u\in B(N)} (u^{-1}+s)\big)$.  %which is the first part in Equation (\ref{eq:zhang2015integridb_5}). 
%Then the data owner applies changes to $acc(B(N))$ and $f(N)$. 
%This alleviates the need for the data owner to store the AIT-trees locally. 
%Instead, the data owner can communicate with the SP to retrieve and update the path in {\em AIT}.

%Although IntegriDB \cite{zhang2015integridb} provides support for extensive types of queries, it sustains two drawbacks. First, the verification preparation cost at the data owner side is significant, as it requires to construct an AIT-tree for each pair of attributes. Second, in the verification process of join queries, it is indeed the client who joins two sets of tuples from two datasets that have the same value on join attribute. 

\nop{
%integriDB
Zhang et al. \cite{zhang2015integridb} define a variation of accumulation values to facilitate the verification of {\em sum} operation in constant time. In specific, given a set $S=\{x_1, \dots, x_n\}$, the accumulation value is defined as $acc(S)=g^{\Pi_{i=1}^n (x_i^{-1}+s)}$. In this way, we have $\Pi_{i=1}^n (x_i^{-1}+s)=s^n+a_{n-1}s^{n-1}+\dots +a_1s+a_0$, where $a_1=\Sigma_{i=1}^n (\Pi_{j\neq i} x_j^{-1})$ and $a_0=\Pi_{i=1}^n x_i^{-1}$. Naturally, the sum of $S$ equals $a_1a_0^{-1}$. In \cite{zhang2015integridb}, a new authenticated data structure named {\em Authenticated Interval Tree (AIT)} is also proposed to store a set of key-value pairs. Given a dataset, for each pair of attributes, the client constructs an $AIT$ over the accumulation values and keeps the root signature locally. The combination of $AIT$ and accumulation values enable the client to verify the integrity of multidimensional range queries (based on the set intersection protocol \cite{papamanthou2011optimal}) and the Group-by Sum operations (based on the accumulation values). 
When inserting/deleting a tuple, the data owner needs to update the accumulation values in every $AIT$, which incurs $O(m^2 \log n)$ complexity, where $m$ is the number of attributes.
However, a major drawback is the expensive verification preparation cost. Moreover, when there exist duplicated values in the dataset, the verification would be more complicated.

Papadopoulos et al. \cite{papadopoulos2014taking} further extends the application of set membership verification protocol to support authenticated range queries on arbitrary dimensions. On any dimension $A_i$, given a range $[l, u]$, let $k_l$ and $k_u$ be the first and last attribute value that fall into the range respectively, and $R_i$ denote the set of tuples that reside in the range on $A_i$. 
The authors rewrite $R_i=P_i(k_u)\setminus P_i(k_l-1)$, where $P_i(k)=\{t\in D| t.A_i \leq k\}$ is named the {\em prefix set} of $k$, and $\setminus$ is the set difference operator. 
The multi-range query result $R$ on attributes $A_1, A_2, \dots, A_k$ is equivalent to $R_1\cap R_2 \cap \dots \cap R_k$. 
Following this rationale, the authors first propose an efficient {\em set difference verification protocol} based on the accumulation values (Equation \ref{eq:acc}). 
Then the authors propose a verification protocol that consists of three phases. 
In the setup phase, the data owner constructs a {\em Merkle hash tree} or an {\em accumulation tree} from the prefix sets of each dimension. An accumulation tree differs from the ordinary {\em Merkle tree} in that each internal node stores the accumulation values produced over the values of its children, rather than the hash digest.
In the proof preparation phase, the SP first prepares proof of $R_i$ from $P_i(k_u)$ and $P_i(k_l-1)$, for each query dimension. Then the SP constructs the set intersection proof of the query results $R$ from $R_1, \dots, R_k$. 
Correspondingly, the client verifies the range query integrity in two steps: verify that $R_i=P_i(k_u)\setminus P_i(k_l-1)$ based on the set difference verification protocol, and that $R=R_1\cap R_2 \cap \dots \cap R_k$ based on the set intersection verification protocol.
In addition to verifying multidimensional range queries, \cite{papadopoulos2014taking} facilitates efficient data update by updating the accumulation values in every dimension. The authors further improve the update efficiency by organizing attribute values in buckets.

%Accumulators schemes
Nguyen \cite{nguyen2005accumulators} defines a new collision-resistant dynamic accumulator scheme \ref{eq:acc} based on bilinear pairing. The scheme provides signatures whose size is at most one half of the existing approach \cite{ateniese2000practical,camenisch2002dynamic}. In the meantime, another attractive of this scheme is that it is completely trapdoor-free. This improves efficiency and manageability, as various groups can share the same initial set-up parameters. This scheme is proved to be secure under the q-strong Diffie-Hellman assumption.
The author demonstrates the application of this accumulator scheme in ID-based ring signature scheme, a group signature scheme with membership revocation, and an identity escrow scheme with membership revocation.

%Tamassia set member
Since then, the bilinear pairing-based accumulator scheme has been widely used to authenticated outsourced computations. Papamanthou et al. \cite{papamanthou2011optimal} present the first scheme to reduce the verification of set operations (intersection, union, subset and set difference) to be linear with the answer size. In particular, for each input set, an accumulation value is calculated following \cite{nguyen2005accumulators}. Next, the data owner constructs a Merkle tree from the accumulation values, and takes the root signature as the digest of the tree structure. Then the tree and the sets are sent to the SP. Given a request that asks for the intersection of a collection of sets $\{S_1, \dots, S_t\}$, the SP returns the intersection $I$ as well as the cryptographic proof $\Pi$ to demonstrate its correctness. The proof consists of four parts (Section \ref{sc:set_protocol}). The first two components are meant to show that the SP does not tamper with the accumulation values $\{acc(S_1), \dots, acc(S_t)\}$. Whereas the last two parts prove that $I$ is included in every set $S_i$, and that there is no common element shared by $\{S_1, \dots, S_t\}$ except for $I$. An attractive property of the set operation verification protocol lies in that the verification complexity is only dependent on the final result size, and irrelevant from the total number of elements in the collection. This protocol can be directly applied to the authentication of database queries, which boil down to set operations.

%\Boxiang{I am not sure if we need \cite{nguyen2005accumulators,papamanthou2011optimal,papadopoulos2015practical}, as they are not particularly used for query integrity verification.}

%web crawler
An authenticated web crawler is proposed by Nguyen et al. \cite{nguyenverification, goodrich2012efficient}. In this model, the SP stores all webpages in the collection, while the client issues {\em conjunctive keyword search queries} over a set of pre-defined keywords to the SP. An authenticated data structure that supports efficient web content update is constructed by a web crawler based on the two building blocks: accumulators and Merkle tree. The empirical study suggests that the 5,000 content updates can be finished within 1 second. 
Considering that the web contents evolve at a high rate, a nice property of the propsoed authenticated data structure is that it supports efficient data update (with $O(\log n)$ complexity). However, the efficiency of verification at the client side is limited by the execution of the extended Euclidean algorithm to calculate the coefficients of B$\acute{e}$zout's identity.

%taking authenticated range queries to arbitrary dimensions
Papadopoulos et al. \cite{papadopoulos2014taking} further extends the application of set membership verification protocol to support authenticated range queries on arbitrary dimensions. On any dimension $A_i$, given a range $[l, u]$, let $k_l$ and $k_u$ be the first and last attribute value that fall into the range respectively, and $R_i$ denote the set of tuples that reside in the range on $A_i$. 
The authors rewrite $R_i=P_i(k_u)\setminus P_i(k_l-1)$, where $P_i(k)=\{t\in D| t.A_i \leq k\}$ is named the {\em prefix set} of $k$, and $\setminus$ is the set difference operator. 
The multi-range query result $R$ on attributes $A_1, A_2, \dots, A_k$ is equivalent to $R_1\cap R_2 \cap \dots \cap R_k$. 
Following this rationale, the authors first propose an efficient {\em set difference verification protocol} based on the accumulation values (Equation \ref{eq:acc}). 
Then the authors propose a verification protocol that consists of three phases. 
In the setup phase, the data owner constructs a {\em Merkle hash tree} or an {\em accumulation tree} from the prefix sets of each dimension. An accumulation tree differs from the ordinary {\em Merkle tree} in that each internal node stores the accumulation values produced over the values of its children, rather than the hash digest.
In the proof preparation phase, the SP first prepares proof of $R_i$ from $P_i(k_u)$ and $P_i(k_l-1)$, for each query dimension. Then the SP constructs the set intersection proof of the query results $R$ from $R_1, \dots, R_k$. 
Correspondingly, the client verifies the range query integrity in two steps: verify that $R_i=P_i(k_u)\setminus P_i(k_l-1)$ based on the set difference verification protocol, and that $R=R_1\cap R_2 \cap \dots \cap R_k$ based on the set intersection verification protocol.
In addition to verifying multidimensional range queries, \cite{papadopoulos2014taking} facilitates efficient data update by updating the accumulation values in every dimension. The authors further improve the update efficiency by organizing attribute values in buckets.

%Pattern Matching
The application of the accumulation values is extended to the authentication of pattern matching by Papadopoulos et al. \cite{papadopoulos2015practical}. Given a text document $T$ and a pattern $p$, pattern matching checks the existence of $p$ in $T$. One key observation is that there is a match of $p$ in $T$ if and only if there exist two suffixes of $T$, $S_i$ and $S_j$, such that $S[i]=pS[j]$. In order to provide an authentication scheme with optimal proof size, in the setup phase, the data owner calculates the accumulation values of all the suffixes, and builds an accumulation tree from them. The SP can promptly constructs the proof by assembling the readily-computed accumulation values in the tree. Theoretical analysis demonstrate that the proof contains at most $10$ accumulation values, and thus is of size $O(1)$.

% Zero-Knowledge proof for range query based on Hierarchical Identity Based Encryption
Ghosh et al. \cite{ghosh2016efficient} proposed a protocol based on {\em Hierarchical Identity Based Encryption (HIBE)} \cite{boneh2005hierarchical} to answer zero-knowledge verifiable one-dimensional range queries. A HIBE is a public key crypto system where the identity can be used as the public key, and an identity at level $k$ can issue private keys to its descendant, whereas it cannot decrypt messages intended for other identities. 
In \cite{boneh2005hierarchical}, Boneh et al. design a new HIBE based on bilinear pairing to limit the ciphertext size to merely three group elements. Based on the Bilinear Diffie-Hellman Exponent assumption, the proposed HIBE is proved to be chosen plaintext secure (CPA-secure) and chosen ciphertext secure (CCA-secure). 
In \cite{ghosh2016efficient}, given a range $[a,b]$ and the query answer $\{v_1, \dots, v_m\}$, the protocol leverages the classic signature scheme to prove that every element $v_i$ belongs to the dataset $D$, while it takes advantage of HIBE \cite{boneh2005hierarchical} to prove that there are no other values in $D$ that belong to the query interval. Compared with existing signature chaining schemes \cite{narasimha2006authentication}, \cite{ghosh2016efficient} provides more privacy it guarantee in that the client learns nothing about the dataset besides the query answer.
}

\vspace{-0.05in}
\subsection{Trusted Hardware based Verification}
%\vspace{-0.15in}
Most of the existing work discussed query verification through software mechanisms. An alternative approach is to rely on some tamper-proof, trusted hardware that is deployed on the untrusted server. 
Sumeet Bajaj et al. \cite{bajaj2013correctdb} is the only work  that we are aware on using trusted hardware for query authentication. They utilize the trusted hardware such as the IBM 4764 \cite{IBMSCPU} co-processor (SCPU) to provide secure execution in an untrusted environment. The main idea is to build MHT based ADS and store the root hash of the ADS inside SCPU. This can eliminate the generation of digital signatures of 
root node hash, thereby saving a signature operation on each
update, and signature verifications for each query. 
The original query will be rewritten into sub-queries, ensuring that the processing within the SCPU is minimized, and any intermediate
results generated by the SP can be validated
by the SCPU using the ADS. Any operation executed by the SP that cannot be authenticated should be processed inside the SCPU.  \cite{bajaj2013correctdb} provided the query parsing scheme for various types of queries, including range queries, projections, joins, aggregations, grouping and ordering. When there is any data update, it leads to the updates on 
the MHT based ADS used by the SCPU for verification, which is modified only by the SCPU. 

\nop{
After the SCPU validates the query result, it signs the final query result with a random value sent by the client within the query request
and sends the signature to the client. The client can verify the result by using the public key of the SCPU. 
In particular, for different query type,the SCPU use different software verification mechanism, as we discussed in section \ref{sec:treebased}, to verify the query result from server. Sometime, the SCPU needs further computation to get the final query result. But before it proceeds to the final result, it will verify the intermediate result from server first.
Take the range query for example, the SCPU receive query result and reconstruct the root digest of the MHT like the client does in \cite{devanbu2003authentic}. And for the join query, the SCPU use sort-merge join for two relations which is similar method discussed in \cite{yang2009authenticated}. Overall, the SCPU relieves the client from burdensome verification compare with the software method. Since everything that needed for verification can be transfer locally within server, this method saves a lot of communication cost between server and client and it also supports efficient updates because data structures are modified only by the SCPU which is assumed to be secure.
}

\vspace{-0.05in}
\subsection{Dealing with Data Updates}
For most of the existing deterministic authentication methods, the 
data owner is responsible to perform all update operations on the data, and update the ADS according to the updates. 
For the tree-based methods, one way to update the ADS is to first update (insert, delete or modify) the leaf nodes and their hash values. The root signature can be updated by propagating the update on the leaf nodes along their paths up to the root. 
The updates of both data and ADS are sent back to the SP. 
In addition, the tree-based
approaches require the re-distribution of the digest of ADS root signature to
all clients. 
The naive approach for handling batch updates
would be to do all updates to the ADS one by
one and update the path from the leaves to the root once
per update. Obviously this naive method may lead to expensive cost of performing unnecessary hash function computations on the predecessor path when a large number of
updates affect a similar set of nodes (e.g., the same leaf). 
To optimize the update cost on the ADS, \cite{li2006dynamic} suggests to 
recompute the hashes of all affected nodes only
once, after all the updates have been performed on the tree. 

For the signature-based verification methods, any update to a tuple requires to re-compute the signed digests of the tuple as well as its neighboring ones. \cite{narasimha2006authentication} provides a detailed protocol to support updates on the digests. 
The accumulation value based authentication methods deal with updates in the similar way. When inserting/deleting a record, the data owner needs to update the accumulation values of these records \cite{zhang2015integridb}. 

When using the trusted hardware, the data owner can
issue an update query directly to the SCPU. All updates
are then performed by the SCPU, and thus eliminating the update overhead at the side of the data owner. 
\vspace{-0.05in}
\subsection{Discussion}
\nop{
\begin{table}[!htbp]
\begin{tabular}{|c|c|c|c|c|c|}
\hline
Approach & Preparation & VO Construction & VO Verification & VO Size & Query Variability \\\hline
Tree-based & Medium & Low & High & High & High \\\hline
Signature & High & Low & Low & Low & Low \\\hline
Accumulation & Medium & High & High & High & High \\\hline
Interactive & High & ? & High & ? & High \\\hline
Hardware & ? & ? & Low & Low & High \\\hline 
\end{tabular}
\end{table}
}

One of the major concerns of the proof-based authentication methods is the verification overhead, which includes the size of the proof, the computational overhead in terms of execution time to construct ADS and the VO, and the verification time by the client. 
The tree-based, signature-based and accumulation value based authentication methods require one-time setup. At the setup phase, the tree-based methods have to construct ADS, while the signature-based and the accumulation value based methods construct the  signatures/accumulation values of individual records.  
The VO size of the tree-based methods is determined by the size of the query results, which in turn increases both query verification time and data communication overhead. 
As shown in \cite{chen2011cloud} the communication between the client and SP (e.g., the cloud) can cost up to 3500 picocents/bit, 2-3 orders of magnitude higher than processing costs (1 US picocent = \$$1\times 10^{−14}$) \cite{bajaj2013correctdb}. 
On the other hand, the proof size of the signature-based methods is a constant. Hence, they do not incur the high transmission costs as the tree-based approaches. However, 
the construction of the signature-based proofs is more expensive
than the tree-based approaches, since computing
even a single cryptographic trapdoor requires a high number
of CPU cycles costing up to 30,000 picocents \cite{chen2011cloud,bajaj2013correctdb}.

\nop{
The tree-based authentication approaches allow for efficient VO construction in which the SP simply traverses the ADS, however, it suffers from high verification cost and communication overhead. Typically, the VO incorporates certain portion of the tree. In the verification procedure, the client needs to recover the root digest from the VO.
The signature-based authentication dramatically reduces the VO size, as a single aggregate signature guarantees result authenticity. A major drawback is the limited support for various queries. 
The accumulation value based approach can verify a diversity of queries at the cost of high verification overhead. The main reason is the expensive primitive set operations.
The interactive proof based verification demands high preparation cost for the construction of arithmetic circuits. 
The trusted hardware based solution alleviates the verification cost as most verification is performed by the server-side SCPU. 
}

\vspace{-0.05in}
\section{Probabilistic Authentication Methods}
\label{sc:prob}

Unlike the authentication approaches that use proofs, the authentication without using proofs only can return a probabilistic integrity guarantee, i.e., how likely the returned query results are sound and complete. 
The existing
probabilistic authentication methods can be classified
into two types: (1) the checkpoint-based solutions, and (2) the
interactive-proof based solutions.
\vspace{-0.05in}
\subsection{Checkpoint-based Solutions}

Sion \cite{sion2005query} designed the first probabilistic verification approach for query execution on outsourced databases. Given a batch of $b$ queries $\{Q_1, \dots, Q_b\}$, the client first generates $r>1$ random numbers $\{x_1, \dots, x_r\}$, where $1\leq x_i\leq b$. Then, according to the random indexes, the client computes the query results $\{\rho(Q_{x_1}), \dots \rho(Q_{x_r})\}$ for the picked queries. 
%\Bo{how could client computes the query results, since the client does not have the dataset} 
For each query, it generates a challenge token $C(Q_{x_i})=(H(\epsilon||\rho(Q_{x_i})), \epsilon)$, where $\epsilon$ is randomly generated. The client sends the query batch as well as the challenge tokens to the SP. 
The SP returns the query results and the query execution proof $\{x_1', \dots, x_r'\}$. The results pass the verification only if $\{x_1, \dots, x_r\}$ matches $\{x_1', \dots, x_r'\}$. Due to the one-way non-invertible cryptographic hash function, the SP has to execute the queries over the target dataset to obtain the correct random indexes. 
The probability that the SP gets all the random indexes by executing $w<b$ queries is 
\[P_c(w,r)=\frac{\binom{w}{r}}{\binom{b}{r}}.\] 
The client can control the escape probability that the SP passes the verification to be below a given threshold by 
 adjusting $r$, i.e., the number of queries with challenge tokens. According to the experiment results, an assurance level of 5\% escape probability requires the client to prepare 25\% queries with challenge tokens. 

%The author reduces the number of query executions at the client side by allowing the client to construct $f$ fake challenge tokens without executing any query. The SP cannot distinguish the real and fake tokens before processing all the queries because the one-way hash function is applied to produce the tokens. This probabilistic approach is advantageous in that it can support the verification of the results of arbitrary queries. Whereas, one obvious limitation is that the client has to process some queries locally to generate the challenge tokens. And to ensure a small escape probability, the number of local query executions can be proportional to the batch size.  According to the experiment results, an assurance level of 5\% escape probability requires the client to run 25\% queries locally. It does not work well on dynamic databases. After each data update operation, the client has to calculate the challenge token over the new data.

Xie et al. \cite{xie2007integrity} proposed a probabilistic verification framework to check the soundness, completeness and freshness of the query results on outsourced encrypted databases. 
It is worth noting that the authentication approach in \cite{xie2007integrity} also can protect data privacy as it is applied on the encrypted data.
The key idea is that the client creates a small set of fake tuples $\Delta$, and inserts $\Delta$ into the original database $D$. 
The SP processes queries over the database with $\Delta$, assuming it cannot distinguish real tuples from the fake ones. The correctness of a query result $R_Q$ is evaluated by checking if the returned result $R^S$ contains $Q(\Delta)$, where $Q(\Delta)$ is the results of evaluating $Q$ on $\Delta$. If it does not, the client believes that the results violates the correctness requirement with 100\% certainty. Otherwise, the client trusts the correctness of $R^S$ with a probability. Formally, 
assume the original dataset $D$ has $N$ tuples, and $K$ fake tuples are inserted into $D$. The probability that an attacker can delete $m$ tuples without being caught is 
\begin{equation*}
 \prod_{i=0}^{m-1} \frac{N-i}{K+(N-i)}
\end{equation*}
The empirical study in \cite{xie2007integrity} shows that for a dataset of $N = 1, 000, 000$ tuples, when the fake tuples are more than 10\% of the original
data, and more than 50 tuples are deleted, it is close to impossible for the attacker to escape from being caught by the probabilistic approach.

In the reasoning of probabilistic guarantee, the authors only consider tuple deletion since tuple modification  or insertion can be easily detected through the following data authentication.
Before outsourcing, the data owner inserts a special digest value for each (real/fake) tuple. For any tuple $t=\{tid,a_1, \dots , a_m\}$, where $tid$ is the unique tuple identifier, the digest is calculated as 
\begin{equation}
\label{eq:h}
h(t) = 
	\begin{cases}
		H(tid\oplus a_1 \oplus \dots \oplus a_m) & \text{if } t \text{ is real} \\
		H(tid\oplus a_1 \oplus \dots \oplus a_m)+1 & \text{if } t \text{ is fake}
	\end{cases}
\end{equation}
On receiving any tuple $t$ from the SP, the client can tell if it is a real or fake tuple from the digest value.
The way to generate digest values facilitates efficient data update operation, as the digest of new tuples can be readily calculated at the client side.

\nop{
This approach alleviates the overhead at the client side in two ways. First, the client does not have to store any fake tuple locally in order to evaluate $C_C(Q)$ for any query. Instead, the client only keeps the description of the deterministic generation function $\mathcal{F}$, which is used to generate all the fake tuples in the preparation phase. 
Second, $\mathcal{F}$ is carefully designed so that $C_C(Q)$ can be efficiently calculated without querying the fake tuples. 

However, this lightweight probabilistic authentication framework suffers from three drawbacks. 
In the first place, in order to provide a high probabilistic guarantee to catch any incorrect query result, the number of fake tuples can be proportional to the original data size. This incurs significant query processing overhead at the SP side. 
Moreover, even though the authors proved the indistinguishability of the fake tuples based on the underlying pseudorandom encryption function, any SP with the knowledge of digest generation (Equation \ref{eq:h}) can easily identify the fake tuples. 
Last, it is unclear how the underlying encryption function impacts the query accuracy.
}
% Preparation, query overhead, verification: O(\Delta n), where \Delta n is the number of fake tuples

%Another paper use probabilistic method
Di Vimercati et al. \cite{di2014optimizing,de2016efficient} 
applied the probabilistic authentication method to join queries. The outsourcing framework considers the join of two relations, $B_l$ and $B_r$ hosted at two storage SPs $S_l$ and $S_r$. 
 The integrity guarantee of $B_l \Join B_r$ is verified by using three types of fake tuples inserted into the outsourced dataset: 
 (1) {\em markers} tuples inserted to both $B_l$ and $B_r$; 
 (2) {\em twins} tuples, which are a portion of real records that each of the storage SPs duplicates locally before sending it to the SP; 
 (3) {\em salts/buckets} tuples that are inserted to destroy recognizable frequencies of combinations in one-to-many joins. Salts are used on the tuples at the many-side of the join so that occurrences of a same value become distinct. Meanwhile the salted replicas are created at one-side 
of the join to create the corresponding matching. Bucketization allows multiple occurrences of the same (encrypted) value at the many-side of the join, requiring that all the values have the same number of occurrences. 
The client can verify the result correctness regarding the markers and twins. The probability that no marker is omitted is 
  $(1-\frac{o}{f})^m$,  
  where $f$ is the cardinality of a relation  with $t$ twin pairs and $m$ markers, and  $o$ is the number of original tuples that are omitted without being detected. The probability that for each twin pair, either both tuples are omitted or both are preserved is 
$(1-2\frac{o}{f}+2(\frac{o}{f})^2)^t$. 
%The empirical study shows that for a synthetic database containing 1,000 tuples in both join operands, 

\nop{
The computational SP receives the perturbed and encrypted relations from the storage SPs and applies the join query to them. The perturbed tuples returned in the query result as the verification object for the client to verify the completeness of the query results. Since the relations are encrypted, so the authenticity of the query results is preserved.

One of the drawbacks of this method is the large communication cost between the storage SP and the computation SP.
}

\vspace{-0.05in}
\subsection{Interactive Proof based Verification}

Yupeng Zhang et al. \cite{vSQL} designed {\em vSQL}, a system for verifiable SQL queries over dynamic outsourced databases. vSQL combines two different approaches, namely information-theoretic interactive proof system \cite{cormode2012practical} and  
a novel scheme for verifiable polynomial delegation that can provide auxiliary inputs for the proof.
vSQL first translates SQL queries as arithmetic circuits.  Then it relies on an information-theoretic interactive proof system named the CMT protocol \cite{cormode2012practical}, which allows a client to verify that $y = C(x)$, where $x$ is the client’s data, and $C$ is a circuit corresponding
to the client’s query. 
\nop{The building block for the CMT protocol is the sum-check protocol \cite{lund1992algebraic}. The common input of the prover and verifier is an $\ell$-variate polynomial 
$g(x_1, \dots, x_\ell)$ over a field $F$; the prover’s goal is to convince the verifier that
\[H = \sum_{b_1\in\{0,1\}}\sum_{b_2\in\{0,1\}}\dots\sum_{b_\ell\in\{0,1\}}g(b_1, b_2, \dots, b_\ell).\]
Then let $C$ be a depth-$d$ arithmetic circuit
over a finite field $F$ that is layered, i.e., for which each gate of $C$ is associated with a layer, and the output wire from a gate at layer i can only be an input wire to a gate at level $i − 1$. The CMT protocol processes the circuit one layer at a time, starting
from layer 0 (that contains the output wires) and ending at
layer $d$ (that contains the input wires). The prover $P$ first provides a value $y$ for the output of the circuit $C$ on input x. Then, in the i-th round, $P$ reduces a claim (i.e., an algebraic
statement) about the values of the wires in layer $i$ to a claim
about the values of the wires in layer $i + 1$. The protocol
terminates with a claim about the wire values at layer $d$ (i.e.,
the input wires) that can be checked directly by the verifier $V$
who knows the input $x$. 
}
However, it is not feasible to apply the CMT protocol directly since the client may not be able to store the data $x$ (i.e., the input database). To address this problem, vSQL designs a new polynomial-delegation protocol, which allows the client to be able to evaluate a certain multivariate polynomial $p_x$ that
depends on $x$ (but not on $C$) at a random point. 
In particular, the data owner treats the database $D$ as an array of $|D|$ elements, and computes the multi-linear extension $D'$ of $D$. The data owner uses the polynomial-delegation protocol to generate the commitment $com$ of $D'$, and sends $com$ to the client and sends $D$ to the SP. At the query evaluation phase, the client verifies the query result following the $CMT$ protocol. 
The verification of all layers except the last one follows the CMT protocol between the client and the SP. In the last step of the CMT protocol,  it has to evaluate at a random point of $D'$ that is picked by the SP. Since the client does not have access to $D'$, it verifies the last step by using the commitment $com$ via the polynomial-delegation protocol. Since the point of the last step is not the same as used for all previous steps, vSQL only provides a non-deterministic guarantee of the query results. If the result returned by the SP is incorrect, the client will reject it with overwhelming probability. 

\nop{
At the last step the protocol, the client verifies that both polynomial relationship between the input and the layer $d-1$ and the digest $com$ are correct.

 by an interactive protocol between the server and client. The key building block of vSQL is the information-theoretic interactive proof (IP) system due to Cormode et al. \cite{Cormode:2012:PVC:2090236.2090245} that allows a client to verify that $y=C(x)$ for some circuit $C$ and input $x$ known to the client. In particular, the evaluation of the query can be translated to $d$ layers of circuit consists of addition and multiplication operation, and the layer $d$ is the input of the data in polynomial form and the layer $0$ is the output. Each layer of the circuit has a polynomial relationship, the client can check the relationship of the two polynomials at random point. If the server returns a claimed output to the client, the client starts to choose a random challenge at layer $1$ and verify the relationship with previous layer until to the layer $d$ (input polynomial). If the client knows the authenticity of the input, then he/she can verify the correctness of the output from the server. This process is called $CMT$ protocol. But the client cannot store $x$ locally in the outsource setting, the authors developed a polynomial-delegation scheme that allows the client to be able to outsource a certain multivariate polynomial $p_x$, where the $p_x$ depends on the dataset $x$. Before outsourcing $p_x$, a short digest $com_x$ of $p_x$, which is used to verify claimed results about the evaluation of $p_x$, will be generated and stored at the client side. Thus the client does not need to store the dataset locally.

At the preprocessing phase, the data owner views the database $D$ as an array of $|D|$ elements and computes the multilinear extension $D'$. The multilinear extension of arrays has the ability to efficiently combine two multilinear extensions. The data owner uses the polynomial-delegation protocol to generate the commitment $com$ to $D'$, sends $com$ to the client and upload the database $D$ to server. At the query evaluation phase, the client verifies the query result follow the $CMT$ protocol. At the last step the protocol, the client needs to verify that both polynomial relationship between the input and the layer $d-1$ and the digest $com$ are correct.

The vSQL protocol can support expressive updates on the database because it supports auxiliary input. Particularly, the server can construct the polynomial of some input, and he/she can commit the input by sending the digest to the client. At any time, the client can send a query request to the server, the server needs to return the evaluation of the query and proof. The client has the ability to check the correctness of the query result based on the proof for polynomial that the server commits. Thus, the client doesn't need to ask the server to send the whole database back and update the database locally.

In particular, the digest $com_x$ consists of two parameters $(c_1,c_2) $from the bilinear pairing group, where $c_1=g^{f(s_1,...s_l)}$ and $c_2=g^{\alpha f(s_1,...s_l)}$. Assume the output a query is y, then the proof for y is $\pi:=(g^{q_1(s_1,...,s_l)},...g^{q_{l-1}(s_1,...,s_l)},q_l)$. The client verifies the output y by parsing the proof $\pi$ as $(\pi_1,...,\pi_{l-1},q_n)$, then checking if $e(c_1/g^y,g) \stackrel{?}{=} e(g^{s_l-t_l},g^{q_l(s_l)})\cdot \prod_{i=1}^{l-1}e(g^{r_i(s_i-t_i)+s_{i+1}-t_{i+1},\pi_{i}})$ and $e(c_1,g^{\alpha}=e(c_2,g)$. If they are equal, the client accepts the result, otherwise rejects. However, not all the SQL queries can be easily transfered to arithmetic circuit, for non-arithmetic gate like comparison, the authors augmented the CMT protocol which supports auxiliary inputs from the server side. The auxiliary inputs will reduce the the overhead induced by arithmetic circuits.
}

\vspace{-0.05in}
\subsection{Dealing with Data Updates}
The checkpoint-based methods can deal with data updates efficiently. The data owner can choose new checkpoints in the updated records. The digest values of these new checkpoint records can be readily calculated at the client side.
The interactive-proof based method \cite{vSQL} supports efficient data updates by separating the computation of the update from its verification. The SP has to update the  digest $com$. The client can provably evaluate if $com$ is updated by using the polynomial-delegation scheme. 
\vspace{-0.05in}
\subsection{Discussion}

The existing checkpoint-based approaches have different assumptions of the SP. \cite{sion2005query} requires that the SP to be aware of the verification protocol.  \cite{xie2007integrity} relaxes the assumption and makes the verification approach to be transparent to the SP. The SP is not required to be involved in the verification protocol at all. This reduces the computational overhead by the SP. According to the empirical study  of \cite{xie2007integrity}, the fake tuple based approach   \cite{xie2007integrity} incurs much cheaper verification setup overhead at the data owner side than the challenge token based approach \cite{sion2005query} and MHT tree based deterministic approach \cite{pang2005verifying}. However, the computational cost at the SP side by the fake tuple based approach \cite{xie2007integrity} can be higher than the two aforementioned approaches, due to the large number of inserted fake tuples to achieve high probabilistic guarantee. 

The alternative interactive proof based approach \cite{vSQL} aims to make the expensive theoretical interactive proof system to be practical. Compared with the generic verification approaches \cite{bitansky2012extractable,gennaro2013quadratic}, it leads to a 
significant performance speedup at the SP side. Its performance is also comparable to a deterministic approach \cite{zhang2015integridb} that only supports  a restricted subclass of SQL queries.

\vspace{-0.05in}
\section{Freshness Authentication}
\label{sc:freshness}

%\Wendy{Include the following paper: 1. A Completeness and Freshness Guarantee Scheme for Outsourced Database, in ICNDC 2011. } \Bo{I wrote it at the bottom. This is a very short paper, does not discuss in details. Also I think it only supports updates rather than the freshness of the the database}

Most of the works discussed so far only consider the verification of soundness and completeness of the query results. In this section, we review the existing studies on verification of freshness of the query results, i.e., these  results are obtained by executing the queries over the most up-to-date data. Similar to the verification of soundness and completeness, the freshness verification methods can be classified into deterministic and probabilistic approaches. 

\noindent{\bf Deterministic approaches.}
A straightforward solution is to extend the proof-based solutions to provide freshness verification. The key challenge of such extension is to ensure that the signatures attached to individual tuples are indeed constructed from the most up-to-date data. Then the client can verify the freshness of the query results from the constructed proof of the query results in the similar way as the existing proof-based methods, as long as the proof is constructed from the up-to-date signatures, for example, either by aggregated signature or by the tree-based ADS. 
Li et al. \cite{li2006dynamic} first raises the issue of query freshness. They provide a number of solutions, including publishing a list of revoked
signatures, including the time interval of validity as part of the signed message and reissuing the signature after the interval expires, and using hash
chains to confirm validity of signatures at frequent intervals.
They point out that any of these approaches can be applied directly on the tree-based solutions to update the single signature of the root of the tree. Each data update will require re-issuing one signature only.
Xie et al. \cite{Xie:2008:PFG:1353343.1353384} designs a scheme that associates the signature $S_t$ (e.g., the aggregated signature or the root signature of MHT tree)  with a certificate $Certificate_{C_A}(S_t, t)$ in the outsourced database, to indicate that the current signature at time $t$ is $S_t$. When a client retrieves
the query result, it also retrieves the signature and verifies if it is for timestamp $t$. Pang et al. \cite{Pang:2009:SVO:1687627.1687718} combined signature chaining with timestamps. 
The key idea is that the timestamps are embraced into the calculation of hash values of tuples. Thus the signature of a tuple $t_i$ is calculated as $\sigma_i=sign(H(i||t_i||t_{i-1}||t_i.ts))$, where $sign()$ is a signature function (e.g. RSA \cite{rivest1978method}), $H()$ is a collision-free hash function, and $t_{i-1}$ is the immediate left neighbor of $t_i$ along an ordering attribute, and $ts$ denotes the timestamp of the last update for tuple $t_i$. Based on the this signature scheme, the proof is constructed in the similar way as \cite{Narasimha2006}. The proof can be used to verify authenticity, soundness, completeness, and freshness. 
\nop{
For each tuple $t_i$, the data owner produces its signature as
\begin{equation}
\label{eq:Pang:2009:SVO:1687627.1687718_1}
\begin{split}
sig(t_i) = & sign(H(t_i.tid||t_i.A_1||\dots ||t_i.Am||t_i.ts||\\
& t_{i-1}.A_{Ind}||t_{i+1}.A_{Ind})),
\end{split}
\end{equation}
where $sign()$ is the signature function like RSA \cite{rivest1978method} or BGLS \cite{boneh2003aggregate}, $H()$ is a collision-free hash function, and $t_{i-1}$ and $t_{i+1}$ are the left and right tuples of $t_i$ along the indexing attribute $A_{Ind}$. %The data owner also creates a $B^+$-tree from the tuples and signatures. Each leaf node entry is in the format of $<t_i.A_{Ind}, sig(t_i), t_i.tid>$, while the internal nodes have the same form as standard $B^+$-tree \cite{comer1979ubiquitous}.
The data owner sends the signature of every tuple to the SP.
Consider a range query $\sigma_{A_{Ind}\in [l,u]}(R)$, the data owner collects the result $R^S=\{t_a, \dots, t_b\}$ and constructs the proof that consists of: (1) the aggregate signature $\sigma$ from all tuples in $R^S$ \cite{mykletun2006authentication}, and (2) $t_{a-1}.A_{Ind}$ and $t_{b+1}.A_{Ind}$, i.e., the indexing attribute values for the left and right boundary tuples. The authenticity verification of $R^S$ is performed by computing the hash values for every tuple $t_i\in R^S$, and comparing them against the aggregate signature $\sigma$ following Equation (\ref{eq:mykletun2006authentication_2}). 
The correctness verification is straightforward. The client only needs to check if for every tuple $t_i\in R^S$, $t_i.A_{Ind}\in [l,u]$. 
The completeness of $R^S$ is affirmed by checking if $t_{a-1}.A_{Ind}<l$ and $t_{b+1}.A_{Ind}>u$.
}

\noindent{\bf Probabilistic approaches.}
The key challenges to extend the checkpoint-based authentication methods to support freshness verification are as follows. First, the client must know what are the most up-to-date 
fake tuples in the outsourced database. This means that the 
client must be aware of all insertions and deletions of the fake tuples starting from the beginning. Second, it is important to ensure the fake tuple based scheme is provably secure, i.e., 
the attacker cannot distinguish fake operations from real
operations. To address these challenges, \cite{Xie:2008:PFG:1353343.1353384}  devise a mechanism to make all fake operations deterministic so that the client can derive the latest status of the fake tuples in the outsourced database. Specifically, the fake operations scheme is defined as $(FS,T,H)$, where $FS$ is a set of functions to generate the fake tuples, $T$ is a function to decide when to submit the fake operations and $H$ is a function to decide when a function in $FS$ is used to create fake tuples. The paper shows that the scheme is a provably secure scheme whose security can be reduced to the underlying encryption primitives. 

\nop{
%\cite{Xie:2008:PFG:1353343.1353384} requires the client to update the fake tuples frequently and should maintain some deterministic functions. Besides, it only provides probabilistic guarantee, 
\cite{Xie:2008:PFG:1353343.1353384} requires the client to update the fake tuples frequently. To improve the efficiency, Jiazhu et al. \cite{6047135} design a new method for freshness verification. The main idea is to insert  fake attribute into the tables and then verify the query result by the aggregation value on the fake attribute. In particular, given a table $T$ with attribute $\{A_1,A_2,...A_m\}$, the data owner inserts a fake attribute $A_f$ into table $T$, and make each tuple has different value on the fake attribute so that the server cannot distinguish the fake attribute from the real ones. The authors considered the database is encrypted, before outsourcing the database to the server, the data owner also has to create a cryptographic check value $sum$  based on all the fake values on each tuple:
$sum=k(f_1)+k(f_2)+k(f_3)+...+k(f_n)$, where $k$ is the key for data encryption which supports homomorphic summation. Then this check value is sent to a trusted third party for verification at  query time. For the verification, the client will send two requests search, one for the real query range and one for the summation of the fake values $sum'$ in that corresponding query range. The authors assume that the server cannot distinguish this two requests, meanwhile the client will get the summation value $sum$ for the corresponding fake values from trusted third party. At last the client verifies the completeness and freshness of the query result by checking if $sum=sum'$. Once there is a update, the data owner will add different fake attribute values into the new tuples and the generate a new cryptographic check value and also update the the check value at trusted third party. Because the trusted third party always has the most updated check value, then it guarantees that the server must send the correct sum value from the most updated table, otherwise it cannot pass the verification.
}
\nop{
$SUM = f_1$,$f_2$,$f_3,...$, $f_n$ into the
original tuples, and then encrypts the tuples
and creates an accumulated signature based on the values from the fake attribute by a homomorphic encryption. The data owner computes the cryptographic check value $sum=k(f_1)+k(f_2)+k(f_3)+...+k(f_n)$ and sends it to a trusted third party. The client sends the query request to both the SP and the third trusted party. The SP returns the query result for both real tuples and the fake tuples (assume SP cannot distinguish them) to the client, the client can verify the query result by comparing the accumulation value of the fake tuples to the check value from the trusted third party. If it matches, then proves the freshness and completeness of the query result. The solution support efficient dynamic update, since the deletion and insertion of fake tuple is simple operation, and only the check value has to be updated at the trusted third party.
}

\nop{
%Only one client can update 
assume there is only one client $C_A$ can update the database at time $t$ that results in a signature $S_t$. $S_t$ is certificated by the client and stored at the untrusted server. As a untrusted server, he/she may not update the database and return the obsolete query results when receive queries from other clients. In order to defend against such attack, the client who updates the signature needs to certificate $S_t$ every $\delta$ units of time no matter the signature has been changed or not. Thus $S_t$ is valid until $t+\delta$, which ensures a client can see all the updates that within the $\delta$ units of time. Obviously, small $\delta$ has high freshness guarantee, but high updates cost. If the client updates the data twice in $\delta$, the sever may also return outdated results to a client. 
%Multiple clients can update
If multiple clients can update the signature, the clients needs to provide a reference to the latest signature. For example, if client $C_A$ update the signature in the last $\delta$ units of time at $S_1$, and client $C_B$ updated again at $S_2$, the client $C_B$ need to certificate the signature $(S_2,S_1)$. In this case, other clients who get the query results will know which signature is most updated one.

As for those protocols that insert fake tuples into the database and provide probabilistic verification guarantee of the query result, the verification of the freshness follows the similar way. The authors define the fake insertion and deletion about the database, but the client can derive the status of the fake tuples in the database at anytime, so the freshness of the database can be checked by finding if some certain fake tuples exist in the query result or not. In specific, the fake operations scheme is defined as $(FS,T,H)$, where $FS$ is a set of functions to generate the fake tuples, $T$ is a function to decide when to submit the fake operations and $H$ is a functions to decide when a function in $FT$ is used to create fake tuples. According the scheme, the fake insertion and deletion will change the outsourced database in a deterministic way. So the client can always derive the the status of the fake tuples in the database and use those fake tuples to verify the query result as well as the freshness.

One drawbacks of the method for keeping the freshness of the tree-based ADS is that the client needs to store the data locally which against the tuition for outsourcing operation.
}
\vspace{-0.05in}
\section{Query Authentication with Other Security Features}
\label{sc:additionalsecurity}
Other security goals can be considered in parallel with query authentication in the DaaS framework. For example, how to efficiently authenticate query results without violating access control policies, and how to verify the integrity of results on privacy-preserving query evaluation. In this section, we review the existing works that integrate query integrity authentication with other security features. 
\vspace{-0.05in}
\subsection{Integrity and Privacy}
%\vspace{-0.1in}

Sarvjeet Singh et al. \cite{Singh:2008:ECO:1353343.1353402} consider a two-party scenario, in which the {\em data owner} has a private database, and the external querier would like to obtain some information from the private database by sending the queries to the data owner. The external querier desires to obtain a proof that shows the results returned by the database owner indeed reflect the correct evaluation of the submitted query over an uncorrupted version of the database.
On the other hand, the data owner does not reveal any private information to the querier except the query results. 
The goal is to design a query authentication method for private databases, aiming to minimize the amounts of data revealed to the untrusted third party, except the query results. The paper included several solutions with various
degrees of privacy, computational complexity and data
bandwidth. The recommended solution is to use Merkle tree for query authentication. The key idea is that the data owner constructs a Merkle tree, and sends the query results with a proof constructed from the Merkle tree. This can preserve database privacy as it reveals only the hashes of tuples that are not part of the result.  
The integrity of query results is verified by  reconstructing the root signature of the hash tree, similar to \cite{devanbu2003authentic}. 

Wang et al. \cite{wang2008dual} consider a three-party scenario that contains the data owner, the SP, and the client who is not necessarily to be the data owner. To protect the private data from the untrusted SP, the data is encrypted by the methods that support encrypted queries over encrypted databases. To provide query verification guarantee, they design a novel encryption method called {\em Dual Encryption}. The key idea of dual encryption is to encrypt the entire data $D$ using a primary
encryption key $k$, and encrypt a selected, small subset of
$D$ using a secondary encryption key $k'$.  The two encrypted datasets are merged and stored at the SP as a single
dataset.  The purpose of dual encryption is to allow for sophisticated cross examination on query results. 
Assume a client has a batch of queries $Q = \{q_1, \dots, q_u\}$. For each query $q_i \in Q$, the client sends its encrypted query $q_i^k$ to the SP and gets its result $\rho(q_i^k)$. To verify the integrity of $\rho(q_i^k)$, it sends $q_i^{k'}$ to the SP, where $q_i^{k}$ and $q_i^{k'}$ are semantically identical (i.e., they return the same set of records). The client analyzes both 
$\rho(q_i^k)$ and $\rho(q_i^{k'})$ by checking if for each replicated tuple $t\in \rho(q_i^k)$, $t$ also appears in $\rho(q_i^{k'})$. 
To help identify how the tuples are encrypted, the data owner attaches {\em dual
information} to each record, which is generated by secret key and one-way hashing. 
Regarding the security guarantee of the dual encryption schemes, 
the authors investigated two possible attacks, namely the {\em query correspondence attack} and the {\em distribution attack}. The query correspondence attack considers the case that if an adversary can find out that two encrypted queries $\rho(q_i^k)$ and $\rho(q_i^{k'})$ are indeed semantically identical, it can break the verification. To defend against this attack, the dual encryption scheme is revised by  requiring 
that $\rho(q_i^k)$ and $\rho(q_i^{k'})$ should not be identical but indeed have certain overlap. The distribution attack 
considers an adversary who may discover data correspondence with high probability by studying the distribution of query results. 
The solution to defend against this attack is to issue the checking query based on the answer size of the query to be verified. Only when the answer size is large enough the checking query can be issued.  

\nop{
The DO needs to build two different hash-trees. The first hash tree is to verify the authenticity of the query result, and the second the hash tree is to verify the correctness of the query result. The verification of the authenticity of the query result is to reconstruct the root signature of the hash tree which similar to as we discussed in paper \cite{devanbu2003authentic}. For the verification of correctness, they only focus the the equality selection and equality join query, therefore the client only needs to check if the hash values of the tuples matches VOs. Specifically, to verify if the selections were performed correctly, the client checks if $\Phi(A_j^i)=hash(a_j,||S_{i,j})$, where $\Phi(A_j^i)$ denotes the $i$th value from attribute $j$ and $S_{i,j}$ is the salt known to the client which aims to protect the privacy of the data. If $\Phi(A_j^i)=hash(a_j,||S_{i,j})$, then tuple $i$ should be present in the result. For the join queries, the client needs to check if the values from different tables that $\Phi(A_j^{i_1}=\Phi(A_j^{i_2})$, then tuples $i_1$ and $i_2$ are include in the result.
}
\vspace{-0.05in}
\subsection{Integrity and Data Confidentiality}
\label{sec:withac}
%\vspace{-0.1in}
Pang et al. \cite{pang2005verifying} is the first work that considers query authentication with the respect of the access control policy. They consider the completeness verification for selection-projection queries, with the assumption that the original selection-projection query has been re-written as a new selection-projection query. The challenge is to make sure that during the construction of completeness proof, the left and right
boundaries of the query results that are used for VO do not violate the access control policy. To address this challenge, 
the data owner inserts two fictitious entries, a left delimiter $r_0 \in (L, U)$ and a right delimiter $r_{n+1} \in (L, U)$ 
 into $D$.
The two fictitious entries ensures that the boundary records used in the proof do not violate the access control policy. For completeness verification, the digital signature of the real tuples in the query results are constructed. While for the two delimiters, their signatures are constructed by aggregating with the boundary records. Formally, 
\[sig(r_0) = s(h(h(L) | g(r_0) | g(r_1))),\]
\[sig(r_{n+1}) = s(h(g(r_n) | g(r_{n+1}) | h(U))).\]
Based on the signatures, the completeness verification is performed in the similar way as in \cite{pang2004authenticating}. 

Kundu et al. \cite{kundu2008structural} consider the scenario that the data organization structures encode sensitive information (such as in XML
documents). The goal is to provide both integrity and confidentiality not only for the content, but also for the structure.  The concept of confidentiality is interpreted as requiring a user to receive only those nodes and the structural information that the user is allowed to access,
according to the stated access control policies. A user
should not receive nor should be able to infer any information
about the content and presence of nodes and structural
information that the user is not authorized to access.
Based on this, \cite{kundu2008structural} formally defined the concept of {\em structure confidentiality}. The aim is to enable an user to authenticate a subtree $S$ from a tree $T$ without leaking any other nodes from $T−S$ ($−$ refers to the “cut” operation). They designed a novel concept called {\em structural signature}, which is based on the structure of the tree defined by tree traversals - postorder, pre-order and in-order. In particular, the structural signature of a node $x$ in tree $T$ consists of a hash of the structural position, the content of $x$, and the
(salted) signature of the tree, where the structural position  of a node is defined as a pair of randomized post-order
(RPON) and randomized pre-order (RRON). To share the tree with a third-party that has the authorization to access 
the subtree $S_z$, $S_z$ can be shared 
according to two different strategies: (1) by sharing the
signed subtree - its nodes and the structure or (2) by sharing
the signed nodes in the subtree and letting the third-party reconstruct
the subtree using the RPON’s and RRON’s of the nodes. The authors  considered both options for integrity verification, with the same goal as building the correct signature of the tree to show correctness. They provide the security analysis which shows the structural signatures do not lead to any leakage of (1) node signatures, and information about the (2) existence of nodes, (3) structural relations or (4) structural order among nodes.

\vspace{-0.05in}
\section{Conclusion and Future Work}
\label{sc:conclusion}
% summary
In this survey, we reviewed the existing studies on integrity authentication of  SQL query evaluation for the DaaS paradigm. We categorize these works by how the verification is performed and the integrity guarantee that these works can achieve. There exists the trade-off between the integrity guarantee and the performance of the verification methods. The deterministic integrity guarantee can be achieved but with possibly expensive  setup cost and verification overhead; such overhead can be improved by relaxing the integrity guarantee to be probabilistic. 

%: the flexibility for diverse queries, the verification objective, the fundamental methods, and the support for dynamic databases. These are the decisive criteria to evaluate the applicability and effectiveness of an authentication system. We present our analysis and comparison of existing work in a systematic way.

In the rest of this section, we discuss several interesting
directions for the future work.
\nop{
\noindent{\bf Design of hybrid approaches.}
The existing works show that the simple solutions (e.g., tuple signatures only \cite{pang2005verifying}) effectively verifies the authenticity of results returned by the server. The incorporation of authenticated data structure \cite{cheng2006authenticating} remarkably promotes the efficiency of range query verification, while the fusion of accumulation values \cite{zhang2015integridb} enables constant aggregation verification time complexity.
%flexibility
On the other hand, there are still urgent demands for a practical authentication approach that can efficiently verify various query operations. So far, \cite{vSQL,zhang2015integridb,ghosh2016efficient} are the only three approaches that can provide deterministic integrity guarantee for range, join and aggregation queries. However, none of them is satisfying from every perspective. In particular, \cite{vSQL} only verifies limited queries; \cite{zhang2015integridb} incurs prohibitive verification preparation cost at the data owner side. Moreover, it does not work well for duplicated data values. \cite{ghosh2016efficient} can only verify one-dimensional range queries, plus it does not work on dynamic databases.
Considering the popularity of cloud database solutions and the severe diminishing public trust on remote database service, it is essential to make improvement over the state-of-art to support efficient verification of flexible queries. 

%authentication of distributed query processing

%privacy-preserving authentication
}
\vspace{-0.05in}
\subsection{Authentication of Query Evaluation on Big Data}
%\vspace{-0.1in}
The three well-known properties of Big data is 3Vs (volume, variety and velocity). Volume refers to the amount of data, variety refers to the number of types of data and velocity refers to the speed of data processing. 
The three properties raise different challenges to the authentication of query evaluation. 
The big data volume requires that the authentication methods must be scalable; the setup and verification should not add significant overhead on query evaluation itself. 
The fast velocity requires that the authentication methods must support frequent updates efficiently. 
The large variety  requires that the authentication methods should be able to support various data types. A naive method is to construct an ADS for each data type, and thus a proof for each data type too. But this may bring overwhelming verification overhead. An interesting research direction that worths to explore is how to address the challenges of volume, variety and velocity of Big data for the design of efficient and scalable authentication solutions. 

MapReduce has emerged as an important computational
model for data intensive applications. The MapReduce infrastructure consists
of two primitives: (a) a {\em map} function that distributes the input
to multiple mapper nodes to generate intermediate key-values
in parallel, and (b) a {\em reduce} function that allows multiple
reducer nodes to merge the intermediate pairs associated with
the same key and then to generate the final outputs. 
Intuitively, all mappers and reducers can be compromised
and return wrong intermediate and/or final results.
Therefore, cloud-based query evaluation raises two main challenge: (1) how to authenticate the query results in the format of key-value pairs; and (2) how to authenticate both the {\em intermediate} and {\em final} results for MapReduce execution. 
A few papers have considered integrity verification of MapReduce execution \cite{6008723,7363785,ghosh2016efficient}. Intuitively, the intermediate results can be authenticated by applying the existing solutions on each mapper. However, this requires to construct an ADS independently for each mapper, which can be extremely expensive. The main challenge is how to address the verification overhead issue for the query results in a distributed fashion. In particular, how to construct the ADS to support efficient verification of both intermediate query results by the mappers  and final query results by the reducers. 
\vspace{-0.05in}
\subsection{Authentication for Various Outsourcing Paradigms}
%\vspace{-0.1in}

In general, the outsourcing paradigms can be classified into three types:
\begin{itemize}
\item {\em Infrastructure-as-a-Service (IaaS)} paradigm: 
the data owner can deploy and run arbitrary software including operating systems and applications.  
The SP provides the computing infrastructure, e.g., data storage and hardware for the
computation. A typical IaaS example is Amazon EC2
Web Service \cite{amazon}. 
%The client configures the data analytics settings as the input for the computations. 
\item {\em Platform-as-a-Service (PaaS)} paradigm:  The SP delivers a computing platform, typically including operating system, execution environment, database, and web server. Some well-known PaaS offers include Microsoft Azure \cite{windowsazure} and Google Cloud SQL \cite{googlecloudsql}. 
 \item {\em Software-as-a-Service (SaaS)} paradigm: The clients use the SP's applications running on an outsourced infrastructure (e.g., the cloud). The applications are accessible from various client devices through either a thin client interface, such as a web browser (e.g., web-based email), or a program interface. 
\end{itemize}

Different computational paradigms bring different challenges of authentication. For example, compared with the IaaS paradigm, the SP of Paas and SaaS paradigms has more cheating power on the computations, as the client has neither knowledge of the computations nor how these computations are configured and executed. Therefore, the verification of result integrity for the SaaS and PaaS paradigms is more challenging, especially for those computations whose outputs highly depend on the setup of the computations (e.g., the initial centroids of $k$-means clustering \cite{liu2013integrity}). 

\vspace{-0.05in}
\subsection{Authentication of Various Types of Outsourced Computations}
This survey focuses on SQL query evaluation only. In practice, the outsourcing paradigm provides a broader range of data analytics services, ranging from simple aggregation to Web search to sophisticated data mining and learning, to name a few. Below we present a few examples of outsourced computations, and discuss the challenges and potential research directions for the verification of these types of outsourced computations. 

{\bf Keyword search on Web.} 
Web search engines are typical outsourced computations. The search service providers may return incorrect search results due to various reasons (e.g., to make profits from advertisements) \cite{goodrich2012efficient}. Furthermore, the service providers may return the results in a wrong ranking (e.g., to promote some websites). Therefore, the authentication goals are two-fold: (1) to verify that the returned keyword search results are sound, complete, and fresh, and (2) to verify the ranking of the returned results is accurate. \cite{goodrich2012efficient,nguyenverification} have initiated the research on authentication of soundness, completeness, and freshness of the Web-search content. How to verify the ranking of the results is still unsolved. Verification of top-k query evaluation has been studied under different contexts, e.g., for sensor networks \cite{zhang2010verifiable} and location-based services \cite{chen2013authenticating}. It will be interesting to explore if these top-k query authentication methods can be adapted to keyword search queries.  

{\bf Outsourced data mining and machine learning.}
Big corporations like Amazon, Google and Microsoft are providing cloud-based data mining and learning services in various forms. Amazon Web Services (AWS) provides
computation capacity and data storages via Amazon Elastic Compute Cloud (EC2) \cite{amazon} and Simple Storage
Service (S3) \cite{amazons3}. Google provides cloud machine learning engines \cite{googleml} for various machine learning applications, including customer purchase
prediction and spam detection, to name a few. Microsoft provides big data analytics  services 
 on Windows Azure cloud \cite{windowsazure}. This raises the issue of how to evaluate the result integrity of these outsourced machine learning computations. There are several interesting research questions. First, can any existing ADS structure (e.g. MB-tree \cite{li2006dynamic} and VB-tree \cite{pang2004authenticating}) be adapted to the authentication solutions for outsourced machine learning and data mining computations? Second, given the high complexity of machine learning computations, the deterministic approach may not be appropriate as it can bring expensive overhead. On the other hand, it is not straightforward how to adapt the existing probabilistic approaches \cite{xie2007integrity} to the outsourced machine learning computations, since inserting some fake tuples can change the machine learning results (e.g., classification and clustering) of the original data. The challenge is how to design those fake tuples for authentication while keeping the original data analytics results unchanged. 
%\Wendy{Boxiang, can you write a paragraph summarizing all of our work on authentication of data mining here?} 
The existing authentication solutions include various types of data mining computations, including association rule mining \cite{dong2013result,dong2016trust,dong2013integrity}, outlier mining \cite{dong2017authenticated,liu2012audio}, clustering \cite{liu2013integrity}, Bayesian network construction \cite{liu2014result}, and collaborative filtering \cite{vaidya2014efficient,tang2016protect}. Many of these solutions (e.g. \cite{liu2012audio,liu2013integrity,liu2014result})  consider the probabilistic approaches. One of the weaknesses of these approaches is that they may be vulnerable against the attacker who is aware of the authentication mechanism, in particular, how the fake data points are constructed. Then the attacker can distinguish the fake data points from the real ones. An interesting research direction is to design robust probabilistic verification methods that can address the trade-off between efficiency and the robustness of the authentication methods. 

{\bf Spatial query evaluation.} 
A large body of work has considered query evaluation on spatial databases (e.g. 
\cite{jing2014authentication,papadopoulos2011authenticated,hu2013spatial,hu2013verdict,yung2012authentication}). The goal of authentication is to verify if the returned locations satisfy the given constraints (e.g., within a range, one of the nearest neighbor, etc.). The main challenge of spatial query authentication is how to compute the distances efficiently for verification. Various techniques (e.g., Voronoi diagram and R-trees) are used to represent the underlying spatial data set. The remaining technical issues to be resolved is how to design efficient, {\em online} authentication methods that can deal with dynamic location data that are updated frequently. This requires that the authentication methods can support fast updates on the ADS, quick construction of the proof of the query results, and cheap verification at the client side who may use resource-constrained devices such as mobile phones for verification.

\section{Acknowledgements}
This material is based upon work supported by the National Science
Foundation (NSF) under Grant No. 1350324 and 1464800.

%\input{others}

%\input{relatedwork}

%\Wendy{Make the format of all reference items to be consistent}
\bibliographystyle{plain}
\bibliography{bib}

\vspace{-0.6in}
\begin{IEEEbiography}
[{\includegraphics[height=1in]{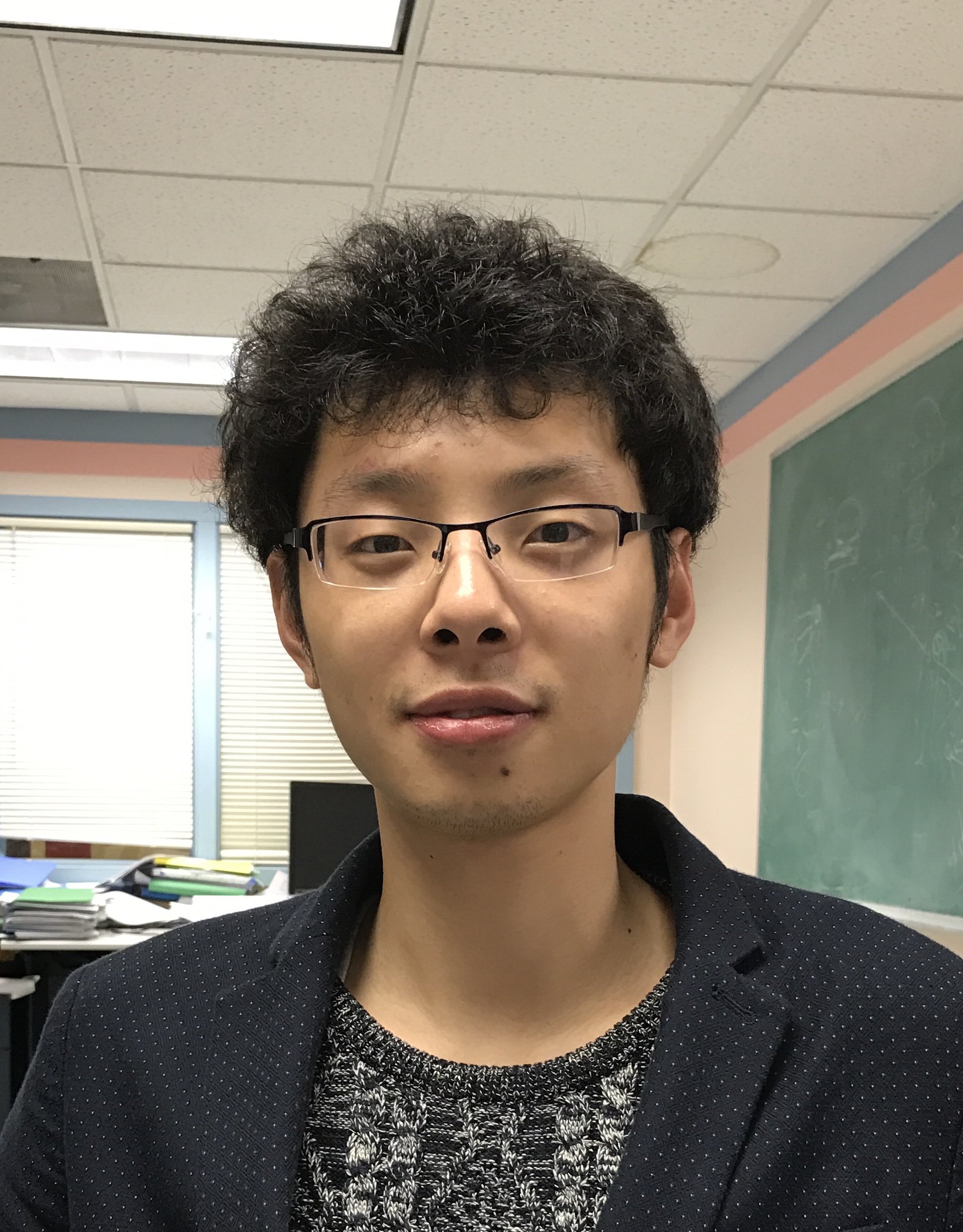}}]{Bo Zhang} was born in China in 1992. He received the B.S. from Wuhan University of Technology of China, Hubei, China.
Currently he is a Ph.D. student of Stevens Institute of Technology, Hoboken, NJ, USA since 2015.
His research interests include database, verifiable computation and data mining.
\end{IEEEbiography}
\vspace{-0.65in}

\begin{IEEEbiography}[{\includegraphics[height=1in]{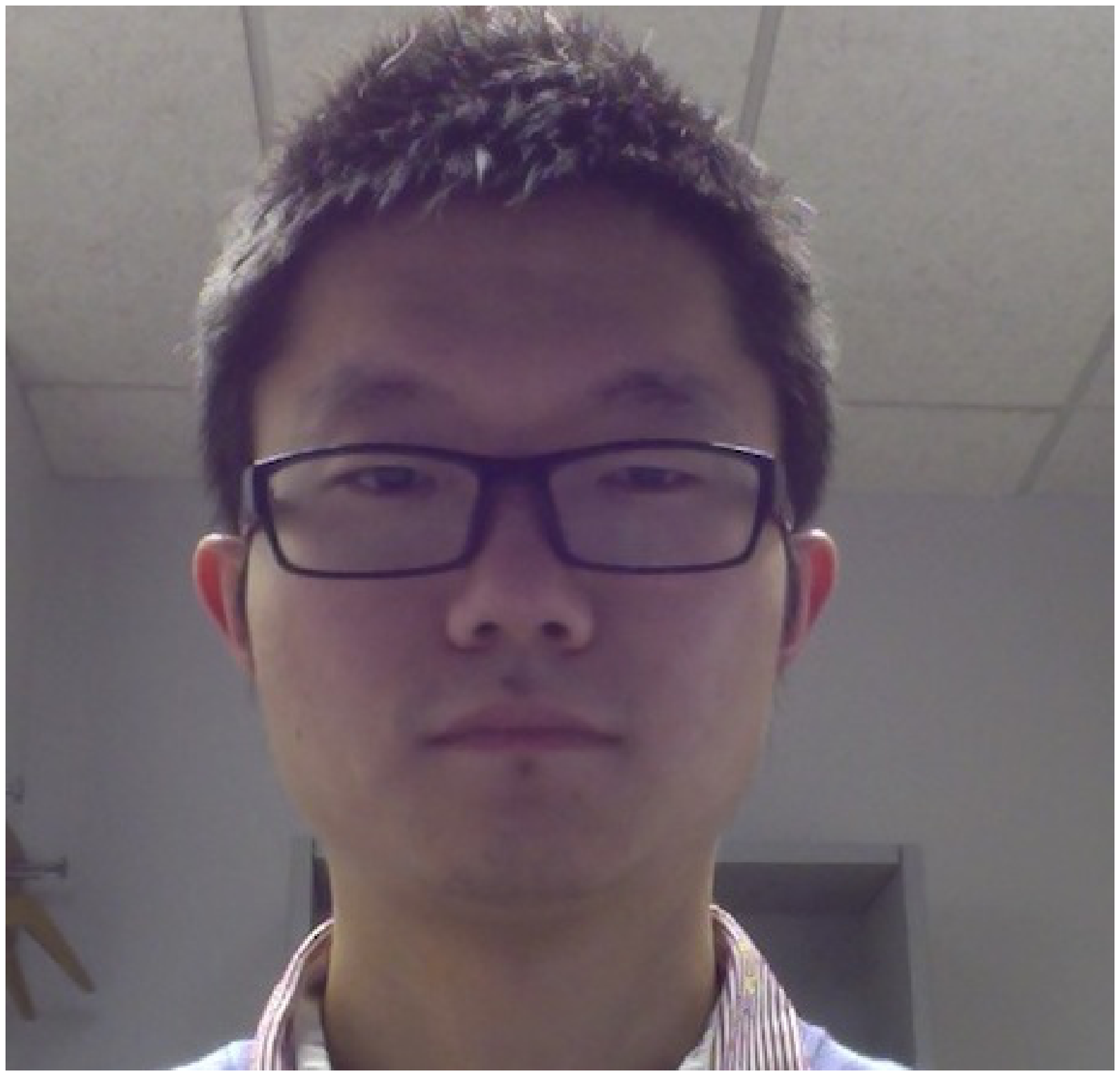}}]{Boxiang Dong}
is an assistant professor in the Computer Science Department, Montclair State university, New Jersey. He received his Ph.D degree in computer science from Stevens Institute of Technology, New Jersey.
He is dedicated to facilitating the integration of big data analysis and cybersecurity.
His research interests include verifiable computing, data mining, anomaly detection, data security and privacy.
\end{IEEEbiography}
\vspace{-0.6in}

\begin{IEEEbiography}
[{\includegraphics[height=1in]{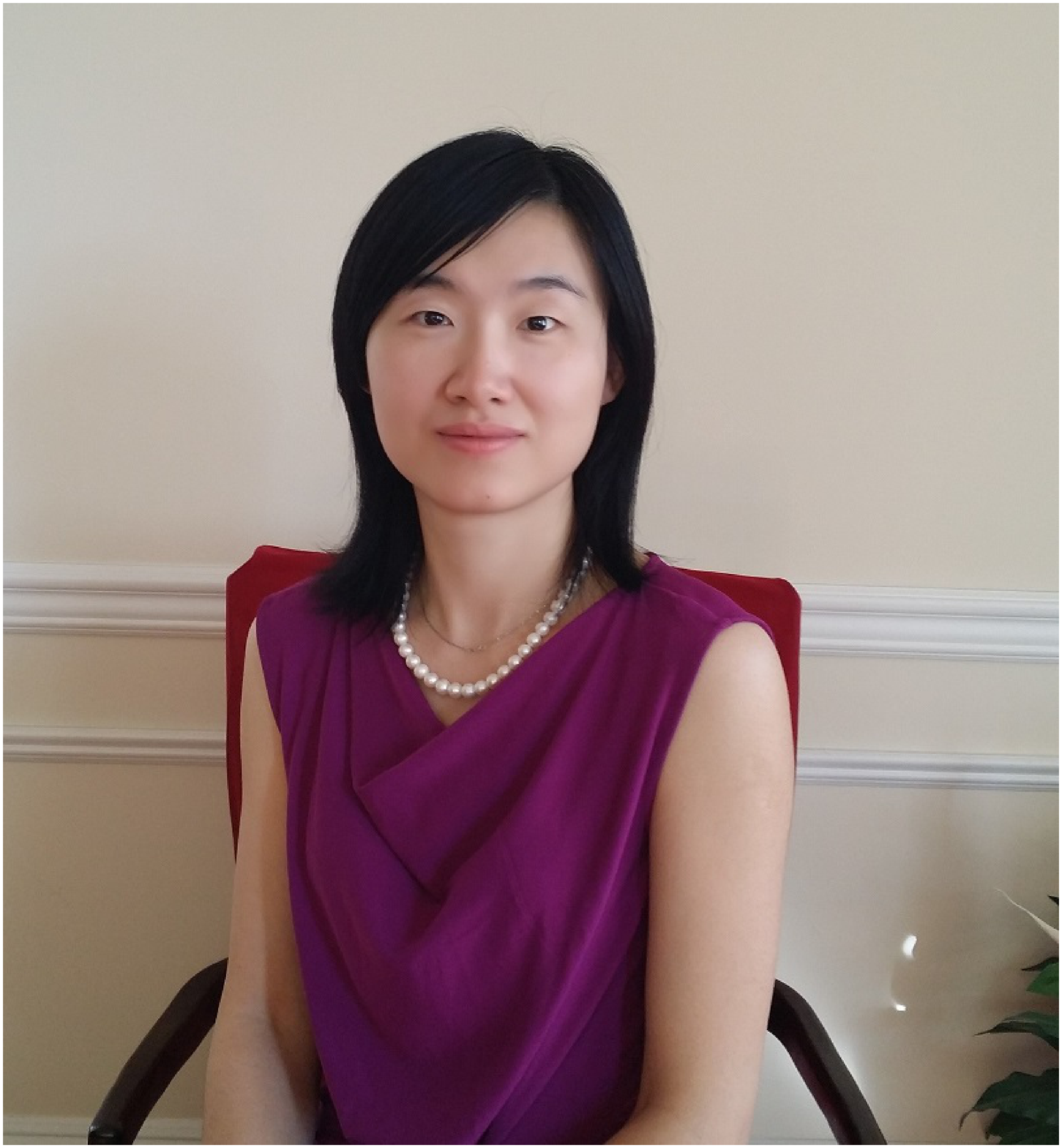}}]{Dr. Wendy Hui Wang} 
is an associate professor in the Computer Science
Department, Stevens Institute of Technology, New Jersey. She received
her PhD degree in computer science from University of British
Columbia, Vancouver, Canada. Her research interests include data
management, data mining, database security, and data privacy. She is a
member of the editorial boards of Journal of Information Technology
and Architecture and Journal of Information Systems.
\end{IEEEbiography}

\end{document}